\documentclass[twocolumn,twocolappendix]{aastex63}
\usepackage{amsmath,amstext,natbib,comment}
\usepackage{color,soul}
\graphicspath{{figures/}} 
\graphicspath{figures}

\shorttitle{}
\shortauthors{Su et al.}

\begin{document}

\title{Imaging of the Vega Debris System using JWST/MIRI}

\author[0000-0002-3532-5580]{Kate Y.~L.~Su}
\affiliation{Steward Observatory, University of Arizona, 933 N Cherry Avenue, Tucson, AZ 85721--0065, USA}
\affiliation{Space Science Institute, 4750 Walnut Street, Suite 205, Boulder, CO 80301, USA; ksu@spacescience.org}

\author[0000-0001-8612-3236]{Andr\'as G\'asp\'ar}
\affiliation{Steward Observatory, University of Arizona, 933 N Cherry Avenue, Tucson, AZ 85721--0065, USA}

\author[0000-0003-2303-6519]{George H.\ Rieke}
\affiliation{Steward Observatory, University of Arizona, 933 N Cherry Avenue, Tucson, AZ 85721--0065, USA, also Department of Planetary Sciences}

\author[0000-0002-1226-3305]{Renu Malhotra}
\affiliation{Lunar and Planetary Laboratory, The University of Arizona, Tucson, AZ 85721, USA}

\author[0000-0003-4705-3188]{Luca Matr\'a}
\affiliation{School of Physics, Trinity College Dublin, Dublin 2, Ireland}

\author[0000-0002-9977-8255]{Schuyler Grace Wolff}
\author[0000-0002-0834-6140]{Jarron M.\ Leisenring}
\affiliation{Steward Observatory, University of Arizona, 933 N Cherry Avenue, Tucson, AZ 85721--0065, USA}

\author[0000-0002-5627-5471]{Charles Beichman}
\author[0000-0001-7591-2731]{Marie Ygouf}
\affil{Jet Propulsion Laboratory, California Institute of Technology, Pasadena, CA}

\correspondingauthor{Kate Su}
\email{ksu@spacescience.org}

\accepted{accepted for publication in ApJ}

\begin{abstract}

We present images of the Vega planetary debris disk obtained at 15.5, 23, and 25.5 $\mu$m with the Mid-Infrared Instrument (MIRI) on JWST. The debris system is remarkably symmetric and smooth, and centered accurately on the star. There is a broad Kuiper-belt-analog ring at $\sim$80 to $\sim$170 au that coincides with the planetesimal belt detected with ALMA at 1.34 mm. The interior of the broad belt is filled with warm debris that shines most efficiently at mid-infrared along with a shallow flux dip/gap at 60 au from the star. 
These qualitative characteristics argue against {\it any\ } Saturn-mass planets orbiting the star outside of about 10 au assuming the unseen planet would be embedded in the very broad planetesimal disk from a few to hundred au. We find that the distribution of dust detected interior to the broad outer belt is consistent with grains being dragged inward by the Poynting-Robertson effect. Tighter constraints can be derived for planets in specific locations, for example any planet shepherding the inner edge of the outer belt is likely to be less than 6 Earth masses. The disk surface brightness profile along with the available infrared photometry suggest a disk inner edge near $\sim$3$-$5 au, disconnected from the sub-au region that gives rise to the hot near-infrared excess. The gap between the hot, sub-au zone and the inner edge of the warm debris might be shepherded by a modest mass, Neptune-size planet.

\end{abstract}


\section{Introduction} 
\label{sec:intro}

Vega was the first-discovered and one of the prototypical planetary debris disks. It opened a broad field of study that now is being used to identify relatively low mass exoplanets beyond the reach of other discovery techniques, as well as to reveal detailed properties of the systems of small bodies in other planetary systems. Debris studies are an important component of our efforts to understand the formation and evolution of planetary systems \citep[e.g.,][]{wyatt_jackson16,hughes18,chen_su_xu20,najita22}.

The Vega debris disk was discovered when IRAS observed Vega for calibration and found it is too bright at 60 $\mu$m by a factor of 15. The temperature associated with the dust responsible for this additional emission was around $\sim$85 K, analogous to temperatures at the inner edge of the Kuiper Belt. IRAS found hundreds of debris disks, and the number has only grown with the contributions of additional cold space telescopes, such as ISO, Spitzer, Akari, Herschel, and WISE. 

Subsequent observations have demonstrated that debris disks are not simply a far infrared phenomenon \citep{hughes18}. IRAS and subsequent observations of the integrated emission of disks (i.e., without resolving them) showed that many have warmer dust, which was attributed to structures analogous to our Asteroid Belt \citep[e.g.,][]{su13}. Although previous space infrared telescopes were too small to image disks in detail, JWST has begun doing so at resolutions that resolve them well, e.g., \citet{gaspar23_fom}, showing that the asteroid-belt analogy needs significant revision.  

The well-resolved infrared images that are now becoming available complement images at other wavelengths to help provide a comprehensive understanding of the debris disk phenomenon. A prelude to the resolved images on many systems in scattered light was the pioneering image of $\beta$ Pic by \citet{smith84}.  Spectacular results on the morphology of disks and its variety have been revealed in high resolution scattered light images from HST \citep[e.g.,][]{schneider14}, JWST \citep[e.g.,][]{lawson23} and with adaptive optics from the ground \citep[e.g.,][]{esposito20}. ALMA has opened up the mm-wave for resolved images of debris disks, which reveal the locations of the ``parent'' bodies (which need be no larger than $\sim$ a millimeter) whose motion is dominated by gravity and whose collisions produce the dust seen at shorter wavelengths \citep[e.g.,][]{macgregor17_fom, matra19b,booth23_epseri_clumps}. All of these studies  provide insights to exoplanetary systems not accessible by any other means, serving as a powerful complement to the detection of exoplanets by transits, radial velocity effects, or even direct imaging. 

Despite its iconic status, the Vega disk has been left behind. The infrared images around Vega have resolutions at best of only $\sim$ 6\arcsec\ \citep{su05, sibthorpe10}. Until very recently there was no deep search for scattered light \citep[but see][]{wolff24}, in part because the face-on aspect of the disk implies low scattering efficiency. The low elevation of Vega seen from the ALMA site has compromised the image with this telescope \citep{matra20}; an image with the Large Millimeter Telescope (LMT) \citep{marshall22_vega_lmt} complements the ALMA image, but its resolution is insufficient for detailed analysis. 

MIRI on JWST has a beam area at 25 $\mu$m nearly 50 times smaller than the MIPS on Spitzer, changing the situation dramatically. Many nearby debris disks can now be imaged well in the infrared. For example, adopting a typical outer ring radius of 75 au \citep{hughes18} and a distance of 20 pc, MIRI at 21 $\mu$m has ten diffraction-limited beams across the diameter of the ring, or $\gtrsim$ 50 resolution elements inside it. There are two dozen known debris disks within this distance; although not all are amenable to detailed observation (e.g., faint disk emission in contrast to the bright stellar photosphere in the mid-infrared), the potential is substantial to achieve resolved images of the warm dust in a representative sample. The most attractive targets are Fomalhaut, Vega, and $\epsilon$ Eridani, at distances of 7.7, 7.7, and 3.2 pc respectively, each with very prominent debris disks that are bright at 24 $\mu$m \citep{hughes18}. This paper is part of a JWST GTO program (PID 1193) to image these systems with both NIRCam and MIRI in detail. The first target was Fomalhaut, with results published in \citet{gaspar23_fom}.  Vega is the second target in our study, with this paper on the MIRI results and an accompanying one in those with NIRCam (Beichmen et al., 2024, submitted).

The Vega system is of particular interest to compare and contrast with the Fomalhaut system, given the similarity of the stars, that they are at nearly the same distance (i.e., the physical resolution on the systems will be the same) and the general similarity of their debris disks based on the available observations. Vega is also of interest because its pole-on orientation results in very poor limits on planets using the radial velocity technique \citep{hurt21}, so probing the potential effects of unseen planets on the debris disk may be the best available way to learn about any planetary system around the star. 

In Section \ref{sec:data}, we briefly describe the JWST/MIRI observations, highlight the necessary steps to produce the final disk-only images but we defer discussing the details to Appendix \ref{fy_section}. The general disk morphology is described in Section \ref{sec:disk_morphology}, in a broader context with other relevant, multi-wavelength data obtained previously with Herschel and ALMA. In Section \ref{sec:analysis}, we detail the disk properties in terms of the radial distributions of surface brightness (\ref{sec:radprof}), temperature and optical depth distribution (\ref{sec:tc_tau}), and quantify the dip/gap properties (\ref{sec:dip}) and the potential asymmetric structure (\ref{sec:azmuthal}), and then constrain the properties of the inner few au zone where the MIRI images cannot directly probe (\ref{sec:inner_radius}). We discuss the insights that the high-resolution MIRI images can provide for the unseen planets in the Vega system, particularly focusing on the dust dynamics by balancing the two competing mechanisms between collisions and Poynting-Robertson (PR) drag (\ref{sec:pr_win}) and highlighting the tight constraints in the planet mass for a PR-dominant disk (\ref{sec:planet_perturbations}). A conclusion is given in Section \ref{sec:conclusion}.

\begin{figure*}
    \centering
    \includegraphics[width=\linewidth]{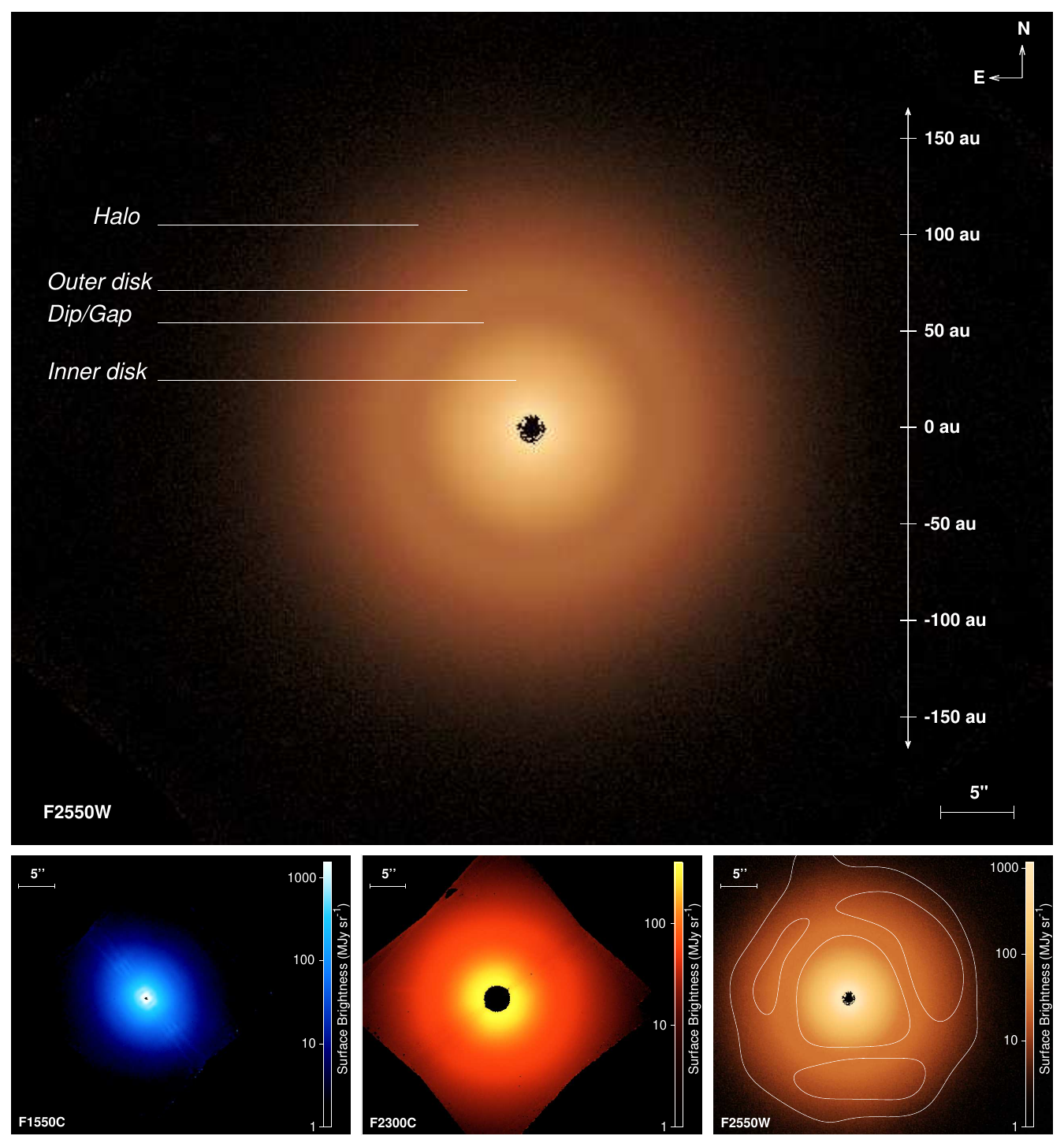}
    \caption{The bottom row shows the three MIRI images of the Vega debris system at the same orientation and field of view.  Slight artifacts from the coronagraph persist in the 15.5 $\mu$m image, while it and the 23 $\mu$m image have the central region eliminated by the coronagraphs; the 25.5 $\mu$m image is missing the core because of saturation. The millimeter disk emission detected by ALMA \citep{matra20} is superimposed as contours on the 25.5 $\mu$m image for comparison. The top panel is the enlarged disk image at 25.5 $\mu$m with the major disk features labeled for reference as they are discussed further in this paper. These features are: (1) the halo, visible at 25.5 $\mu$m to a radius of $\sim$33\arcsec\ ($\sim$250 au); (2) the outer disk (Kuiper-belt analog) extending from $\sim$10\arcsec\ to 22\arcsec\ ($\sim$78 to $\sim$170 au) and the dominant feature in the ALMA image at 1.34 mm; (3) the inner disk extending from the inner edge of the outer disk to as close as the images penetrate to the star; and (4) the dip in surface brightness of the inner disk from $\sim$5\arcsec\ to 10\arcsec\ ($\sim$40$-$78 au). }
    \label{fig:miri_pr}
\end{figure*}

\section{Observations and Data Reduction}
\label{sec:data}

Observations of Vega were obtained on August 18, 2023 with the MIRI instrument \citep{rieke15_miri,wright23_miri} on board JWST \citep{gardner23_jwst} as part of the GTO program PID 1193. Three filters (F2550W, F1550C and F2300C) were used in the imaging and coronagrapic modes for characterizing the debris disk structure around Vega with sub-arcsec resolution. To remove the stellar contribution in the data, a PSF reference star, HD 169305, was also observed in the same modes for PSF subtraction.  The raw data were processed and calibrated with the JWST Science Calibration pipeline \citep{bushouse23_jwstpipeline} and some custom tools for the PSF subtraction. Details about the data set (modes, filters, and integration) are given in Table \ref{tab:JWSTobs}. Reduction of the data to produce final images, including PSF subtraction, required a long series of detailed and demanding steps. We have captured this procedure as a reference for future work in Appendix \ref{fy_section}. The final products are star-free, disk images shown in Figure \ref{fig:miri_pr}. 

Given the slow degradation of MIRI imaging detector response at the long wavelengths, we double checked the current flux calibration using ancillary infrared photometry obtained for the whole Vega system as detailed in Appendix \ref{sec:appendix_photometry}. We confirm that the flux calibration uncertainty is within a few \% using the JWST Science Calibration pipeline version 1.12.3 and the reference files of {\tt jwst\_1193.pmap}. Compared to the infrared photometry obtained in the apertures that are larger than the field of view of our MIRI disk images, we also confirm that the MIRI disk images do not miss   any significant level of low surface brightness emission (see Apendix \ref{sec:appendix_25um}), but there is a small amount of missing flux within $\lesssim$1\arcsec\ from the star due to the coronagraphic mask of F1550C and the saturation in the F2550W filter. The missing central flux was further estimated in Appendix \ref{sec:appendix_constraints_inner_disk} for both the F1550C and F2550W images but not F2300C because of its large inner working angle.

\section{General Disk Morphology}
\label{sec:disk_morphology}

The MIRI disk images (Figure \ref{fig:miri_pr}) have a resolution element of $\lesssim$1\arcsec\ and the debris emission can be traced to the edge of the field of view for all the instrumental modes used. For a better perspective view on the debris distribution in the millimeter, we also show the ALMA observations of the cold, outer belt (the 10\arcsec\ tapered image from \citealt{matra20}) as the overlaid contours in Figure \ref{fig:miri_pr} for comparison. 

The F2550W disk image reveals that the emission is very extended (up to 33\arcsec\ at 25.5 $\mu$m, i.e., $\gtrsim$250 au from the star), azimuthally smooth and axis-symmetric. There is only a weak inflection near the position of the inner edge of the cold outer belt. The most noticeable feature in this image is the dip/gap-like structure at $\sim$8\arcsec\ ($\sim$60 au) away from the star.  The dip/gap is also visible in the F2300C image but is much less pronounced at F1550C. The properties of the dip will be discussed  in Section \ref{sec:dip}. Unlike all three MIRI images, the ALMA results (particularly the high-resolution radial profile) show no significant levels of millimeter emission interior to the outer broad belt, which extends from 9\farcs6 to 22\farcs2 in radius and peaks at 15\farcs3 \citep{matra20}.

To quantify how circular the disk emission appears to be, four different levels of contours on the F2550W disk image were selected for the orbital fitting similar to what has been done for the MIRI Fomalhaut data \citep{gaspar23_fom}. These four contours represent disk radii of 3\farcs4 (region in the inner disk), 6\farcs4 (region near the inner edge of the dip/gap), 18\arcsec\ (region near the peak of the 1.34 mm disk emission), to 22\arcsec\ (at the outer radius of the 1.34 mm disk emission). The results are reported in Table \ref{tab:orbitfits}.  The disk is truly circular with eccentricity $e \lesssim$0.002. In addition, no offset between star and disk was found during the sub-pixel shifting in the PSF subtraction during data reduction. Taking the outer eccentricity of $\sim$0.002 at face value, the inclination of the disk is $i\sim$ 7\arcdeg$-$11\arcdeg. Given a stellar inclination of 6.2\arcdeg\ determined from the ground-based, optical interferometric data \citep{monnier12_chara}, it appears that the disk is closely perpendicular to the stellar rotational axis, a phenomenon that is found in $\sim$80\% of the debris disks \citep{hurt_macgregor23}.

\begin{table}
    \caption{25 $\mu$m disk surface brightness contour fits}
    \centering
    \vspace{-4mm}
    \begin{tabular}{cccc}
    \hline
    \hline
    contour level  & $a$ & $e$ & $i$ \\
    (MJy/sr) &   (\arcsec / au) &  & (\arcdeg) \\
    \hline
    316  & 3.4 / 25.8  & 0.00024$\pm$0.033 & 7\arcdeg$\pm$3\arcdeg\\
    100  & 6.4 / 49.1  & 0.00014$\pm$0.011 & 8\arcdeg$\pm$1\arcdeg\\
    10  & 18.8 / 144  & 0.0018$\pm$0.0002 & 9\arcdeg$\pm$0.2\arcdeg \\
    5  & 21.8 / 168  & 0.0021$\pm$0.0002 & 11\arcdeg$\pm$0.1\arcdeg\\
    \hline
   \end{tabular}
   \\
   
   $a$: semi-major axis, $e$: eccentricity, and $i$: inclination.
  \label{tab:orbitfits}
\end{table}

Not only is the disk very circular in the mid-infrared, but in Section~\ref{sec:azmuthal} we also show that the disk is very azimuthally symmetric. This means that the disk radial profiles lose very little spatial information, and we can enhance the signal-to-noise as well as making features stand out more clearly by radial averaging. A similar approach has been taken by \citet{matra20} to show a high resolution profile at 1.34 mm, with an averaged beam width of 1\farcs35, i.e., similar in resolution to our MIRI images.  Much of the rest of this paper will therefore be based on radial profile analysis.

It is interesting to compare the Vega disk with that of Fomalhaut: the two stars are of similar spectral type, at virtually identical distances, have luminous debris disks of very similar spectral energy distributions (SEDs), which were interpreted as indicating very similar debris disk structures on the basis of Spitzer and Herschel $\sim$6\arcsec\ images \citep{su13}. Well resolved MIRI images destroy this simple picture (see \citet{gaspar23_fom} for Fomalhaut). Both disks have significant mid-infrared emission from inside of their outer planetesimal belt, but for Fomalhaut this mid-infrared emission is highly structured whereas in Vega it is very smooth. The Fomalhaut disk is clearly offset from the star; the Vega system is well-centered. The outer planetesimal belt around Fomalhaut is narrow (a width of $\sim$13.5 au, \citealt{macgregor17_fom}), whereas for Vega it is very broad, $\sim$70 au \citep{matra20}. Comparing these systems is thus a fascinating exercise not only in understanding each one, but in probing the degeneracies in SED modeling.

In summary, JWST's superior resolution and sensitivity resolve the inner debris emission around Vega and reveal a dip/gap structure at $\sim$60 au for the first time. The details revealed by the MIRI images support much more insightful models than were possible previously.

\section{Analysis}
\label{sec:analysis}

\begin{figure}
    \centering
    \includegraphics[width=\linewidth]{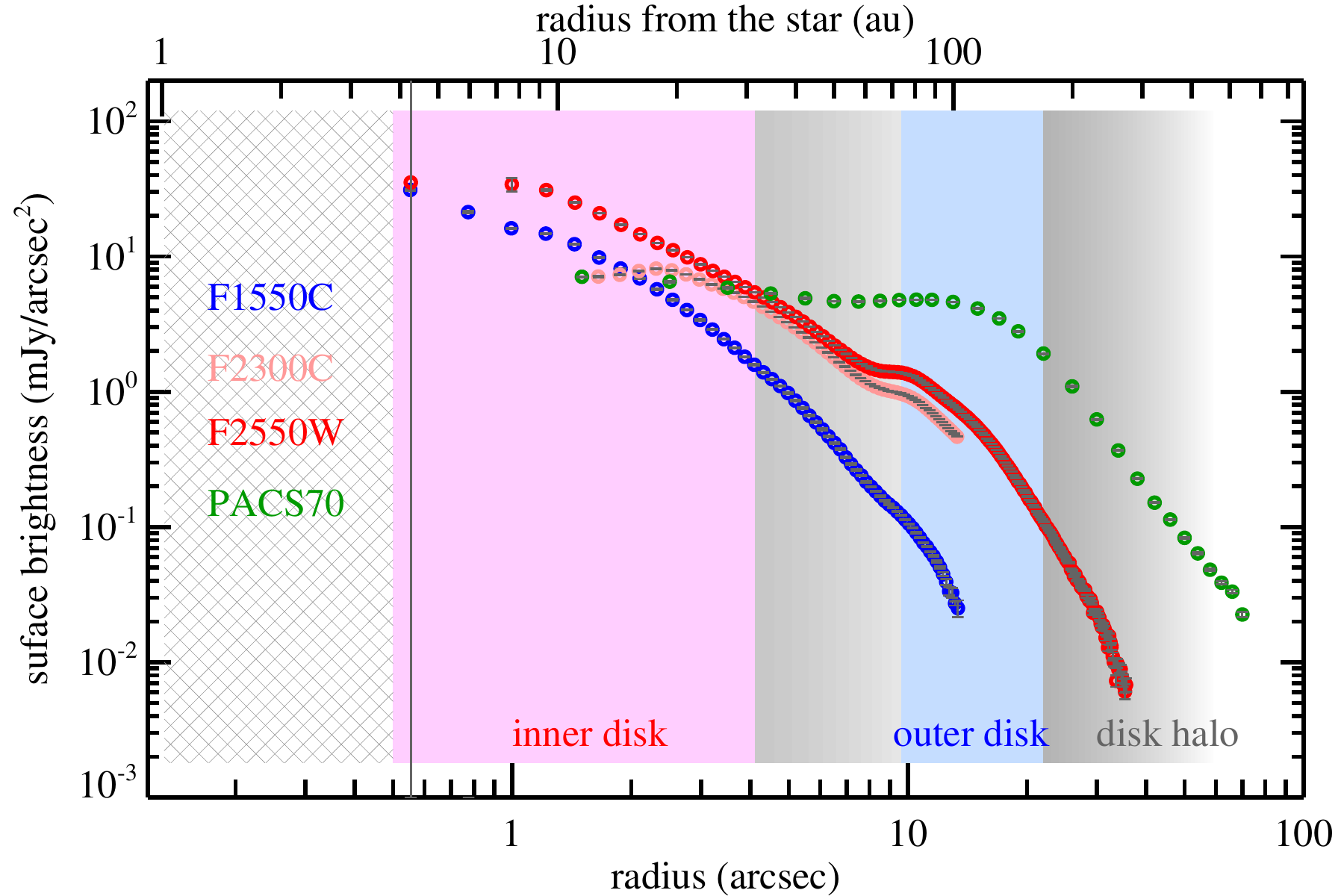}
    \caption{Vega disk surface-brightness radial profiles observed at four different wavelengths, F1550C (15.5 $\mu$m), F2300C (23 $\mu$m) and F2550W (25.5 $\mu$m) from JWST, and PACS70 (70 $\mu$m) using Herschel, that are well described as broken power laws (i.e., linear behaviors in log-log plot). The shading shows radial zones of different optical depth (see Figure \ref{fig:tc_tau}). }
    \label{fig:radprof}
\end{figure}

\subsection{Radial Profile of the Disk}
\label{sec:radprof}

Given the face-on disk geometry and azimuthal symmetry, we use the radial profile\footnote{Azimuthally average radial profiles are computed at a series of concentric rings with a width of 2 pixels ($\sim$0\farcs22) centered at the star. The disk brightness is the average value of all the pixels that fall in each ring, and the measurement error at each radius is the standard deviation of all pixels in that ring, divided by the square root of the number of pixels in the ring.} to further characterize the disk radial structure. For the purpose of characterizing multi-wavelength disk emission, we also include the 70 $\mu$m disk radial profile published by \citet{su13} (at a resolution element of 6\arcsec) for the analysis. Figure \ref{fig:radprof} shows the multi-band, disk radial surface brightness profiles, which are the product of the vertical optical depth ($\tau_{\perp}$, purely geometrical) and the dust temperatures ($T_d$) manifest in the Planck function ($B_{\nu}(T_d)$) for a face-on, optically thin debris system. The disk radial profiles can be best described as broken power-laws (i.e., linear lines in a log-log plot) for regions that are well resolved outside a few arcsec for MIRI wavelengths and $\sim$10\arcsec\ for PACS70). It is interesting to note that at a given wavelength the Planck function behaves like broken power laws for a given radial temperature distribution ($T_d \sim r^{-0.5}$ for blackbody-like grains), i.e., the power-law indices between the inner and outer radial regions (or the slopes in a log-log plot such as Figure \ref{fig:radprof}) are different. The slope depends sensitively on the observed wavelength and drops much faster in the outer region compared to the inner one for a fixed wavelength and temperature distribution. The transition between the broken power laws is also wavelength dependent, i.e., transitioning at larger radial locations for longer wavelengths as illustrated in Figure \ref{fig:radprof}. In other words, the wavelength-dependent radial profiles such as the one shown in Figure \ref{fig:radprof} are encoded with information about a debris system's dust temperature distribution and vertical optical depth.

\begin{figure}
    \centering
    \includegraphics[width=\linewidth]{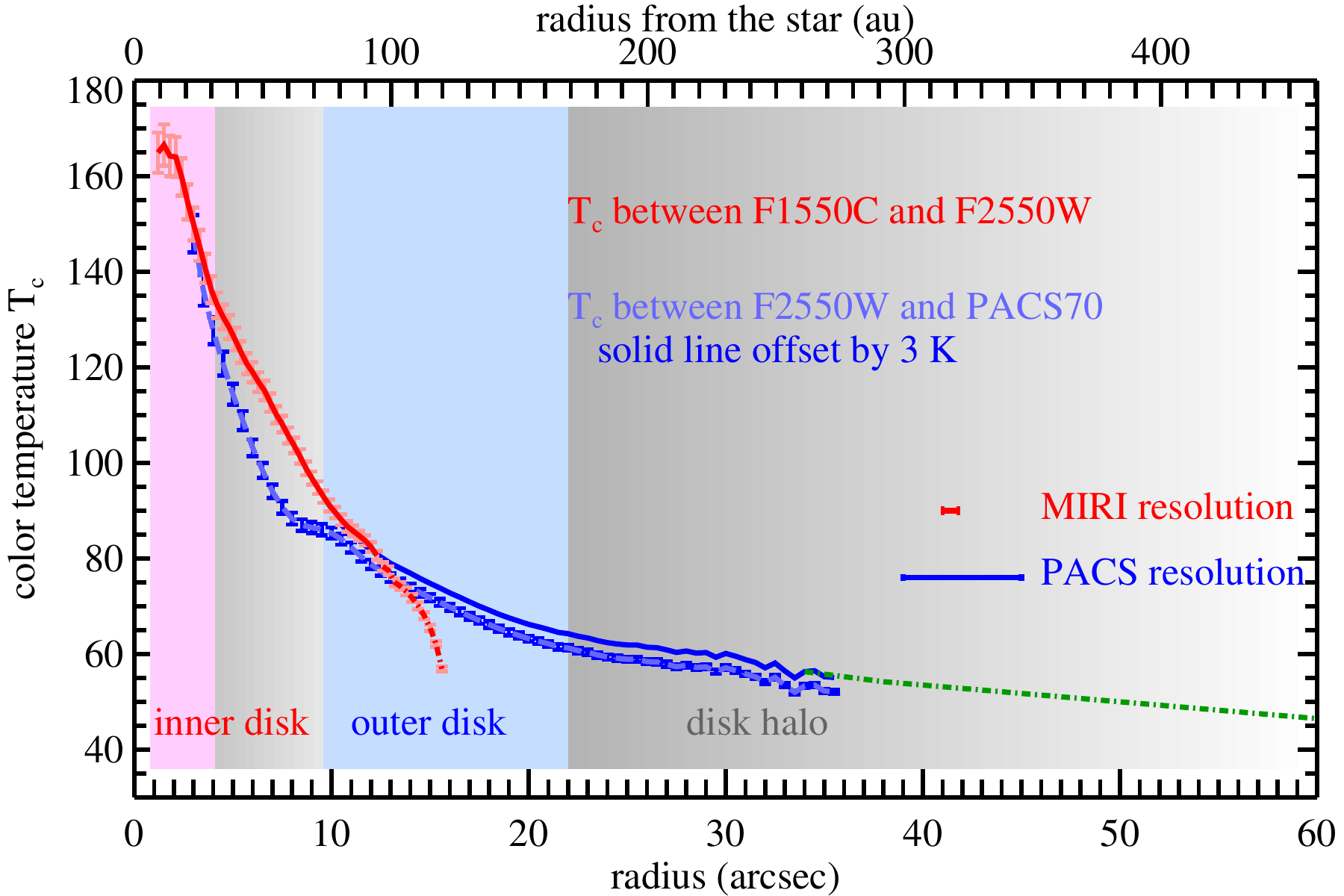}
    \includegraphics[width=\linewidth]{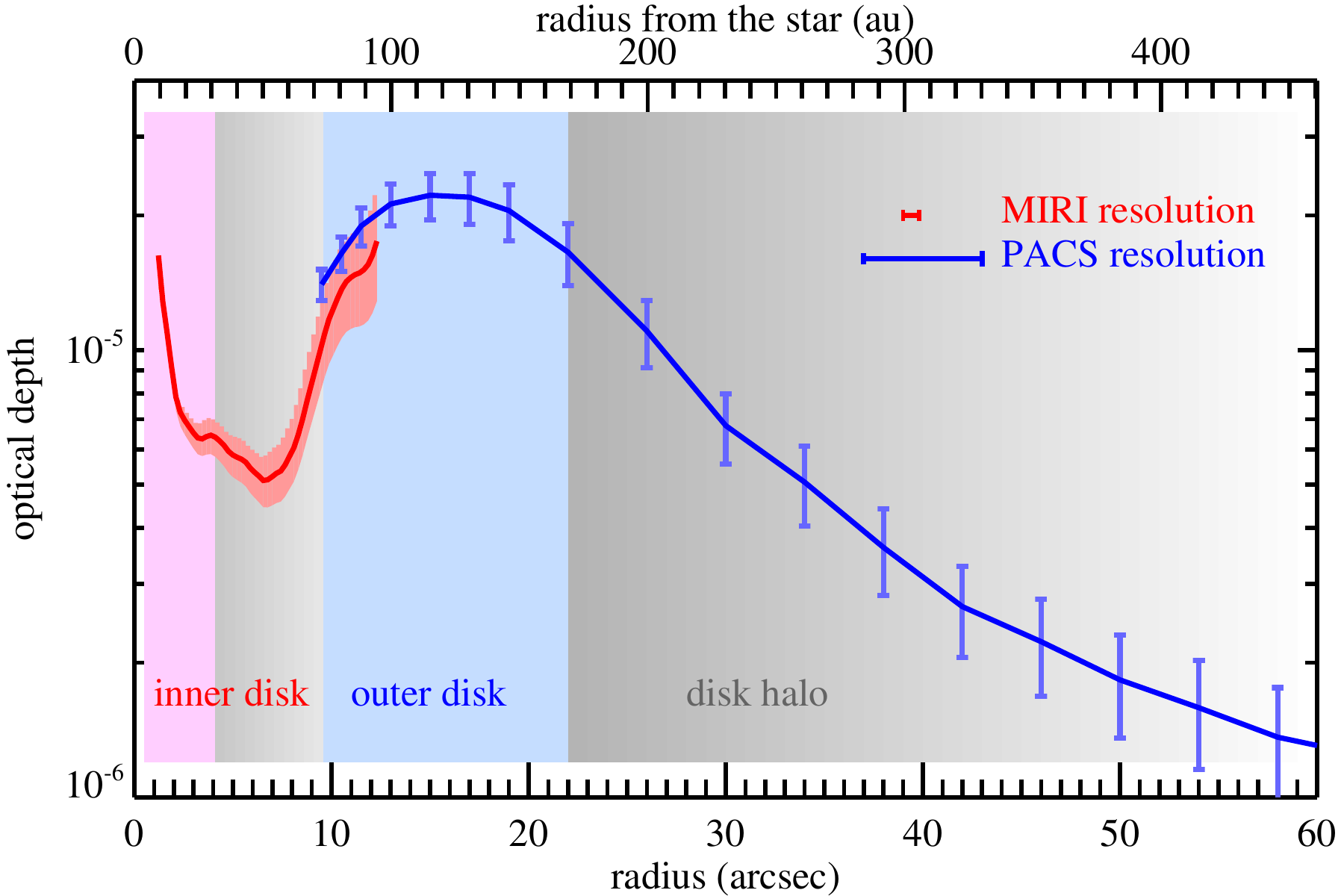}
    \caption{Upper panel: disk color-temperature radial profile derived from the flux ratio of two observed bands (red color using F1550C and F2550W, and blue color using F2550W and PACS70). The spatial resolutions of MIRI and PACS are shown as horizontal bars for reference. Due to the different resolution, limited field of view and flux calibration uncertainty at different bands, the solid lines are the adopted dust temperature profile for the Vega system for the radial range of 1\arcsec$-$33\arcsec, and an $r^{-0.5}$ extrapolation (green dot-dashed line) outside that range. Bottom panel: disk optical-depth radial profile derived from the color temperatures (upper panel), which was used to define several major radial zones (color shaded regions) as detailed in Section \ref{sec:tc_tau}. }
    \label{fig:tc_tau}
\end{figure}

\subsection{Disk Temperature and Optical-depth Distributions}
\label{sec:tc_tau}

Consequently, we will use the multi-wavelength disk surface brightness profiles (Figure \ref{fig:radprof}) to determine the bulk properties of the disk emission, i.e., the radial distribution of the dust temperature and the integrated vertical optical depth, under the assumptions of (1) uniform grain properties (size distribution and composition) at all disk location, (2) the disk being face on, and (3) optically thin.  We use the color temperature derived from the flux ratio of two wavelengths as the proxy for the dust temperature. The temperature uncertainty is directly translated from the error of the observed disk flux in the radial profile (i.e., larger uncertainty for a small radius annulus), which is $\sim$2--4 K inside 6\arcsec\ while $\lesssim$1 K outside. Considering the coronagraphic inner working angle, the resolution and the field of view, the dust temperatures are derived in two different regions. For the inner $\sim$10\arcsec\ region, the flux ratio between MIRI F1550C and F2550W was used, while for the outer ($\gtrsim$10\arcsec\ region), the flux ratio between MIRI F2550W and PACS70 was used. The derived color temperatures are shown in the upper panel of Figure \ref{fig:tc_tau}. Given the different spatial resolution between JWST and Herschel (0\farcs8 at 25.5 $\mu$m vs.\ 6\arcsec\ at 70 $\mu$m), the flux ratio inside $\sim$6\arcsec\ is greatly affected by the Herschel PSF, therefore not truly representative; outside that range, the dust temperatures estimated from F2550W and PACS70 join nicely with the ones estimated from F1550C and F2550W if the former is offset by 3 K (upper panel of Figure \ref{fig:tc_tau}). The uncertainty in the absolute flux calibration across different observatories can easily explain this small offset. Limited by the field of view of the F2550W data, the color temperature outside $\sim$33\arcsec\ cannot be determined; the dust temperatures outside that range are extrapolated as $T_d \sim r^{-0.5}$.

For optically thin debris systems, the overall debris dust temperatures are usually estimated using a system's SED where the dominant component can be identified. For example, a distinct two-component disk structure was inferred for the Vega system based on its global SED and low-resolution far-infrared disk images \citep{su13}. The multi-wavelength resolved disk images provide a substantial improvement: we have been able to determine the bulk dust temperature {\it distribution} in the radial range of 1\arcsec$-$60\arcsec\ ($\sim$8$-$500 au). 

The vertical optical depth distribution, $\tau_{\perp}(r)$, can then be determined by adopting the derived dust temperature distribution under the optically thin assumption ($F_{\nu} \sim \tau_{\perp} \cdot B_{\nu}(T_d)$). To be conservative, we adopt a uniform uncertainty of $\pm$3 K in the dust temperature distribution when calculating the error in the optical depth as shown in the bottom panel of Figure \ref{fig:tc_tau}. The maximum optical depth in the Vega system is $\sim$2$\times10^{-5}$ in the outer disk, further validating the optically thin assumption. 

Using the optical depth distribution, we define several major radial zones as follows. The outer disk (or cold planetesimal belt) defined as $r \sim$10\arcsec$-$22\arcsec\ ($\sim$78$-$170 au)  directly corresponds to the broad, Kuiper-belt analog observed by ALMA \citep{matra20}. Outside the cold planetesimal belt lies the disk halo, a surprise discovery from Spitzer \citep{su05}. The inner disk defined as $r \lesssim$4\arcsec\ is occupied by warm debris emission predominantly at $T_d \sim$140$-$170 K (i.e., the inner warm component as suggested by \citealt{su13}). The transition between the inner and outer disk is where the dip/gap is located in the MIRI images. The detection of the outer disk in the millimeter wavelengths  \citep{hughes12,matra20,marshall22_vega_lmt} shows it has a large population of mm-size grains. It is likely to be a dynamically active planetesimal belt that produces $\mu$m-size grains through collisional cascades. Unfortunately, current millimeter observations do not have enough sensitivity and resolution to detect the presence of mm-size grains in the inner disk \citep{matra20,chavez23_LMT_Vega}. Note that even with deep ALMA observations no inner millimeter emission is detected in Fomalhaut either \citep{su16_fomalhaut,chittidi24}, which is subject to future investigation.
The nature of the inner disk and the transition region is relatively unexplored, and we will discuss different scenarios in Section \ref{sec:discussion}.

\begin{figure}[tbh]
    \centering
    \includegraphics[width=\linewidth]{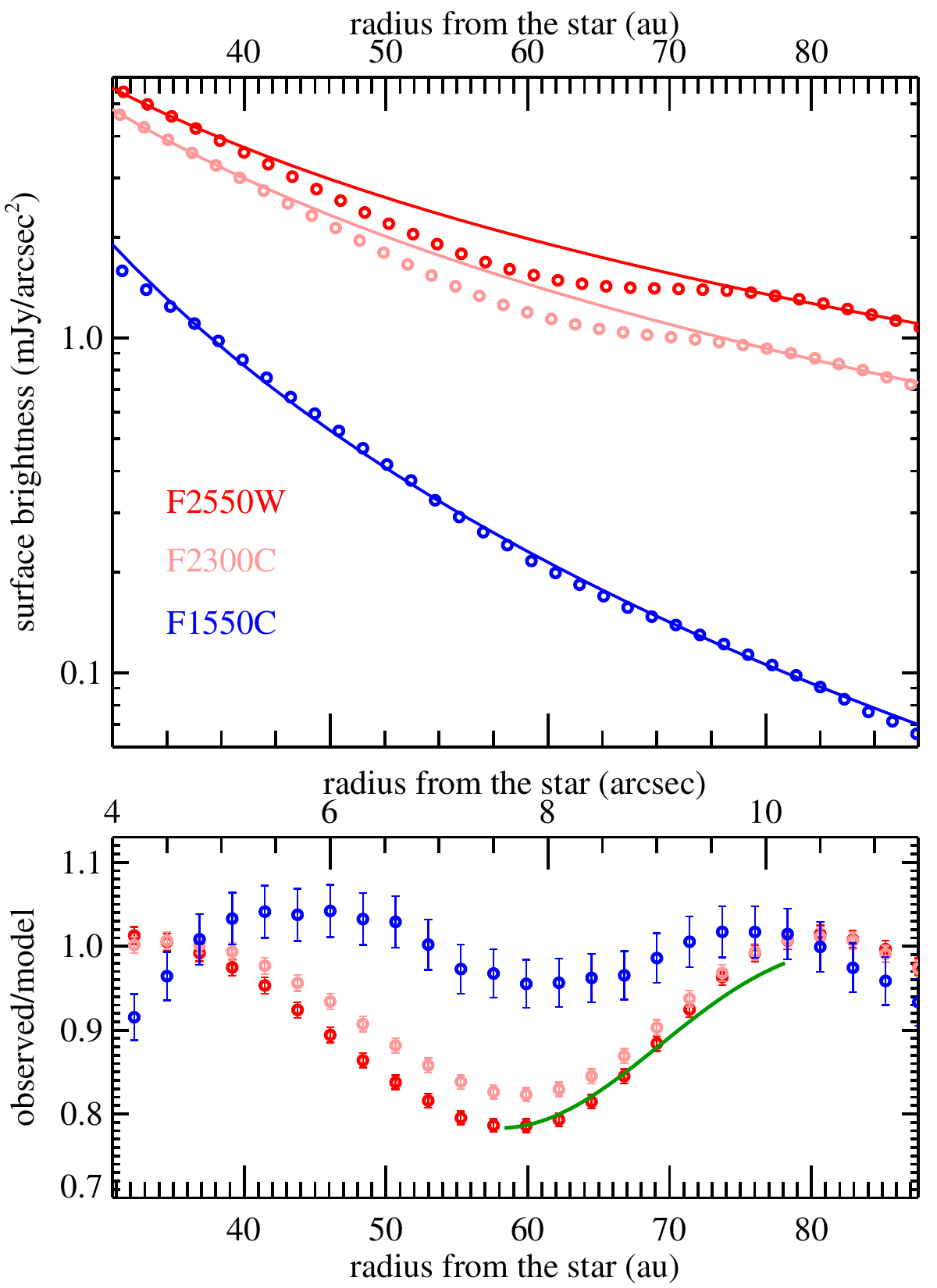}
    \caption{The upper panel shows the disk radial profiles in the $r\sim$4\arcsec$-$12\arcsec\ region using the three MIRI filters, in comparison to power-law profiles (solid lines). The bottom panel shows the dip profiles defined as the observed divided by the power-law model (i.e., the flux deficit). The dip is broad ($\sim$20 au) and of similar depth ($\sim$20\%) in both F2550W and F2300C, but narrower and shallower in F1550C. The green solid line is the inner side of the F2550W profile but flipped to allow close comparison with the outer side. It implies that there is a slight asymmetry before and after the bottom of the dip.}
    \label{fig:dip}
\end{figure}

\begin{figure*}[htb]
    \centering
    \includegraphics[width=0.49\linewidth]{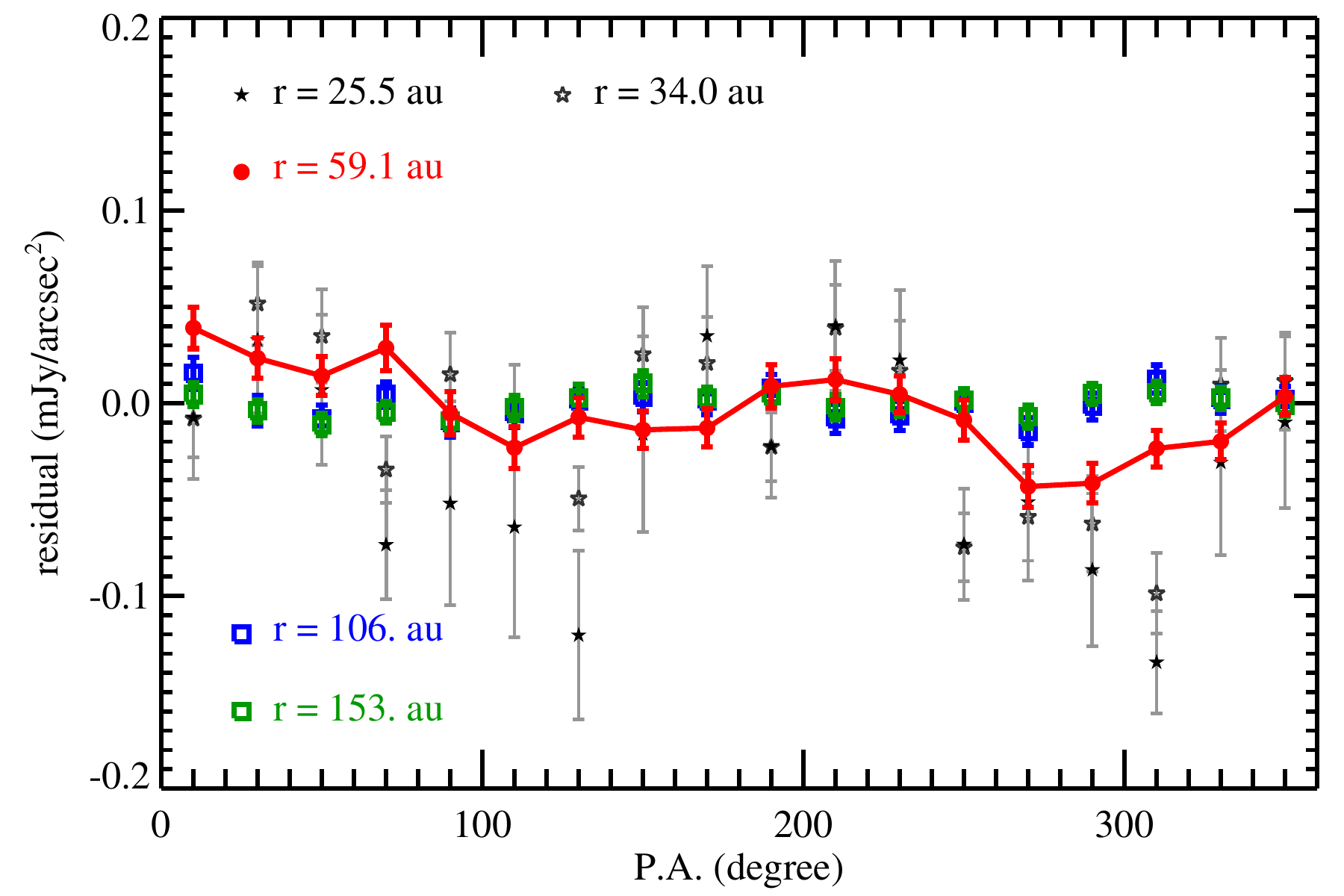}
    \includegraphics[width=0.49\linewidth]{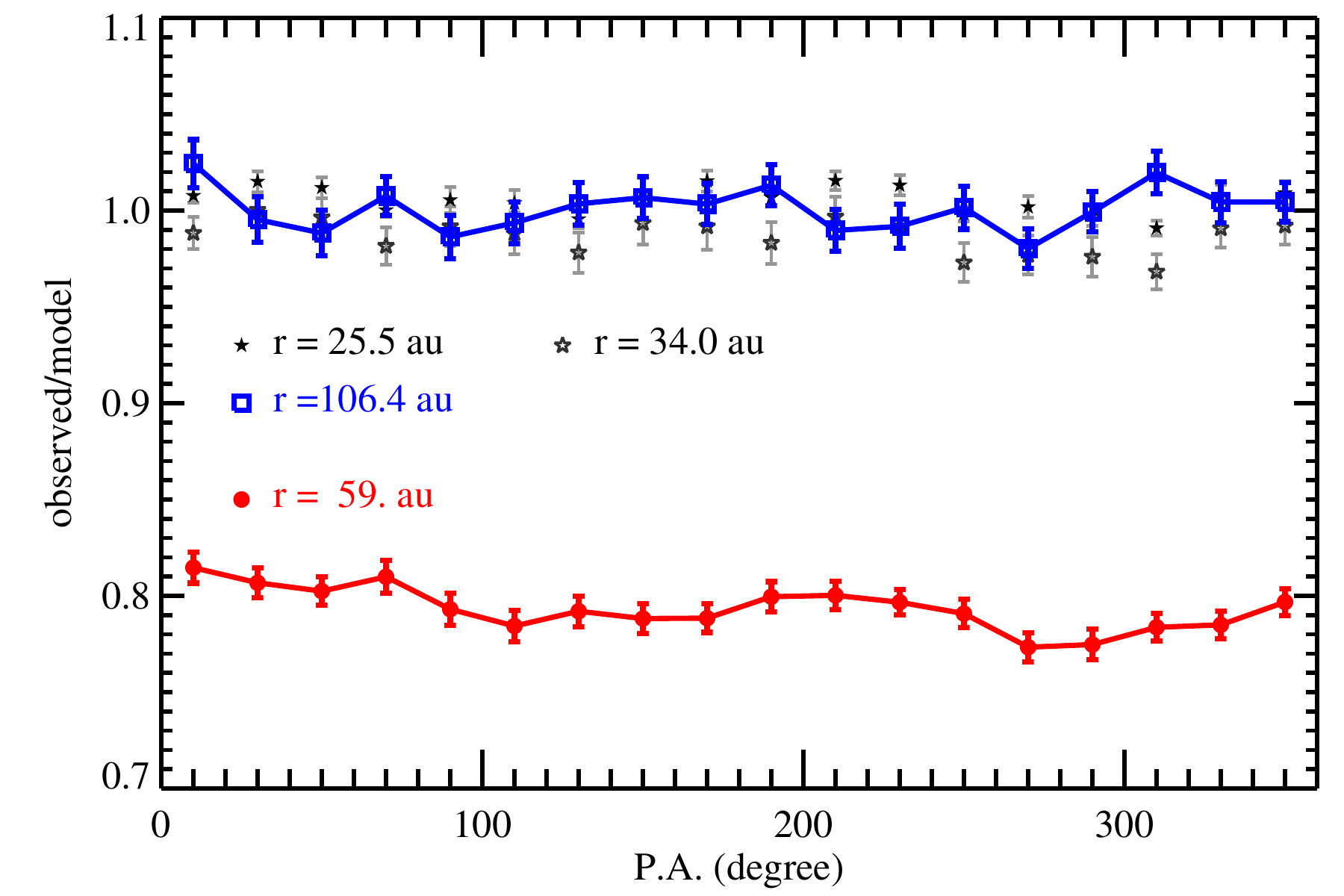}
    \caption{Azimuthal disk brightness distribution in the ``enhanced" F2550W disk image estimated by two metrics: the left panel as the residual flux by subtracting the median value at each of the radii, and the right panel as the ratio between the observed and broken power-law models. An annular width of 10 pixels (1\farcs1 $\sim$ resolution) and azimuthal segments of 20\arcdeg\ are used for the computation (details see Section \ref{sec:azmuthal}). The red points with a connected line represent the gap area while the blue points represent the outer disk.} 
    \label{fig:azimuthal}
\end{figure*}

\subsection{The Dip Properties}
\label{sec:dip}

In addition to the very circular, smooth disk morphology,  the dark gap (i.e., a dip in the disk surface brightness) is the most noticeable feature in the MIRI Vega disk images. The upper panel of Figure \ref{fig:dip} shows the zoom-in disk surface brightness near the dip area along with the model power-law profile that fits the disk brightness well before (closer to the star) and after (further from the star) the dip. Using the model power law, the dip profile (observed divided by the model) is shown in the bottom panel of Figure \ref{fig:dip}. The depth of the dip (flux deficit) is larger at longer wavelengths, reaching $\sim$20\% at 59 au in the F2550W filter but only $\sim$5\% in the F1550C filter. The width of the dip (defined as the FWHM) is $\sim$20 au at 25.5 $\mu$m and shows similar wavelength-dependent behavior (i.e., wider at longer wavelengths). The dip profiles also show tentatively asymmetric behavior before and after the deepest flux deficit: the far side of the dip profile is slightly shallower (i.e., less flux deficit, see the green line in the bottom panel of Figure \ref{fig:dip}). We also searched for the dip in the Herschel 70 $\mu$m image; unfortunately, the poor resolution (6\arcsec) would not allow any meaningful constraint.

\subsection{Quantifying Disk Asymmetric Structure}
\label{sec:azmuthal}

To search for (and constrain) the potential disk asymmetric structure, we constructed the disk azimuthal profiles at several disk radii. An annular width of 10 pixels (similar to the MIRI spatial resolution, 1\farcs1 $\sim$9 au at the distance of Vega) and azimuthal segments of 20\arcdeg\ were used to calculate the disk average value and standard deviation in each of the azimuthal segments along the disk circumference. The calculation was performed using two ``enhanced" disk images at 25.5 $\mu$m by taking out the axisymmetric, broken-power-law trend. The first one is the residual disk flux obtained by subtracting the median value at each of the radii (left panel of Figure \ref{fig:azimuthal}), and the second one is the ratio between the observed and broken power-law models (right panel of Figure \ref{fig:azimuthal}).

A handful of radial locations, from inner region (r=25.5 and 34 au), dip/gap (r=59 au), to outer disk (r=106 au) and disk halo (r=153 au), were used for the assessment. At large radii (outer disk and disk halo region), there is no azimuthal asymmetry in the residual flux, i.e., flat within the uncertainty.  Closer to the star (inner disk), the residual flux appears to show two azimuthal regions with large negative fluxes at position angles (P.A.) of $\sim$70\arcdeg$-$130\arcdeg\ and $\sim$250\arcdeg$-$310\arcdeg.  Unfortunately, these azimuthal regions are along the PSF diffraction spikes\footnote{JWST PSF has six prominent diffraction spikes that separate by $\sim$30\arcdeg. The combination of the PSF structure and the two observations taken with $\sim$10\arcdeg\ sky rotation leaves the imprint of very bright star along these azimuthal regions.} (i.e., the uncertainties in these regions are much larger). Similar azimuthal behavior is also seen in the dip/gap region (although less pronounced compared to the inner ones). A similar behavior is also seen in the flux ratio map (left panel of Figure \ref{fig:azimuthal}), which suggests the current F2550W disk image can rule out (at the 3 $\sigma$ level) any azimuthal brightness asymmetry  $>$10\%  near the dip/gap area. The lack of azimuthal structures is of interest since they are often signposts for the presence of planets (see Section \ref{sec:signpost}).

\begin{figure*}[t]
    \centering
    \includegraphics[width=0.8\linewidth]{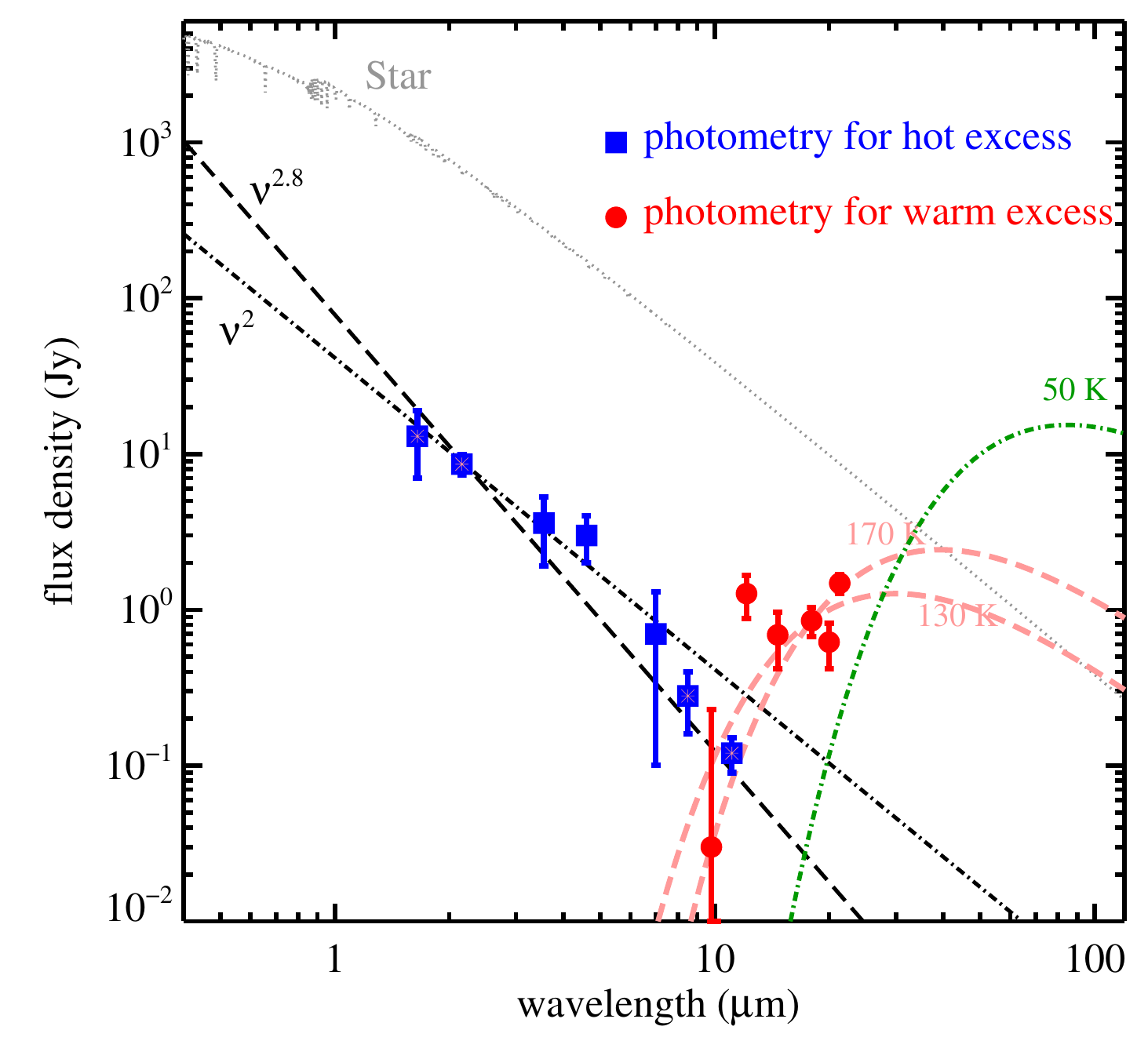}
    \caption{Excess SED of the Vega system using the infrared photometry tabulated in Table \ref{tab:IR_photometry} for constraining the hot/warm excess. Large blue squares are the photometry for constraining the hot excess emission ($\nu^{2.8}$ spectral slope) where the points between 3 and 8 $\mu$m were obtained by subtracting a very accurate photospheric model from integrated measurements of Vega, and the other three points were measured with interferometers. Pink asterisks mark the values from interferometric measurements. Large red dots are the photometry using larger apertures and are well described by the warm dust emission with temperatures of 130$-$170 K, in contrast to the cold ($\sim$50 K) outer disk emission. }
    \label{fig:sed_innerexcess}
\end{figure*}

\subsection{Constraints on the Inner Zones of the Vega Disk}
\label{sec:inner_radius}

\subsubsection{The hot excess}
\label{sec:hotexcess}

Interferometric measurements at 2 $\mu$m show a hot excess at $\sim$1.3 \% above the stellar photosphere \citep{absil06}. As shown in Figure \ref{fig:sed_innerexcess}, measurements of this component follow a $\nu^{2.8}$ spectral shape well up to $\sim$10 $\mu$m ($\nu$ is frequency). That is, they fall more steeply than Rayleigh Jeans, which goes as $\nu^{2}$. This suggests that the hot excess contributes negligibly to the MIRI data presented here. To produce a hot spectrum bluer than Rayleigh Jeans, the emitting dust needs to be confined to very close to the star (close to or within the dust sublimation radius, $\sim$0.2 au as suggested by \citealt{su13} and \citealt{rieke16_magnetictrapping}) and be dominated by small and very refractory grains like graphite and Fe and Mg oxides, as verified in detail by \citet{myrvang18}.

\subsubsection{The warm excess}

We now turn to the disk outside of 0.2 au but within a few au of the star. Limited by the coronagraphic inner working angles (0\farcs49 at F1550C and 2\farcs16 at F2300C) and the saturation at the core ($\lesssim$1\arcsec) of the F2550W data, we cannot directly determine the disk structure inside these boundaries from the disk images. Here we provide two metrics for estimating this inner edge of the disk emission and show that it cannot be smaller than $\sim$3 au from the star. This puts it physically separated from the hot excess component detected by near-infrared interferometric measurements, and also puts it well outside the sublimation radius for typical dust grains (e.g., of astrosilicates).

The first metric relies on infrared photometry to determine the level of infrared excesses by taking the aperture sizes and observed wavelengths into account. This also includes recalibrating Vega's photospheric emission relative to Sirius (a proposed standard in absolute flux calibration) so that the accurate level of infrared excess can be used to characterize the emission properties like spectral slope, dust temperatures and location.  We detail the photometry used in this exercise in Appendix \ref{sec:appendix_photometry}. Figure \ref{fig:sed_innerexcess} gives an overview of these excess measurements as tabulated in Table \ref{tab:IR_photometry}. 
Although many of the photometric measurements were obtained with apertures that are large enough to include the outer disk ($T_d\sim$60$-$80K as shown in Figure \ref{fig:tc_tau}), we show in Appendix \ref{sec:appendix_intep_photometry} that measurements at wavelengths of $\sim$10$-$20 $\mu$m have very little contribution from the cold, outer disk. The red dots in Figure \ref{fig:sed_innerexcess} probe the debris emission with dust temperatures of 130$-$170 K, consistent with the warm emission from the inner disk (1\arcsec$\lesssim r\lesssim$4\arcsec, revealed by MIRI as illustrated in Figure \ref{fig:tc_tau}). Figure \ref{fig:sed_innerexcess} also shows that an even warmer ($\sim$250 K) dust emission might be present in the warm debris, but cannot be the dominant component.   
A series of SED models were built based on the collected infrared photometry and showed that the inner disk edge needs to be beyond 1 au (details see Appendix \ref{sec:appendix_intep_photometry}).

The second metric is more demanding, and is based on modeling the radial profiles. This metric utilizes a forward modeling technique with the assumption that the behavior of the instrumental PSFs is well understood as detailed in Appendix \ref{sec:appendix_parametricmodel}. Such a forward modeling method not only can constrain the inner disk structure, but also can provide an estimate of the missing flux within the coronagraphic masks and the saturated core in the MIRI data as briefly summarized below. 

We fit the F1550C and F2550W radial profiles of the disk emission using pure, geometric parametric models. The basic assumption is that the true disk surface brightness profile inside $\sim$1\arcsec\ follows simple descriptions: either a point source centered at the star, the same broken power laws extended inward from the outer part of the disk profile, or a simple ring-like structure. Because of the face-on disk geometry, the observed profile would be the true model surface brightness profile convolved with the instrumental PSF (i.e., no assumption regarding the grain properties). Details about the computation/fit are given in Appendix \ref{sec:appendix_parametricmodel}. We find that the excess emission cannot be a simple, unresolved (by MIRI) point source centered at the star because it would create observable PSF structure near the region of the first bright Airy ring of the PSF (at radii of $\sim$1\arcsec$-$1\farcs5). An inner cut-off for the disk emission at MIRI wavelengths is necessary,  consistent with the first metric  (i.e., the hot and warm excesses are physically separated). The power-law models suggest that the inner edge of the disk  cannot be smaller than 0\farcs3 nor larger than 0\farcs7, i.e., the inner cut-off needs to be in the range of $\sim$2$-$5 au from the star. Combining with the parameter search in both F1550C and F2550W data, it seems that either a ring at $\sim$4–5 au or a power-law disk with a similar cut-off is consistent with the data, and a total flux of $\sim$0.14 Jy at 15.5 $\mu$m, and $\sim$0.4 Jy at 25.5 $\mu$m for such structures best describes the observed profiles.

\section{Discussion -- Where are the planets?} 
\label{sec:discussion}

\subsection{Direct planet searches}
\label{sec:direct_planet_searches}

Given the proximity of Vega, it has been the subject of many direct searches for planets. \citet{meshkat18_vega} searched with a coronagraphic integral field spectrograph and extreme AO system between 0\farcs5 and 2\arcsec\ (i.e., about 4 au to 15 au). They achieved 5 $\sigma$ limits equivalent to 20$-$30 $M_{\rm Jup}$, with no detection. \citet{beichman24_vega} used the JWST/NIRCam coronagraph to perform the deepest search for planets in Vega. Between 5\arcsec\ and 30\arcsec\ (about $\sim$40 au to 200 au) their detection limit is $\le$0.5 $M_{\rm Jup}$ while at 2\arcsec\ it is $\le$ 2 $M_{\rm Jup}$ assuming an age of 700 Myr, again with no detection. \citet{hurt21} summarize a decade of radial velocity (RV) measurements on Vega. The pole-on orientation of the system is unfavorable for the RV technique except for exoplanets with orbits significantly inclined relative to the stellar rotation. They estimate that the RV method loses sensitivity beyond $\sim$6 au but it would appear to exclude Jovian planets out to this radius. Finally, they searched for planetary transits using the TESS data but found none, not surprising given the pole-on inclination of the star. In summary, it is likely that there are no Jovian mass planets orbiting Vega (with the possible exception of a radial velocity candidate very close to the star, \citealt{hurt21}). 

\subsection{Indirect planet search via debris disks}
\label{sec:signpost}

Since the discovery of circumstellar debris disks, the distribution of the dust structures has been recognized as a pathfinder to detect exoplanets indirectly. The interest in this approach is that it is sensitive to planets down to Neptune masses \citep{liou99} and even significantly below \citep{stark08_resonantsignature}. There are many features of debris disks that are often attributed to planets within them, e.g, disk warps, sharp edges, gaps, clumps, scale height features, center offsets, and azimuthal asymmetries. Of these, the last two are the most reliable to infer unseen planet properties \citep{dong20}.

There are many numerical simulations that aim to predict the observable disk structure under the influence of planets in more detail. Many works mostly focus on the mean motion resonances (MMRs) \citep[e.g.,][]{ozernoy00,deller05,reche08,tabeshian16,regaly18_cavity} while some focus on the disk radial structure and the sharpness of shepherded edges \citep{chiang09,pearce24_sharpness} linked to planets interior and/or exterior to the observed structures. Most of these N-body studies have modeled the disk as an ensemble of particles only under gravitational forces; therefore, they are directly applicable to the observations that trace large, mm- to cm-size grains and have been systematically applied to the ALMA resolved disk images \citep{pearce22}. 

However, applying these simulations to the infrared resolved disk images is not straightforward due to the extremely large dynamical range of particle sizes ($\sim$10$^3$ between grains and pebbles and $\sim$10$^6$ between pebbles and planetesimals) in the disk where observations only probe a limited range of sizes ($a_{\rm crit}$) at a given wavelength ($\lambda_{\rm crit}\sim 2 \pi a_{\rm crit}$, \citealt{backman93_PPIII}). Small grains are sensitive to non-gravitational forces (stellar radiation pressure and winds, and drag forces like PR and stellar wind drag), and their dynamics are significantly altered, often beyond the effects of gravitational potential of the planet \citep{moromartin02}. However, the greatly improved thermal infrared images of the Vega disk reported in this paper allow a much deeper search for indirect evidence of planets than was possible previously. Given the small size of the grains emitting prominently at mid-infrared, the models need to include not only gravitational forces but also the non-gravitational forces such as PR and stellar wind drag and radiation pressure for grains larger than the blow-out size.

In the following subsections, we first discuss the indirect evidence and constraints for the presence of planets using the well-resolved, high quality of the MIRI disk images, and also the potential for an inner planet or other alternative explanations to be responsible for the structure inside the region where our MIRI images cannot directly probe (within the saturation core  and the coronagraphic mask, Section \ref{sec:innerplanet}). For the first part, in Section \ref{sec:gap-opening-planet} we explore the unseen planet properties by assuming the dip presents a physical gap in a very extended distribution of planetesimals.
Our simple, qualitative assessment argues against any Saturn-mass planets orbiting Vega outside of about 10 au. We then explore the possibility that the Vega's inner disk emission could be dominated by the dragged-in grains interior to an outer Kuiper-belt-like belt (Section \ref{sec:pr_win}). In Section \ref{sec:planet_perturbations}, we investigate to what degree an unseen planet in this region would create detectable deviations in the drag-dominant inner disk (Section \ref{sec:planet_perturbations}), and show that a planet of much lower mass than half Saturn would still create readily detectable perturbations in the debris system. The shallower gap is consistent with the presence of a few Earth masses in the drag-dominated disk.

\subsubsection{Qualitative Constraints on a gap-opening planet}
\label{sec:gap-opening-planet}

As discussed by \citet{matra20}, the current ALMA observation is not sufficiently deep to confirm or rule out mm emission for the warm debris in the Vega system, while the far-infrared images from Herschel lack necessary resolution to push the limit on such emission further. In other words, it is possible that the Vega system has an extended planetesimal disk from a few au to $\sim$100 au, and that the dip/gap is caused by an unseen planet in the parent-body population as the MIRI 25.5 $\mu$m disk image apparently suggests. 

Under this assumption, we can put some qualitative constraints using the global disk properties and that of the dip/gap observed by MIRI. First, the entire disk is extremely circular and the center of the disk circumference is at the star position within the instrumental pixel (0\farcs11 $\sim$1 au scale). Millimeter observations obtained by both single dish and interferometer show no evidence of clumpy structure in the cold disk for scales as small as 1\farcs35 ($\sim$10 au across) and as large as a few 10s\arcsec\ \citep{marshall22_vega_lmt,matra20}.  Furthermore, ALMA data show no detectable offset ($\Delta a \sim$10 au) between the star and the center of the cold broad belt, a typical signpost for an eccentric shepherding planet. All the evidence suggest the planet that maintains the inner edge of the broad planetesimal belt needs to have low ($\lesssim$0.1) eccentricity. As we discuss the general mass limits next, we will return to this topic in Section \ref{sec:planet_perturbations} where we develop an example showing that, in contrast to the general limits, more stringent limits can be placed on embedded planets that are in specific interesting locations. 

For a planet on a circular orbit, the most prominent planet signature is a gap in the disk, which corresponds to the chaotic unstable zone surrounding the planet \citep{wisdom80,morrison_malhotra15}, where the width of the chaotic zone is a function of the planet-to-star mass ratio and the semi-major distance of the planet from the star. Taking the width ($\sim$20 au) of the dip/gap from Section \ref{sec:dip} at face value, an $\sim$1.6 $M_J$ mass planet at the location of the dip ($\sim$59 au from the star) would be responsible for the wide gap. This is contrary to our mass limits for planetary objects in the system (see Section \ref{sec:direct_planet_searches}). Not only should such Jovian planets have been detected by the JWST NIRCam observation \citep{beichman24_vega}, but also they would have created two stable co-rotating resonance structures inside the gap while no detectable azimuthal structure is seen (Section \ref{sec:azmuthal})\footnote{The 1:1 MMR Trojan population is expected to be the most prominent feature for planets with low (0--0.1) eccentricity \citep{regaly18_cavity}.}. Since the properties of the dip/gap are derived in the mid-infrared, one might argue that the true width of the gap might be narrower for mm-cm-size grains. Assuming a width of $\sim$10 au, a half Saturn mass planet is required, which is right at the current detection limit \citep{beichman24_vega}. A recent study by \citet{friebe22} argues that the same width of the gap can be carved by either a massive, barely migrating planet or low-mass, migrating planet with a very different dynamical state of the resulting planetesimal disk. In the latter case, the planetesimal disk is expected to be highly stirred and possesses a scattered disk component (i.e., fuzzy edge and weak/non-existing Trojans). It is interesting to note the low-mass, migrating planet case is consistent with the broad ($\Delta r / r \sim 0.56$), outer planetesimal belt detected by ALMA \citep{matra20} and the bright disk halo indicative of high collision rate in the planetesimal population \citep{su05,wolff24}. 

Broad shallow gaps can also be created away from the orbit of a planet by MMRs \citep{tabeshian16}. Assuming the shallow dip/gap at $\sim$60 au is opened by a planet's 2:1 MMR as modelled by \citet{tabeshian16}, the most likely location of the perturbing, low-eccentricity planet would be at $\sim$95 au for creating an interior gap, or $\sim$38 au for creating an exterior one, although in both cases general asymmetries in the gap are also predicted that are not reflected in the Vega disk.

In addition to the caveat described before, perhaps the largest shortcoming in these numerical simulations (and application to observations) is neglecting collisions. \citet{thebault12_planetsignature_collisionallyactivedisks} explored further different planet configuration (mass and eccentricity) and collisional levels including proper treatment of radiation pressure. They concluded that collisions always significantly damp planet-induced spatial structures (termed ``blurring effect"), but the gap should remain detectable in face-on orientation (like Vega) if the planet mass is larger than Saturn mass in a collisionally active system ($\tau \sim 10^{-3}$). The Vega disk, being a lower optical depth system, should experience less blurring effect, pushing the planet mass limit even lower, i.e., sub-Saturn mass. 
Based on these qualitative comparisons and the assumption of an extended planetesimal disk, the upper limit in the mass of single planet that could be responsible for the features detected in the Vega system is below the Saturn mass and at low ($e<$0.1)  eccentricity.   

Lastly, the mass of planetesimals might also impact the observable planet-induced structures \citep{sefilian23}. Using the observed dust mass derived from the sub-mm and mm observations \citep{hughes12,holland17_SONS,marshall22_vega_lmt}, the planetesimal mass is in the range of $\sim$10--30 $M_{\oplus}$ (0.5--2 Neptune masses) if extrapolating to 10--100 km-size planetesimals using the nominal Dohnanyi size distribution ($a^{-3.5}$, \citealt{dohnanyi69}). Therefore, for low-mass planets (a few Neptune masses), the disk gravity might also affect the observed gap properties in the Vega disk. Future investigation including all observed properties and proper treatment of the collisions, particularly in an extended planetesimal disk configuration, is needed to put tighter constraints on the mass of the planet that might be responsible for the observed disk features.

\subsubsection{Dust transport to the inner disk regions}
\label{sec:pr_win}

To go beyond these general considerations, we need to understand the dust transport within the disk. In the absence of giant planets in the solar system, \citet{liou99} pointed out that the column density of the interplanetary dust particles (IDPs) would be constant interior to the Kuiper belt, under the influence of PR and stellar wind drag alone. Even for a collision dominated planetesimal belt, detailed numerical calculations including self-consistent treatments of collisions and drag forces do show that the small amount of dragged-in grains could be the most prominent feature in the mid-infrared interior to a collisionally-dominated planetesimal belt \citep{lohne17}. The smoothness and filled-in structure of the Vega disk suggest a drag-dominated disk. It is also interesting to note that a dip/gap structure might be present near the inner edge of the belt due to the strong temperature effects in the mid-infrared (see Figure 13 in \citealp{lohne17}). In other words, these disk features revealed by our JWST observations might originate from a PR-dominated phenomenon.

In this subsection, we present preliminary models to help understand the dominant physical effects that determine the location and distribution of the dust particles we observe at the MIRI wavelengths inwards of $\sim$90 au. Although collisions typically dominate the grain evolution and dynamics in relatively high density regions, i.e., the peak of the mm disk emission ($\sim$118 au \citealt{matra20}), at radii inside the planetesimal belt where the optical depth drops (see Figure \ref{fig:tc_tau}), PR drag will become important. To address the tradeoff, we will model the hypothesized  parent-body planetesimals at the inner edge of the planetesimal belt and calculate the grain population within which particles will migrate into the inner zone by balancing the collisional and PR timescales, then follow their evolution into the star. Throughout the discussion, we use a parameter called $\beta$ denoting the ratio of radiation pressure to gravity \citep{burns79}. The $\beta$ value depends on grain size, shape, composition and star type, with smaller grains generally having larger $\beta$ values. We show that this model can create a surface brightness distribution very similar in radial behavior and symmetry to that observed at 25.5 $\mu$m. The discussion below provides a summary with details provided in Appendix~\ref{transport}.

\begin{figure}
    \centering
    \includegraphics[width=0.99\linewidth]{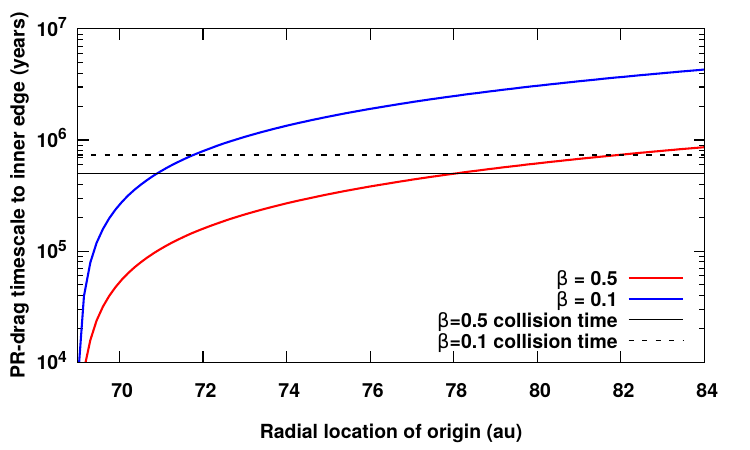}
    \caption{The PR drag timescale of particles with $\beta$ values of 0.1 and 0.5 to the inner edge of 69 au. The black solid and dashed lines mark the estimated collisional timescales, 0.5 My for the $\beta=0.5$ and 0.7 Myr for the $\beta=0.1$ particles within the model region,  highlighting that they are able to travel to the inner edge unimpeded from 78 au distance (details see Section \ref{sec:pr_win}). Note that the collsional time scales are inversely proportional to $\sqrt r$, which is $\sim$10\% different in this narrow model region. }
    \label{fig:PRtimeKBA}
\end{figure}

\begin{figure*}
    \centering
    \includegraphics[width=\linewidth]{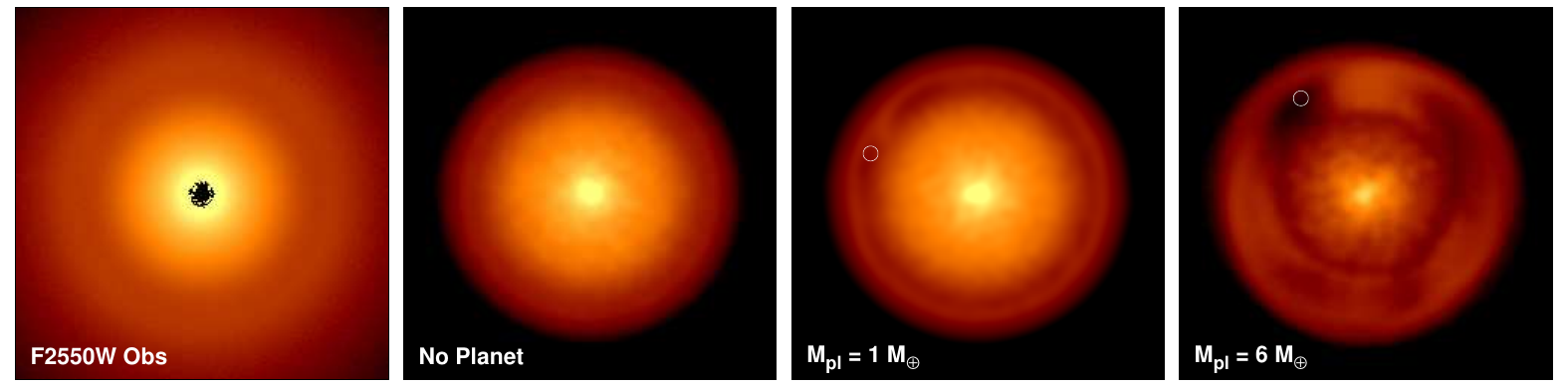}
    \caption{The model 25.5 $\mu$m disk morphology at 2 Myr for a PR dominated disk, compared to the observed Vega F2550W image ({\it first panel}), all in the same display scale. The second panel assumes no planet (details see Section \ref{sec:pr_win}), while the third and
    fourth panels assume a 1 and 6 M$_{\oplus}$ mass planet at 65 au (marked by white circles), respectively (details see Section \ref{sec:planet_perturbations}). The models do not include the entire outer planetesimal disk and halo, hence the difference with the observed images at large radii.
    \label{fig:Evolved}}
\end{figure*}

We use our dynamical numerical code DiskDyn\footnote{https://github.com/merope82/DiskDyn} to
estimate the dust distribution between 69 and 85 au where the derived optical depth decreases inward. Beyond this region (the main planetesimal belt and disk halo), the dust distribution is significantly altered by collisions and radiation pressure, i.e., sensitive to the initial conditions that require constraints from all available data. We analytically estimate the collisional timescale for the ``smallest" dust particles not affected by the removal of radiation pressure (i.e., $\beta$= 0.5) in the narrow 69$-$85 au model region by matching the current observed 25.5 $\mu$m flux ($\sim$0.17 Jy). This approach is valid because mid-infrared emission probes the small, $\mu$m-size end of the particle size distribution. With the adopted properties in the collision swarm (details see Section \ref{transport}), we found that the collisional timescale of the $\beta=0.5$ grains is $\sim$0.5 Myr within the model region, assuming a swarm collisional velocity of $\sim$10\% of the local Keplerian velocity. In order to calculate the emitting flux, a fixed dust composition and grain size distribution for the particle swarm are adopted (details see Appendix \ref{transport}). Because these grains with $\beta$ close to 0.5 are the dominant emitters in the mid-infrared wavelengths, the estimated collisional timescales only weakly depend on the dust composition (for typical steep dust size distributions, i.e., a power law index of $\sim-$3.5). Larger particles generally have longer collisional timescales for two reasons: (1) lower particle density in a cascade system and (2) the threshold for particle sizes that can efficiently destroy them is also larger. For $\beta=$ 0.1 grains, the corresponding collisional timescale is $\sim$0.73 Myr within the model region.

We can now compare the estimated collisional timescale to the PR drag timescale within the model region. The time for PR drag to transport a particle from an outer ($R_{\rm out}$) to an inner ($R_{\rm in}$) location can be expressed as \citep{burns79}
\begin{equation}
    t_{\rm PR}(a) = -\frac{\rm c}{4{\rm G}M_{\ast}\beta(a)}\left(R_{\rm in}^2-R_{\rm out}^2\right)\;,
\end{equation}
where c is the speed of light and G is the gravitational constant. For a $\beta$ = 0.5 dust particle transported to the inner radius of 69 au within its collisional timescale, $R_{\rm out}$ is $\sim$78 au (assuming $M_{\ast} = 2.135 M_{\odot}$). Figure \ref{fig:PRtimeKBA} shows the estimated PR timescale for two different $\beta$ (i.e., size) values. Larger particles will also be able to traverse to the inner edge of the disk, although from closer regions. For example, the $\beta=$0.1 particles can migrate to 69 au from $\sim$72 au within their collisional timescale. It appears that the PR timescales are comparable to (or marginally shorter than) the collisional timescales for particles with $\beta \sim$0.1$-$0.5 as shown in Figure \ref{fig:PRtimeKBA}. In other words, the dust dynamics inside the model region ($\sim$69 au) are dominated by the PR drag.

We now model the resulting surface brightness of the disk image at 25.5 $\mu$m assuming that particles reaching the inner edge of 69 au will migrate inward under PR drag,  unimpeded by collisions. In this model, particles are neither being continuously produced in the model planetesimal belt nor are they collisionally interacting (i.e., the model disk flux inside the model planetesimal belt will drop with time). A complete collisional/dynamical simulation for the entire system would be computationally prohibitive and outside the scope of this paper. Using the DiskDyn code with the same setup as earlier, we now simulate the system to evolve in time. Particles traverse inward with velocities roughly inversely proportional to their $\beta$ values. Our simulation included realistic accelerations for different grain sizes and evolved the dynamical integrations with a numerical integrator. For example, it takes $\sim$1.8 Myr for $\beta$ = 0.5 grains to reach the inner $\sim$au region from 69 au, just $\sim$3 times longer than it took to travel from 78 to 69 au. We ran dynamical simulations up to 4 Myr of evolution to allow larger particles to reach the inner au region. The model images were convolved with a JWST F2550W WebbPSF model \citep{webbpsf_tool} for morphological comparison inside 85 au. Our simulations do not include dust sublimation effects (grains are continuously migrating to the star without truncation); therefore, the inner core of the model images would appear centrally peaked once the $\beta$ = 0.5 grains reach the inner $\sim$au region. The second panel of Figure \ref{fig:Evolved} shows the 25.5 $\mu$m model images after 2 Myr of evolution in comparison with the real data (the first panel). Not only is the morphological agreement excellent between the unperturbed (no planet), PR dominated disk and the observation, but also the ``dip'', a transition between the planetesimal belt and the PR-drag dominated disk, is clearly visible after 2Myr of evolution.

\subsubsection{Planetary perturbations in a PR-dominated disk}
\label{sec:planet_perturbations}

A planetless PR-dominated disk (as shown in the second panel of Figure \ref{fig:Evolved}) would have constant and axisymmetric dust surface density, providing a blank canvas on which inner planets orbiting Vega can paint perturbations. We illustrate this potential by simulating the situation suggested by \citet{matra20} where a shepherding planet of $\ge$ 6 M$_{\oplus}$ is invoked to maintain the inner edge of the observed ALMA planetesimal belt. 

Our simulations place a planet at 65 au with masses of 1, 3, and 6 M$_{\oplus}$. The evolution of the disk morphology up to 4 Myr for these three cases is detailed in Appendix~\ref{transport}. In Figure \ref{fig:Evolved}, we compare the observed JWST/MIRI F2550W image to the 2 Myr snapshots of the 1 and 6 M$_{\oplus}$ models. Even the 1 M$_{\oplus}$ model shows small-scale azimuthal enhancements within the dip, with a deeper overall inflection than the simulation without a perturber. 
The 6 M$_{\oplus}$ simulation results in considerable asymmetric structures within the PR-dominated disk and a secondary gap at 41 au corresponding to a 2:1 MMR. In addition, the inner edge of the planetesimal belt is disrupted by this planet. This latter model is clearly inconsistent with our observations. 

Addition of collisions to our simulation would tend to smooth out the structures, but this effect should be modest for two reasons. First, the face-on aspect of the disk is optimal for detecting such structures; even in scattered light (i.e., grain size of sub-microns), where perturbations due to a Saturn mass planet should be detectable \citep{thebault12_planetsignature_collisionallyactivedisks}. The dominant particle size for the 25.5 $\mu$m emission is $>$10 times larger, and much less susceptible to non-gravitational forces, so the scattered light simulation is consistent with detection of significantly less massive planets in our image. Second, for the specific conditions in the Vega disk, the collisional timescale in the inner regions is much longer than the PR-drag timescale, so collisions would not smooth out features strongly. Nonetheless, more complete simulations that include not only grain collisions but also consider a range of possible orbital configurations are needed to confirm our result.

Our models raise the question: ``If a massive planet cannot be located within the dip (or nearby), what shapes the inner edge of the ALMA observed planetesimal belt?" Solving this riddle is the most obvious next step in understanding the Vega system. One possible solution is that a planet as small as 1$-$3 M$_{\oplus}$ may be able to maintain the inner edge without noticeably disrupting the disk morphology and creating the shallow gap through collisional blurring effects \citep{thebault12_planetsignature_collisionallyactivedisks} at the same time. Another, more difficult to assess, is that a past re-arrangement of the Vega planetary system could have left a previously well shepherded debris belt intact, while also removing the more massive planet responsible for its architecture. \citet{faramaz17_vega} addressed whether transport from the outer belt material inward could be achieved with comets released by interaction with an outer planet. They modeled the situation with a Jupiter-mass planet on a slightly eccentric orbit at a distance of 185 au (although such a planet is ruled out by the deep JWST search, \citealt{beichman24_vega}). In this instance, the 5:2 MMR coincides with the planetesimal belt at 100 au, allowing the MMR to be the site of a significant flow of comets inward. A modification of this model with lower planet masses is also worth exploring.

\subsubsection{Summary of indirect evidence for the unseen planets}
\label{sec:summary_indirect_planets}

Assuming the underlying planetary architecture of the Vega disk is similar to what the mid-infrared disk images suggest (i.e., an extended inner planetesimal disk from a few au to $\sim$100 au), we qualitatively explored the possibility that the observed dip is caused by one single gap-opening planet in Section \ref{sec:gap-opening-planet}. Given the fact that there is no detectable offset between the disk center and the star and the extreme disk circularity, any shepherding planet needs to have low ($\lesssim$0.1) eccentricity. Using the classical chaotic zone argument (e.g., \citealt{morrison_malhotra15}) plus the existing predictions for a collisionally active planetesimal disk (e.g., \citealp{thebault12_planetsignature_collisionallyactivedisks}), 
a planet with mass higher than the Saturn mass can be ruled out if the planet locates in the gap at $\sim$60 au. Our dynamical modeling in the preceding section indicates that the mass limit must be much lower to avoid imposing detectable structures on the inner disk. Similarly, a shallow gap can be opened by either an internal or external planet. In such cases, the low-eccentricity planet is expected to be at $\sim$38 au or $\sim$95 au, respectively, if the gap corresponds to the planet's 2:1 MMR. If the gap is a result of a migrating planet, the mass limit can be pushed to even lower values.   

In Section \ref{sec:pr_win}, we present the possibility that the extended inner disk emission originates from a PR-dominated phenomenon, and explore the potential constraints that such a drag-dominanted disk can provide on the mass of the unseen planets in Section \ref{sec:planet_perturbations}. We show that the collisional timescales are longer than that of PR drag for $\beta\lesssim$0.5 particles and that the apparent, axisymmetric disk morphology with a shallow dip/gap inside the ALMA detected planetesimal belt can be explained by either a planetless PR-dominated disk or such a disk with a planet with masses of $\lesssim$3 M$_\oplus$ within the gap. 

It is clear that the extreme disk circularity and the lack of azithmual structures rule out the presence of Saturn-mass planets in the Vega system despite the nature of the extended, inner disk. Whether the migrating planet or MMR hypotheses, either in an extended inner planetesimal disk or a PR-dominanted inner disk (Sections \ref{sec:planet_perturbations}), or something else that can account for the dip, explanations have to contend with the fact that the radial structure is smooth and featureless in the mid-infrared. This situation makes it clear that sophisticated modeling will be needed to tease out the driving process for the structure observed in the Vega system, without predicting additional structures. The largest shortcoming in the current modeling application to observations is neglecting collisions, which have a significant impact on the observable, planet-induced structures as they always significantly damp planet-induced spatial structures \citep{kuchner_start_2010,thebault12_planetsignature_collisionallyactivedisks,stuber23}. 
Future investigation including all observed properties and proper collision treatment will put tighter constraints on the mass of the planet or planets that might be responsible for the observed disk features.  Given the high physical resolution and quality of the MIRI data, the Vega system is ideal for advancing the theory and pushing the understanding of its planetary system to a higher level.

\subsubsection{An inner planet? and alternative explanations}
\label{sec:innerplanet}

We now direct our attention closer to the star, where we find that the warm debris must have an inner edge at $\sim$ 2$-$5 au and is unlikely to be directly connected to the hot excess produced in the sub-au region (Section \ref{sec:inner_radius}). This edge could be maintained by a planet close to the star. It is interesting to note that decade-long radial velocity measurements suggest the possibility of a planet at $\sim$ 0.045 au with a mass of $\sim$0.6 $M_J$ if aligned with the stellar spin  \citep{hurt21}, which is probably too close to affect the inner few-au, region even if it proves to be a real detection. There may be another planet at 0.85 au with a minimum mass of $\sim$0.25 $M_J$, ideally suited for the mechanism separating the hot and warm dust. However, this radial velocity signature can be linked to the stellar activity where the star's rotational modulation is clearly detected both in the radial velocity and TESS photometry, therefore, it is likely to be spurious \citep{hurt21}. Neptune-size planets, with a mass that would probably be sufficient to shepherd the inner edge of the disk, would be undetectable with current observing techniques, and therefore provide a possible mechanism to maintain the inner edge of the PR dominated, inner disk. This hypothetical planet might be the cause for the increasing optical depth behavior inside 
the 2\arcsec\ (15 au) region (see the bottom panel of Figure \ref{fig:tc_tau}), which is contrary to the relatively flat distribution expected from a drag-dominated disk. Further investigation is needed.

One possible explanation for the extra source of dust in the inner 2\arcsec\ region could be the inward scattering of exocomets from the planetesimal belt. This might be driven by a chain of very low-mass planets \citep{bonsor12,raymond_bonsor14}, or by mean-motion resonances with far-away exterior planets on moderately eccentric (e$_p$ $\ge$ 0.1) orbits \citep{faramaz17_vega}. The exocomet scenario also has been suggested to explain some of the bright exozodi detections from LBTI \citep{ertel20_hosts} in addition to the PR drag \citep{rigley_wyatt20_prdrag}. As discussed in the previous subsection, the existence of very low-mass planets in the Vega system cannot be completely ruled out, but they need to be on very low eccentricity orbits, and such a tightly packed configuration, indeed, favors higher scattering efficiency of exocomets \citep{bonsor14}. In this case, the optical depth distribution in the inner region might reflect the process of comet fragmentation and the subsequent evolution of released dust due to collisions and drag forces \citep{rigley_wyatt22_cometfragmentation}. 

Another explanation for the centrally peaked optical depth within 2\arcsec\ from the star is the formation of a dust ring, i.e., pileup of dust due to the interplay of PR drag and grain sublimation in a drag-dominated system as proposed by \citet{kobayashi09}. As a dust grain drifts close to the sublimation radius under PR drag and starts to sublimate (i.e., reducing its size from losing volatile content), the radiation force on the dust grain becomes stronger and temporarily halts its inward migration; therefore, a pileup ring (i.e., density enhancement) near the sublimation radius can form \citep{kobayashi09,vanlieshout14_dustsublimation}. 
The pileup ring can form at extended locations, depending on the sublimating material, such as lower sublimation temperature volatiles (different forms of ices) or higher sublimation temperature refractory materials (such as silicate-like or carbonaceous minerals), as dust grains would be a mixture of such materials. However, the derived color temperatures in the inner 1\arcsec\ region (see Figure \ref{fig:tc_tau} and Figure \ref{fig:inside_Td_constraint}) appear to be too high compared to the typical sublimation temperatures of water ice \citep{kobayashi11_sublimationtemperature}. Further sublimation calculations taking the detail properties of icy grains (i.e., volatile composition and the physical structure and conductivity of the grains) into account might validate the pileup scenario. If confirmed, no planet (even the very low-mass ones) would be needed to explain truncated warm debris in the Vega disk.

\section{Conclusion }
\label{sec:conclusion}

We present sub-arcsec resolution, mid-infrared images of the planetary debris system around Vega at 15.5, 23, and 25.5 $\mu$m obtained with JWST/MIRI. This star, along with Fomalhaut and $\epsilon$ Eri, constitute the three closest and brightest debris disks and together offer a unique opportunity to probe complex dust dynamics due to non-gravitational (radiation pressure, stellar winds, and associated drags) and gravitational (planetary perturbers) effects on debris structures. 

With a factor of $\sim$8 improvement in the spatial resolution over Spitzer, the overall disk morphology around Vega seen with JWST is surprisingly smooth and axisymmetric in the mid-infrared wavelengths. The disk is well centered at the star position within 0.2 pixel ($\lesssim$0.2 au) suggesting an extremely low eccentricity ($e \lesssim$0.002) well constrained, for the first time, from the inner 10 au region up to 100s au. In addition to the extended, symmetric disk morphology, a broad and shallow flux dip/gap at $\sim$60 au from the star is the most noticeable feature in the Vega disk images. The width and the depth of the dip/gap are wavelength dependent -- wider and deeper at longer wavelengths, $\sim$20 au wide with 20\% flux deficit at 25.5 $\mu$m.  Any azimuthal brightness asymmetry $>$10\% level can be ruled out at the 3 $\sigma$ level near the dip/gap area. 

Using the multi-wavelength observations, the distribution of the dust temperature and vertical optical depth, the two most important quantities in debris emission, are directly determined in the radial range of 1\arcsec$-$60\arcsec\ ($\sim$8$-$500 au) around Vega for the first time. Guided by the optical depth distribution, we define several major radial zones in the disk to better understand the nature of debris. The outer disk defined as $r \sim$10\arcsec$-$22\arcsec\ ($\sim$78$-$170 au) directly corresponds to the broad, planetesimal belt observed by ALMA where active collisions among leftover planetesimals continuously generate small dust grains through collisional cascades. Once the newly generated grains become small enough, Vega's intense radiation pressure pushes them either on highly elliptical or hyperbolic orbits, forming an extended disk halo up to hundreds of au. The inner disk, which we roughly defined as $r \lesssim$4\arcsec, exhibits an increased optical depth behavior, in contrast to the transition region (relatively flat) between the outer planetesimal belt and the inner disk.  The dip/gap apparent in all the MIRI images is found in the transition region, representing the lowest optical depth area interior to the outer planetesimal belt.  

Despite the unprecedentedly high-quality MIRI images of the Vega disk, our data cannot directly probe the innermost 1\arcsec\ ($\lesssim$7 au) region from the star due to the saturation and the blockage of coronagraphic masks. Using simple, parametric modeling on the inner disk radial profiles and well-calibrated infrared photometry for constraining the inner dust temperatures, an inner edge at $\sim$3$-$5 au is indirectly inferred. This inner edge is well outside the dust sublimation radius for typical refractory dust compositions and puts it physically separated from the sub-au, hot excess component detected by near-infrared interferometric measurements. The inner boundary of the warm debris might indicate a sub-Neptune mass planet inside and shepherding this edge.

Debris disk structures have been long recognized as a pathfinder to detect exoplanets indirectly, particularly for those below Neptune masses outside the giant planet zone (10s of au) that RV, transit and direct imaging planet detection methods are not sensitive. The extreme circularity and smoothness in the Vega disk morphology qualitatively indicate that there are no massive Saturn-mass planets outside 10 au of the star because such planets are expected to reveal their existence through disk substructures (offset center from the star and azimuthal asymmetries). To push the limits further, we present numerical simulations to model the debris distribution inside the outer planetesimal belt (inside $\sim$90 au from the star). 
We derive that the rate of collisions should not be high inside the outer belt. Assuming this is correct, we show that the debris in the inner region could come from dust particles that are produced in the outer planetesimal belt and migrate inward under the PR drag. A flux dip/gap can form near the inner edge of the planetesimal belt due the combination of strong temperature effects and differential drift rates on grain sizes, without invoking the presence of a planet. We further explore to what degree unseen planets in this region would create detectable deviations in the drag-dominant disk, and find that a planet of mass $\gtrsim$6 M$_\oplus$ at a circular orbit of 65 au (as suggested to maintain the inner edge of the planetesimal belt detected by ALMA) would induce interior asymmetric structures and disrupt the inner edge of the planetesimal belt. Future investigation including all available data and proper collision treatment will validate the nature of the inner debris emission and put tighter constrains on the mass of the planet or planets that might be responsible for the observed disk features.

\appendix 
\restartappendixnumbering

\section{The Whole Vega System Photometry in the Mid-Infrared}
\label{sec:appendix_photometry}

\subsection{Validating JWST photometry and Calibration}
\label{sec:appendix_25um}

Photometry with a large enough field of view to capture all the system flux is useful to: (1) provide assurance that the MIRI images captures all the flux, including low surface brightness components; and (2) as a check that the calibration of the JWST data is correct, given the slow degradation of the signals at this wavelength occurring when our data were obtained. 

We start with the data taken in the $\sim$25 $\mu$m band. 
The Infrared Astronomy Satellite (IRAS), with a detector field of view of 0\farcm75 $\times$ 4\farcm6 \citep{beichman88}, is one source of measurements. However, the calibration of these measurements is dated and potentially difficult to reproduce, so we have generated a new calibration as follows: (1) we identified early-type stars with no 25 $\mu$m excesses (e.g., from WISE) and well measured in the IRAS Point Source Catalog, namely Sirius, Procyon, $\alpha$ Aql, $\alpha$ Gem, and $\alpha$ Leo; (2) we used the Akari 9 $\mu$m measurements of these stars to normalize all of them to the expected flux from Sirius; (3) we averaged these predictions of the Sirius flux at 25 $\mu$m to reduced the errors, obtaining $32.9 \pm 0.6$ Jy; and (4) we used the relative fluxes at $\lambda < 9 \mu$m for Vega \citep{rieke23_absflxIII}  to determine a photospheric flux of 9.1 Jy for Vega at 25 $\mu$m in this system. The IRAS measurement of Vega is 11 Jy, indicating a nominal excess due to the debris system of 1.9 Jy. This value needs a bandpass correction for the large $\Delta \lambda$ of the IRAS filter, which reduces it to 1.77 Jy. Finally, since the IRAS beam does not capture the entire system, we determined the necessary correction from synthetic photometry on our F2550W image, yielding the second entry for IRAS in the table of 1.98 Jy. 

A measurement by the Infrared Space Observatory at 25 $\mu$m is reported by \citet{laureijs02}, using a 79\arcsec\ aperture. They derive an excess of 1.88 Jy, and their analysis of the IRAS data yields 1.73 Jy, in good agreement with our value. 

Yet another determination can be made from \citet{su05} by integrating the 24 $\mu$m radial profile. The flux from the saturated area at the position of the star needs to be added. We do so by matching the disk surface brightness between the MIRI and MIPS profiles at the radii of $\sim$18\arcsec-20\arcsec\ where the disk is well resolved and the potential missing inner flux has very little contribution, resulting in a total flux of 1.9$-$2.1 Jy in the MIPS 24 $\mu$m band.

Finally, integrating the flux in our F2550W image excluding the region within 1.5\arcsec\ that is affected by the saturation yields a total of 1.64 Jy, suggesting that there is a small amount of missing flux inside the saturated core. In Appendix \ref{sec:appendix_parametricmodel}, we derive the missing central flux as $\sim$0.4 Jy, resulting a total flux of $\sim$2.0 Jy in the MIRI F2550W band. Clearly, all of these measurements for the entire system at 25 $\mu$m are in agreement and they confirm both the calibration of our F2550W image and that it represents the total flux correctly. In other words, this comparison also validates the general JWST flux calibration to be within a few \% level.

Similarly, assuming there is no flux within 0\farcs5 in the F1550C disk image, the integrated flux is 0.3 and 0.4 Jy using the apertures of 4\arcsec and 9\arcsec, respectively. These values are $\sim$50$-$70\% lower (beyond the flux calibration uncertainty of $\sim$10 \%) than the large aperture photometry discussed in Appendix \ref{sec:appendix_otherphotometry}. There is also missing flux in the $r\lesssim$1\arcsec\ region in the F1550C disk image, which will be determined in Appendix \ref{sec:appendix_parametricmodel}.

\begin{deluxetable}{lcccccl}
\centering
\tabletypesize{\footnotesize}
\tablecaption{Ancillary Infrared Photometry of the Vega System \label{tab:photometry}}
\tablewidth{0pt}
\tablehead{
\colhead{$\lambda_c$} & \colhead{$F_\nu$} & \colhead{error} & \colhead{Ap.}  & \colhead{Excess} & \colhead{error} & \colhead{Note/Ref.} \\
\colhead{$\mu$m} &
\colhead{Jy} &
\colhead{Jy} &
\colhead{\arcsec}  &
\colhead{Jy} &
\colhead{Jy} &
\colhead{}
}
\startdata
1.65   & 1034  & \nodata  &  0.4  & 13 & 6 & [1,2,3] \\ 
2.16   & 672  & \nodata  & 0.4  & 8.6 & 1.3 &[4,1,3,5] \\ 
3.55   & 377  & 4  & \nodata  & 3.6 & 1.7 &[3,5]  \\ 
4.6  & 170  & 2  & \nodata  & 3.0 & 1.0 &[3,5]  \\ 
7.8   & 77.4  & 1  & \nodata  & 0.7 & 0.6 &[3,5] \\ 
8.5   & 53.1  & 1  & 0.2  & 0.28 & 0.12 &a, [6]  \\ 
9.8   & 40.3  & \nodata  & 12  & 0.03 & 0.2 &b, [7,8]  \\ 
10.1   & 37.9  & \nodata  & 3.9  & -1.25 & 0.9 &b, [9]  \\ 
11.1   & 31.4  & \nodata  & 1.23  &  0.12 & 0.03 &LBTI [10]  \\ 
12.13   & 27.62  & 0.40  & $\sim$18  & 1.27 &  0.39 &b, MSX [11]  \\ 
14.65  & 18.88  & 0.40  & $\sim$18  & 0.69 & 0.27 & b, MSX [11]  \\ 
18  & 12.87  & 0.18  & $\sim$7  & 0.85 &  0.18 &b, AKARI [12]  \\ 
20  & 10.37  & 0.21  & 3.9  & 0.62 & 0.20 & b, [8,9]  \\ 
21.34  & 10.07  & 0.3  & $\sim$18 &1.48 & 0.21& b, MSX [11]  \\ 
24   &   --  &  --  &  -- & 1.9 - 2.1 & -- & b, [13] \\
25   &   9.06  &  0.2  &  23 $\times$ 135  & 1.77 & 0.2 & b, [14] \\
25   &   --  &  -- &  23 $\times$ 135  & 1.98 & 0.2 & b, [15] \\
25  &  9.17  &  0.1  &  39.5  & 1.88 & 0.1 & b, [16] \\
25.5  & --  & -- & outside 1$''$ &1.76& & b, [17] \\
25.5  & --  & -- & total &$\sim$ 2.0   & & b, [18] \\
\enddata
\tablenotetext{}{Reference: [1] \citealt{absil08}, [2] \citealt{defrere2011}, [3] \citealt{rieke23_absflxIII}, [4] \citealt{absil06}, [5] \citealt{rieke22_sirius}, [6] \citealt{mennesson2014}, [7] \citealt{walker2000}, [8] this work, [9] \citealt{tokunaga84}, [10] \citealt{ertel20_hosts}, [11] \citealt{price04}, [12] \citealt{akari_irc_catalog}, [13] integrated from \citet{su05}, [14] IRAS, this work, [15] IRAS, corrected to full system in this work, [16] \citet{laureijs02}, [17] integrated from image in Fig.~\ref{fig:miri_pr}, [18] integrated from image in Fig.~\ref{fig:miri_pr} plus inner model}
\tablenotetext{a}{Leak measurement converted to flux as in \citet{kirchschlager2017}.}
\tablenotetext{b}{Excess relative to the sum of the photosphere and any hot excess}
\label{tab:IR_photometry}
\end{deluxetable}

\subsection{Other Ancillary Infrared Photometry Constraints on the Inner Excess}
\label{sec:appendix_otherphotometry}

There are many past infrared measurements that can constrain the properties of the hot/warm excess in the Vega system (some were discussed in \citealt{su13}). Most historical infrared photometry is calibrated in different filters and standard systems, making it difficult to synthesize measurements together. Progress has been made to build a uniform absolute flux calibration system by using Sirius as the standard \citep{rieke22_sirius,su22_SpitzerPSF,rieke23_absflxIII}. Another caveat for the measured infrared photometry is the vastly different aperture sizes used, depending on the telescope sizes and observing modes that probe different part of the debris emission as now we have a clear view of the Vega system. This subsection aims to provide useful constraints on the properties of the hot/warm emission by evaluating past infrared measurements.  

The accurate comparison across the infrared of Vega with Sirius in \citet{rieke22_sirius}, combined with the absolutely calibrated spectrum of Sirius in \citet{rieke23_absflxIII}, has been used to convert the interferometric excess values at 1.65, 2.2, and 11.1 $\mu$m \citep{defrere2011,absil06, absil13,ertel20_hosts} into fluxes, as tabulated in Table \ref{tab:IR_photometry}. For these interferometric measurements, we use the FWHMs of the interferometry response as the aperture, although they are not exactly the same thing. The interferometers used for the shorter wavelength  measurements have small inner working angles \citep{absil2021}, well below the sublimation radii for plausible dust compositions, so the measurements should be representative of the integrated signal. The values at 11.1 $\mu$m from the Large Binocular Telescope Interferometer (LBTI) have a number of caveats. For example, the inner working angle of the LBTI is $\sim$0\farcs039 (0.3 au at the Vega distance); therefore, it is possible that the hot component seen at $\sim$2 $\mu$m is inside of this radius and hence not fully detected. The overall calibration of the transmitted light may also not be highly accurate because half of the light is suppressed over the field of view.

\citet{rieke22_sirius, rieke23_absflxIII} also provide an opportunity to fill in the spectrum of the hot dust emission between the values from interferometry at 2.2 and 11.1 $\mu$m. To do so, we use the measurement of the hot excess above the photospheric output at 2.2 $\mu$m \citep{absil06,absil13} to determine the bare stellar flux density at this wavelength and use the SED of Sirius to extrapolate to the longer wavelength bands. An effective mean spectral type given the temperature distribution over the surface of Vega \citep{lipatov20} is A2V, but the results are very insensitive to the assigned type because of the insensitivity of A-star infrared colors at wavelengths longer than 2 $\mu$m. To reduce clutter, we have averaged the photometry and estimated the net uncertainties from the rms scatter of the whole suite of measurements \citep{rieke22_sirius}. The resulting values with fiducial wavelengths are provided in the table. 

\citet{walker2000} present a spectrum from 5 to 11.5 $\mu$m, obtained with ISOPHOT. However, the slope of this spectrum differs significantly from expectations for the Rayleigh-Jeans behavior of the star. Their spectrum of Fomalhaut also differs in slope from the one of Vega. To derive a photometric point, we used the accurate photometry at shorter wavelengths to determine a power-law correction to force their spectrum to agree with other measurements of Vega out to 8 $\mu$m and then used this corrected spectrum (with outlier rejection) to derive a photometric point at 9.8 $\mu$m. 

The measurement values from MSX are well defined, but the equivalent apertures are more challenging. \citet{cohen01} used aperture photometry on the MSX measurements, but the signal to noise was compromised in this approach and the final values used PSF fitting. The pixel scale for MSX was 18\arcsec\ and we take this to be the radius of the PSF. 

The 18 $\mu$m value from AKARI was determined by averaging the ratio of fluxes at 9 and 18 $\mu$m from this survey for the well-measured early-type stars without excess: Sirius, Procyon, $\alpha$ Aql, $\alpha$ Gem, and $\alpha$ Leo, and then comparing with the same ratio for Vega. The photometry is through a procedure similar to, but not identical to, aperture photometry, over a radius of 7\arcsec\ \citep{akari_irc_catalog}.

The point at 20 $\mu$m is from the paper establishing a network of mid-infrared calibration stars by \citet{tokunaga84}. At that time, it was customary to ascribe zero color to Vega as the zero point, and the result was that Sirius is shown as being blue by 6\% at this wavelength. We now know that Tokunaga had unknowingly discovered the debris system around Vega simultaneously with the discovery with IRAS \citep{aumann1984} and that the 6\% is an excess due to it.

\section{Constraints on the Inner Disk}
\label{sec:appendix_constraints_inner_disk}

\begin{figure*}
    \centering
    \includegraphics[width=0.49\linewidth]{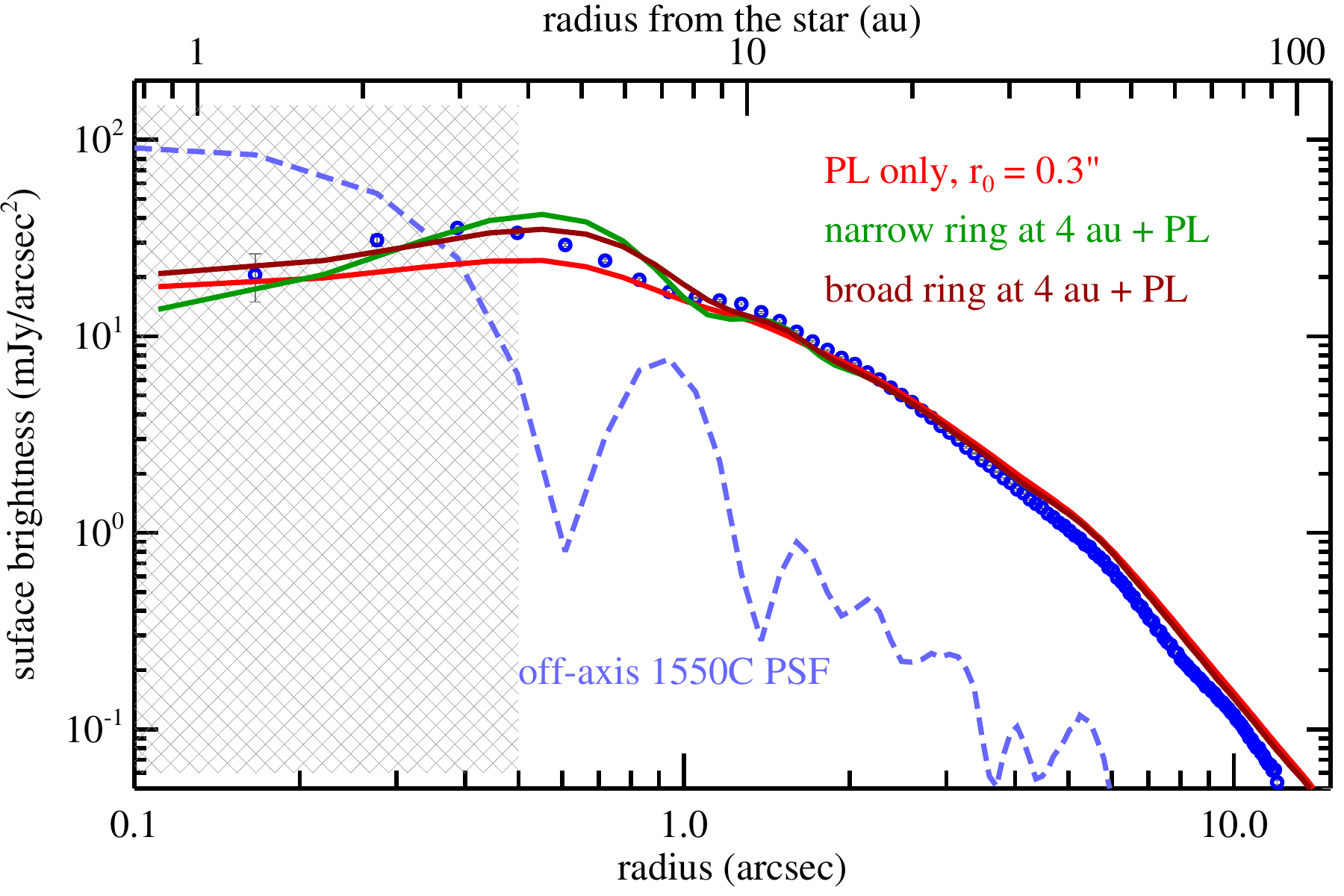}
    \includegraphics[width=0.49\linewidth]{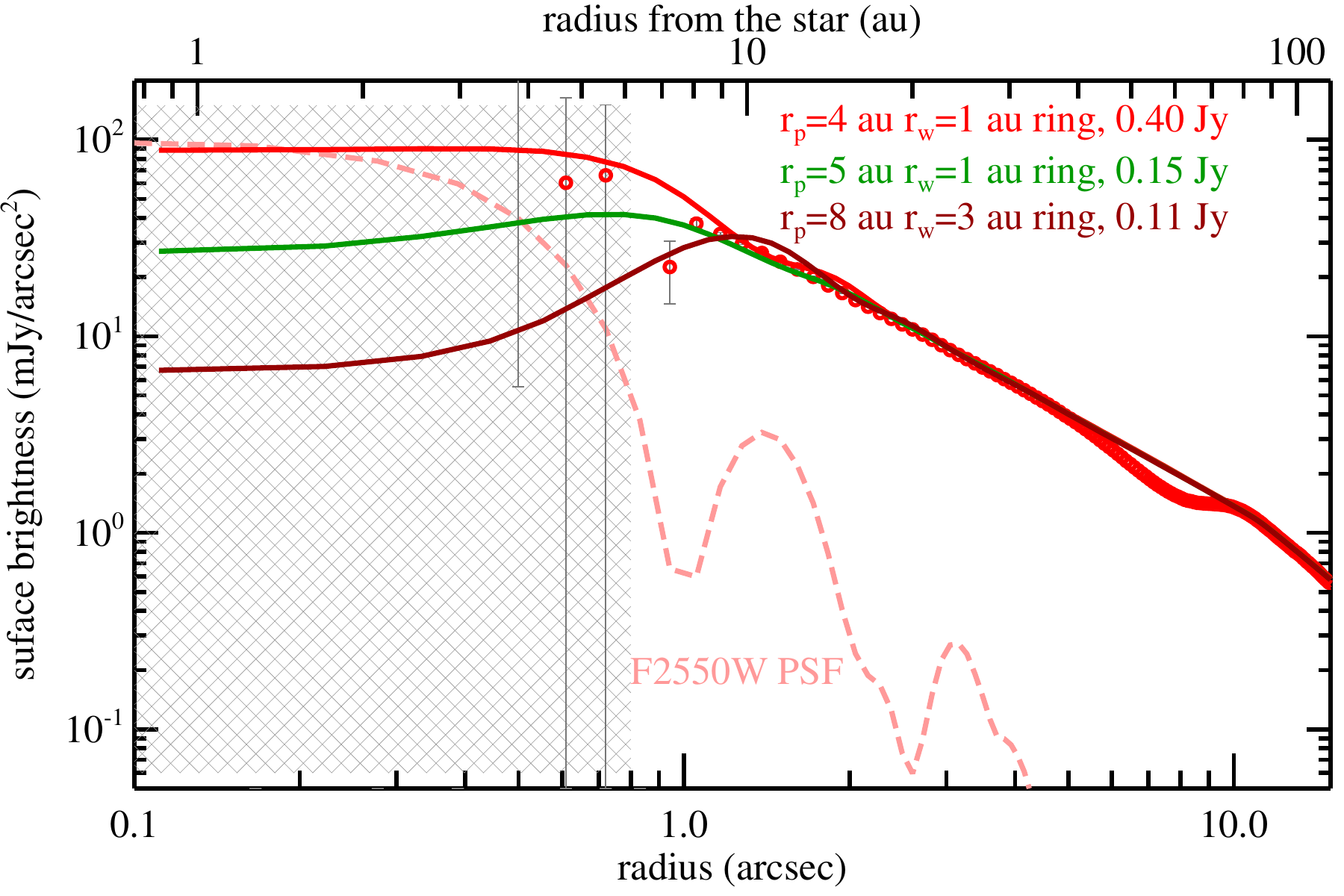}
    \caption{The left panel shows some examples of the acceptable parametric models at F1550C (colored lines) in comparison with the WebbPSF off-axis 1550C PSF (dashed line). The power-law (PL) only model extending to 0\farcs3 under predicts the brightness in the region of 0\farcs6$-$0\farcs8, while the ring plus PL models produce slightly higher values. Both narrow (1 au) and broad (3 au) rings with the same total flux of 0.14 Jy at 15 $\mu$m produce satisfactory fits. The right panel shows some examples of the acceptable parametric models at F2550W for three different peaked ring location ($r_p$) and width ($r_w$) for the inner disk shown on the plot. The hashed areas depict the region either under the coronagraphic mask for F1550C or affected by the saturation for F2550W. }
    \label{fig:parametric}
\end{figure*}

\subsection{Interpretation of the photometry of the inner disk}
\label{sec:appendix_intep_photometry}

Interpreting the photometry at 10$-$30 $\mu$m requires modeling. The first question to be addressed is whether the emission of the outer disk is significant for the shorter wavelengths. To obtain an answer, we took the F2550W flux of the outer zone ($>$10\arcsec\ radius) from this work (a total flux of 0.56 Jy), and far infrared measurements from IRAS, \citet{walker2000, su06, thureau14, pawellek14} and fitted a number of possible models for the outer disk. The model flux was no more than 10\% of the integrated flux at MIRI F2550W and $\sim$0.1 Jy or less at 20 $\mu$m in all cases. Short of 20 $\mu$m, we conclude that the emission of the outer disk can largely be ignored in the large-aperture photometry discussed previously. At longer wavelengths, we can constrain models using the multi-aperture photometry or synthetic photometry from our F2300C and F2550W images. 

To interpret the measurements further, we built a model constrained by (1) the radial profile at 15.5 $\mu$m; (2) the difference in signal at 20 $\mu$m between the SH IRS slit and the four times larger LH one; and (3) the photometry. The model assumed grains of astronomical silicates, with a size distribution of $n(a) \propto a^{-3.5}$, where $a$ is the grain radius, with the minimum grain size, $a_{min}$, a free parameter and the maximum grain size set to 400 $\mu$m (grains this large and larger have no influence on the results). The results were not strongly sensitive to the exact value of $a_{min}$. By varying the inner radius in the SED models, we find that inner radii $<$ 1 au produce too much flux in the 10$-$15 $\mu$m range to be fully consistent with the photometry. In addition, an inner radius of 1 au or smaller approaches the sublimation radius of plausible grain materials. In this case we expect a substantial piling up close to the sublimation radius as grains are eroded and their $\beta$ approaches the blowout size \citep[][and references therein]{kobayashi09,vanlieshout14_dustsublimation}. This effect would significantly boost the emission in the 10 $\mu$m region, putting it well above the photometry, suggesting that the disk has an inner edge well outside where grain sublimation would occur, i.e., beyond 1 au. 

This constraint is based on general physical principles. A more demanding but less direct constraint can be derived by profile modeling as described in the next section.

\subsection{Geometric Parametric Model fits for the inner disk} 
\label{sec:appendix_parametricmodel}

JWST is  diffraction limited for wavelengths longer than $\sim$1.1 $\mu$m  with a strehl ratio of 0.92 at 5.6 $\mu$m \citep{mcelwain2023}. The typical resolutions (FWHM) are thus  0\farcs5 and 0\farcs8 at 15.5 and 25.5 $\mu$m, respectively. This means that the power-law disk profiles (outside a few arcsec) are not affected by the instrument PSFs while the true inner disk profile might be affected by the PSF in addition to the coronagraphic mask and PSF subtraction residual. To provide additional constraints on the inner disk morphology (i.e., the excess emission profile), we fit the observed profiles using pure, geometric parametric models. The basic assumption is that the true disk surface brightness profile inside $\sim$1\arcsec\ follows simple descriptions: either a point source centered at the star, the same power-law extended inward from the outside disk profile, or a simple ring-like structure. Because of the face-on disk geometry, the observed profile would be the true model surface brightness profile convolved with the instrumental PSF (i.e., no assumption on the grain properties). The PSFs used for the convolution are generated using the WebbPSF tool (ver.\ 1.0.0, \citealt{webbpsf_tool}). The model PSF for the F1550C data is assumed to be off-axis at the exit pupil.

\begin{figure}
    \centering
    \includegraphics[width=\linewidth]{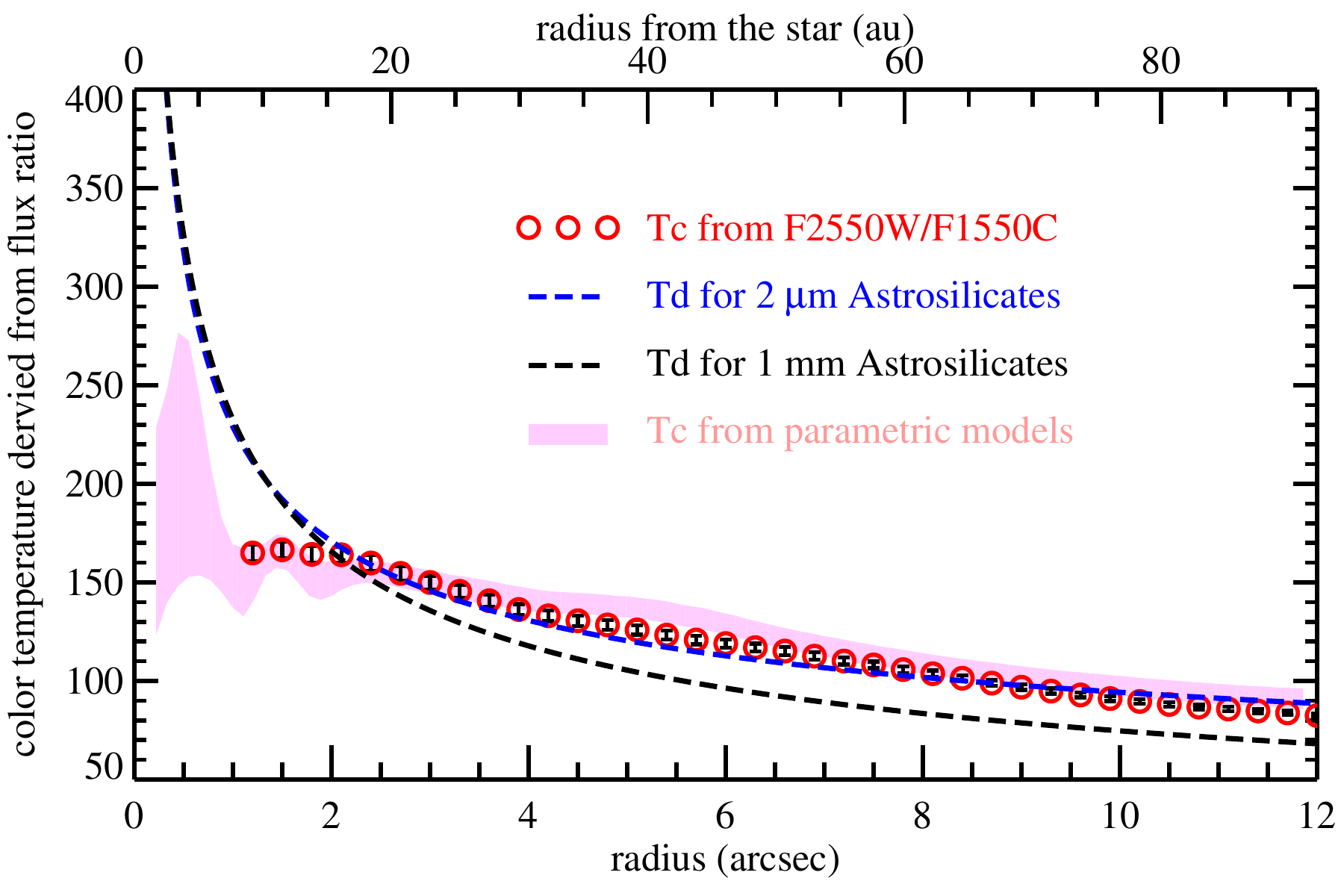}
    \caption{The color temperature distribution inferred from the flux ratio between F1550C and F2550W data. Red circles are derived using the observed F1550C and F2550W radial profiles as shown in Figure \ref{fig:radprof}. The acceptable temperatures using the ring-like parametric models are depicted as the pink color. The maximum dust temperature inside 1\arcsec\ region is $\lesssim$270 K, consistent with the warm excess SED (Figure \ref{fig:sed_innerexcess}) and no significant residual flux after PSF subtraction.  }
    \label{fig:inside_Td_constraint}
\end{figure}

We fit the parametric models in the F1550C and F2550W data separately with some selected fits shown in Figure \ref{fig:parametric}. It is quickly realized that the excess emission cannot be a simple, point source centered at the star; i.e., there needs to an inner cut-off for the excess emission no matter how bright/faint the point source is. This is consistent with the derived dust temperature presented in Sections \ref{sec:analysis} and \ref{sec:appendix_otherphotometry}. The next simplest model is the broken, power-law model which extends inward from the 1\arcsec$-$2\arcsec\ region with an inner cut-off at $r_0$.  The simple power-law model suggests that $r_0$ cannot be smaller than 0\farcs3 (see left panel of Figure \ref{fig:parametric} for an example) and $r_0$ cannot be larger than 0\farcs7 (see right panel of Figure \ref{fig:parametric} for an example). The inner cut-off of the excess emission needs to be in the range of $\sim$3$-$5 au from the star. This suggests that the inner disk (excess emission) is partially resolved at both 15 and 25 $\mu$m. We, then, explore a slightly complex model by assuming the inner disk profile can be best described as a Gaussian ring with a peaked radius $r_p$, a width $r_w$ and a total flux of excess $F_{in,ring}$. Various combinations of $r_p$, $r_w$, and $F_{in,ring}$ are tried, and $r_p$ and $r_w$ are degenerate (i.e., there is not much difference between a narrow (unresolved) and a broad (resolved) ring at most ring locations (except for a large ring at $\sim$8 au radius, only a narrow ring is allowed). Furthermore, the ring location ($r_p$) is also degenerate with the total flux in a way that the larger the ring the lower the total flux (the total 25 $\mu$m flux ranges from $\sim$0.4 Jy for $r_p \sim$ 3 au to $\sim$0.1 Jy for $r_p \sim$ 8 au). Combining with the result in the F1550C data, it seems that a ring at 4$-$5 au with a total flux of $\sim$0.14 Jy at 15 $\mu$m, and $\sim$0.4 Jy at 25 $\mu$m best describes the observed profiles. We also performed consistency check by calculating the inferred color temperatures using the acceptable ring-like models in both F1550C and F2550W. The range of color temperatures is shown in Figure \ref{fig:inside_Td_constraint}, suggesting that grains warmer than $\sim$170 K might exist in the inner disk (region inside the saturation core), but the maximum allowable temperature is $\lesssim$270 K, consistent with (1) the warm excess SED (Figure \ref{fig:sed_innerexcess} and (2) no substantial PSF residual (i.e., existence of barely resolved component close to the star) in the PSF subtracted images. In summary, although we cannot exactly constrain the inner disk morphology, the excess emission needs to have an inner cut-off at 3$-$5 au and can be in a form of power-law or ring-like profile with a total flux of $\sim$0.14 Jy, and $\sim$0.4 Jy at 15 $\mu$m and 25 $\mu$m. 

\section{Details of the dust transport and planetary perturbation simulations}
\label{transport}

\subsection{Collisional timescales in the planetesimal belt}

Particle loss in a planeteismal belt is due to two effects: (1) destruction by collisional cascades that generate smaller grains until they are small enough to be blown out of the system by stellar radiation pressure; and (2) inward migration due to PR drag drawing the particles toward the star. To understand the net effects of grain dynamics, one ought to balance these timescales. Here we estimate the collisional timescales near the inner edge of the outer planetesimal belt around Vega.

\begin{figure*}
    \centering
    \includegraphics[width=0.89\linewidth]{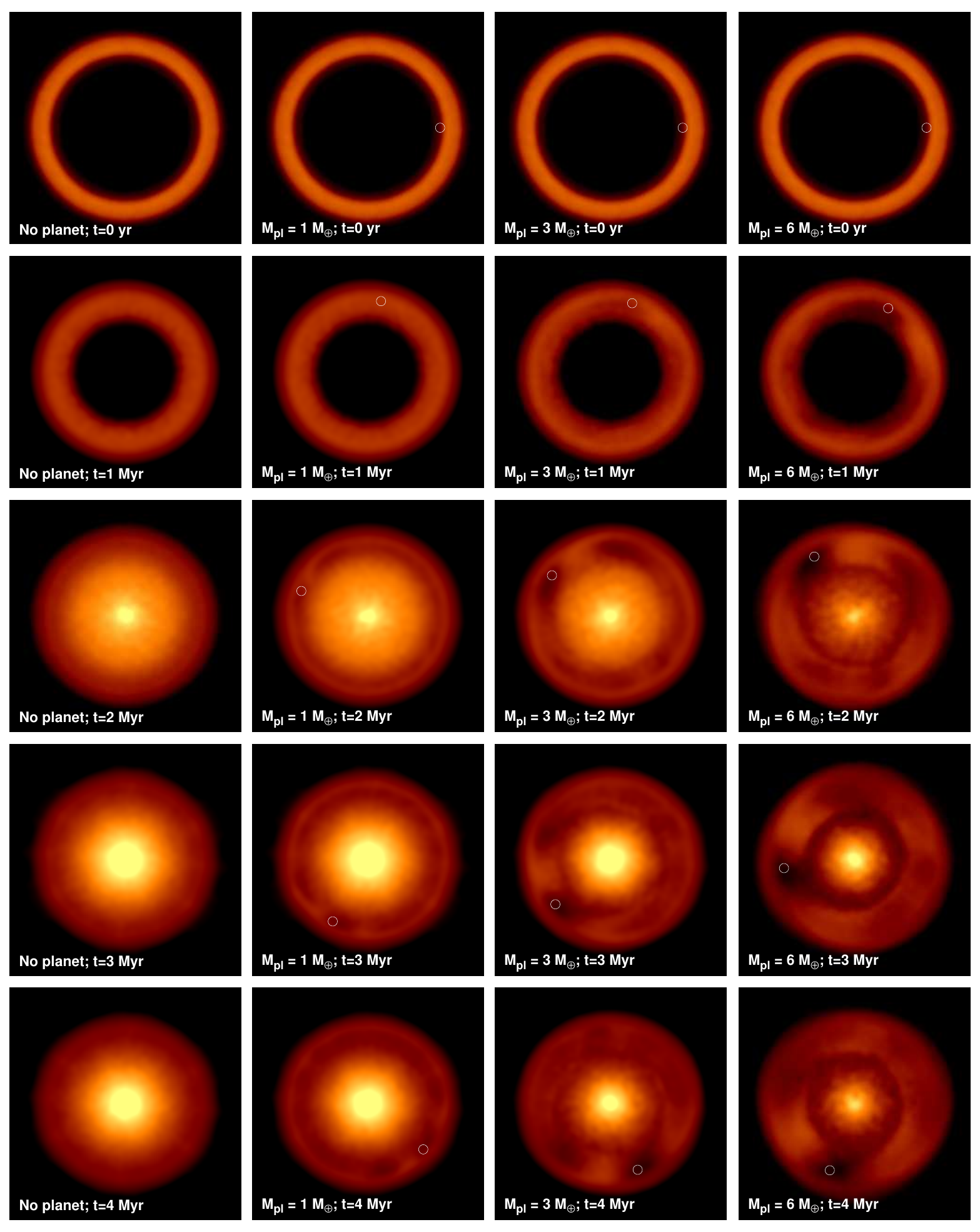}
    \caption{Disk surface brightness simulations highlighting the PR-drag transport of dust particles from the inner zone of the planetesimal belt located between 69 and 85 au. Each row represents different timestamp of the evolution, starting from time zero from the top to 4 Myr on the bottom. The first column has no planet while the following columns show the simulations where a 1, 3, and 6 M$_{\oplus}$ planet, indicated as the white circle, was placed at 65 au. All images were convolved with a F2550W WebbPSF and shown in the same display scale.}
    \label{fig:ALL}
\end{figure*}

For a particle swarm (i.e., a debris disk), a collision for a particle (with a grain radius of $a_1$) will occur if there is another particle (with a radius of $a_2$) within a ``collisional volume'' (V$_{\rm col}$), which is the collisional cross section multiplied by the distance the particle traverses relative to the swarm, i.e.,
\begin{equation}
    V_{\rm col}(a_1, a_2) = \pi \left(a_1 + a_2\right)^2 \ v_{\rm sw} \  t_{\rm col}\;,
\end{equation}
where $v_{\rm sw}$ is the relative velocity of a particle with respect to the swarm and $t_{\rm col}$ is the collisional timescale. 
Multiplying the collisional volume with the number density of particles, $n(a)$, yields the total number
of particles in the volume. Setting that to 1 allows us to calculate the collisional timescale of the smallest particle not removed by radiation pressure, as
\begin{equation}
    t_{\rm col}(a_1) = \left[n(a)\ \pi \ (a_1+a)^2 \ v_{\rm sw}\right]^{-1}\;.
\end{equation}
Integrating this function for the entire size distribution between the minimum and maximum grain sizes ($a_{\rm min}, a_{\rm max}$), we get
\begin{equation}
    t_{\rm col}(a_{\rm min}) = \left[\pi v_{\rm sw}\int_{a_{\rm min}}^{a_{\rm max}}n(a)(a_{\rm min}+a)^2\ da\right]^{-1}\;.
\end{equation}
The relative velocity of a particle in the swarm can be expressed as a function of the eccentricity and inclination deviations in the system \citep{lissauer93}, and is typically 5--10\% of the local Keplerian velocity. The size dependent number density of the colliding particles can be expressed as
\begin{equation}
    n(a) = C a^{-\eta_a}\;,
\end{equation}
where $\eta_a$ ranges from $\sim$3$-$4 \citep{dohnanyi69,pan_sari05,pan_schilichting12,gaspar12b}. The constant $C$ is calculated from the total mass of the dust population ($M_{\rm tot}$) as
\begin{equation}
    C = \frac{3 M_{\rm tot}\left(4-\eta_a\right)}{4\pi\rho\left(a_{\rm max}^{4-\eta_a}-a_{\rm min}^{4-\eta_a}\right)}\frac{1}{V_{\rm disk}}\;,
\end{equation}
where $\rho$ is the bulk density of the dust. $V_{\rm disk}$ is approximately 30,000 au$^3$ for the Vega cold planetesimal belt assuming a geometrically thin disk ranging from $\sim$69 to $\sim$85 au with a thickness of 4 au (an opening angle of $\sim$3\arcdeg).

We use our dynamical numerical code DiskDyn (\url{https://github.com/merope82/DiskDyn}) to estimate
the dust distribution within the model region that yields
a total flux consistent with the observed F2550W flux ($\sim$0.17 Jy). To produce an order-of-magnitude estimate, we simply use astronomical silicates \citep{li_draine01} with a bulk density of 3.5 g cm$^{-3}$ in the DiskDyn code, resulting in a blowout size  ($\beta$ = 0.5) of 6 $\mu$m around Vega ($M_\ast$ = 2.135 $M_\sun$, $L_\ast$ = 48 $L_\sun$), and a total disk mass of 0.28 M$_{\oplus}$ between 6 $\mu$m and 1000 km in the particle swarm, assuming a particle size distribution slope of 3.67 \citep{gaspar12b}. Plugging in these values, which we we derived with DiskDyn to match the observed 25.5 $\mu$m flux for the model region, we find that the collisional timescale of the smallest, non-blown-out grain with $\beta=0.5$ is $\sim$0.5 Myr, assuming a swarm collisional velocity of 500 m s$^{-1}$ (10\% of the local Keplerian velocity). Larger particles will have even longer collisional timescales, as the smallest particle sizes that can catastrophically destroy them is also larger, thereby decreasing the frequency of such collisions (i.e., prolonging the collisional timescales). For $\beta = 0.1$ particles, the estimated collisional timescale is $\sim$0.73 Myr.

\begin{figure*}[]
    \centering
    \includegraphics[width=0.49\linewidth]{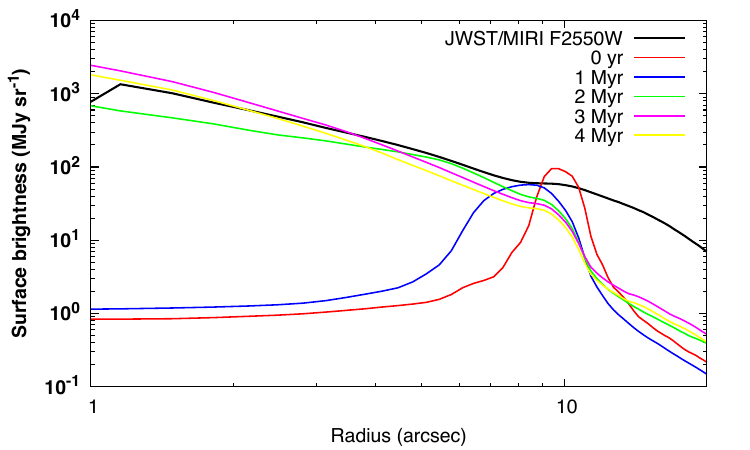}
    \includegraphics[width=0.49\linewidth]{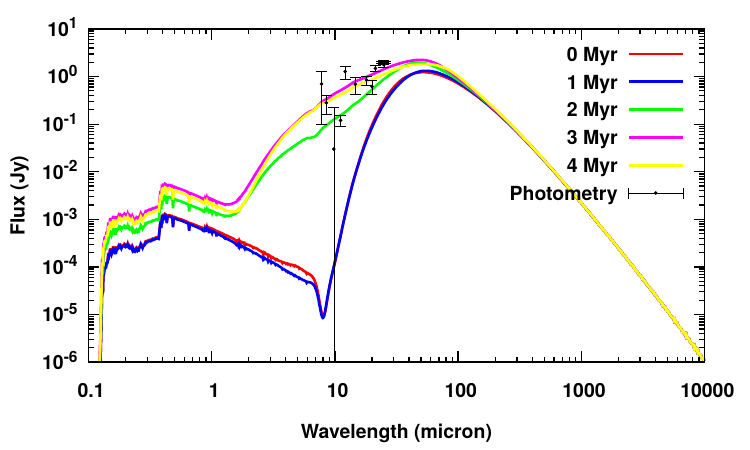}
    \caption{{\it Left panel:} The radial surface brightness profile of the ``no planet'' dynamical model at various evolution times, compared to the observed JWST/MIRI F2550W profile.
    {\it Right panel:} The SED of the inner disk for the 
    ``no planet'' simulation at various stages of evolution, compared to the mid-infrared photometry.}
    \label{fig:Surf}
\end{figure*}

\subsection{Disk morphology evolution with and without planetary perturbations}

As shown in Section \ref{sec:pr_win} (Figure \ref{fig:PRtimeKBA}), the PR timescales for grains with $\beta \lesssim$ 0.5 inside the model region are much shorter than the collisional timescale, i.e., the dust dynamics inside the $\sim$69 au region are dominated by the PR drag. Using the same setup as above, we then follow the dynamical evolution of particle distribution interior to the inner edge of the planetesimal belt using DiskDyn without collisions (i.e., no new grain input from the planetesimal belt nor collisional interaction among them). We ran four simulations on the High Performance Computer cluster Ocelote of the University of Arizona: one without any planetary perturber, and the other three with a planet placed at 65 au with masses of 1, 3, and 6 M$_\oplus$. The largest mass of 6  M$_\oplus$ was chosen to match the lower mass limit given by \citet{matra20} for a shepherding planet, necessary to maintain the inner edge of the observed ALMA planetesimal belt. The model dust distribution was convolved with a JWST F2550W WebbPSF \citep{webbpsf_tool} for morphological comparison inside 85 au.
Because our simulations do not include collisions, they do not completely reproduce the inward flux of particles from the planetesimal belt and beyond. To provide a better understanding of the transport and the perturbations from the planet, we show a full image sequence of the simulations at various timestamps in Figure \ref{fig:ALL} up to 4 Myr.

The structures seen in Figure \ref{fig:Evolved} (at 2 Myr) become even more pronounced by 4 Myr of evolution at all cases. The lack of newly produced particles entering the stream is also evident, as the dip between 40 and 75 au becomes more pronounced at the end of the simulation particularly for the no planet case (left column). Figure \ref{fig:Surf} show the radial surface brightness profile and corresponding SED for the no planet case to better illustrate the unimpeded PR disk evolution. By 1 Myr of evolution, the smallest grains transported inwards by PR-drag make it to $\sim50$ au and the flux distribution starts to flatten out. At 2 Myr (green line), the radial surface profile is only slightly shallower than the observed one (black line) and the transported grains significantly increase the mid-infrared flux in the SED evolution (right panel of Figure \ref{fig:Surf}). This 2 Myr limit is close to the timescale  (1.8 Myr) for the smallest particles to reach the sublimation distance from the inner edge of the planetesimal belt. The agreement between the observations and predicted SEDs for 2$-$4 Myr corroborates that dust transport via PR drag into the inner regions is a viable model.

The right three columns of Figure \ref{fig:ALL} show the evolution of PR-drag disk morphology under the influence of a perturbing planet with three different masses. It is evident that the higher the mass of the planet, the larger the departure from the smooth, symmetric,  PR-dominated disk morphology. Disrupting the inner edge of the planetesmial belt and highly structured MMR features are expected in the planet cases with masses $\gtrsim$3$-$6 M$_\oplus$. Further long-term numerical models with a proper treatment of collisional outcome will be needed to investigate whether such small mass planets are able to shepherd the planetesimal belt over the $\sim$400 Myr age of the system.

\section{Technical Steps for the Data Reduction and PSF Subtraction}
\label{fy_section}

\begin{table*}
\caption{JWST MIRI observations of the Vega system presented in this study \label{tab:JWSTobs}}
\begin{tabular}{clllccll}
\hline
\hline
Obs.\ Id. & Target        & MIRI Mode/Subarray   & PA$^{\S}$& N$_{\rm group}$ & N$_{\rm int}$ & Dither & Time (s) \\
\hline
20 & Background           & F2550W/BSKY           & 133.08   & 5        & 52     & 4pt Cyc$^{\dagger}$  & 1076.4  \\
21 & HD 169305 (PSF Ref.) & F2550W/BSKY           & 132.93   & 5        & 52     & 4pt Ext$^{\dagger}$  & 1076.4  \\
22 & HD 169305 (PSF Ref.) & F2300C                & 132.91   & 74       & 9      & 9pt SGD       & 1965.4  \\
23 & HD 169305 background & F2300C                & 132.96   & 74       & 9      & 2pt           & 436.8  \\
24 & HD 169305 (PSF Ref.) & F1550C                & 132.85   & 143      & 6      & 9pt SGD       & 1861.6  \\
25 & HD 169305 background & F1550C                & 132.87   & 143      & 6      & 2pt           & 413.7   \\
26 & Vega (Rotation 1)    & F2550W/BSKY           & 138.35   & 5        & 52     & 4pt Ext$^{\dagger}$  & 1076.4  \\
27 & Vega (Rotation 1)    & F2300C                & 138.36   & 60       & 49     & None          & 968.1  \\
28 & Vega background      & F2300C                & 133.50   & 60       & 49     & 2pt           & 1936.2  \\
29 & Vega (Rotation 1)    & F1550C                & 138.34   & 90       & 42     & None          & 915.8  \\
30 & Vega background      & F1550C                & 133.43   & 90       & 42     & 2pt           & 1831.6  \\
31 & Vega (Rotation 2)    & F1550C                & 128.34   & 90       & 42     & None          & 915.8  \\
32 & Vega (Rotation 2)    & F2300C                & 128.36   & 60       & 49     & None          & 968.1  \\
33 & Vega (Rotation 2)    & F2550W/BSKY           & 128.38   & 5        & 52     & 4pt Ext$^{\dagger}$  & 1076.4 \\
\hline
\end{tabular}
\tablenotetext{\dagger}{The BRIGHTSKY subarray with 4-point dither pattern starting at \#6 position.}
\tablenotetext{\S}{On-sky position angle of the MIRI aperture.}
\end{table*}

The data acquisition and reduction for this paper draws not only on the Vega measurements but also experience with those of Fomalhaut \citep{gaspar23_fom}. Here, we provide a detailed description of what we consider to be an optimized procedure as a reference. 

The data include coronagraphic imaging using the F1550C and F2300C masks as well as non-coronagraphic imaging with the F2550W filter using the BRIGHTSKY subarray, as detailed in  Table \ref{tab:JWSTobs}. Dedicated background observations were obtained for both imaging and coronagraphic modes for better subtraction of the known instrumental artifacts and the dominant telescope background at 25.5 $\mu$m.  The sequence of observations was optimized to minimize the instrumental latent artifacts at all wavelengths, most importantly at 25.5 $\mu$m, where long exposures on the two bright stars were used. Therefore, we began the entire sequence by first obtaining the background observation with the F2550W filter, thereby ensuring no latent patterns in the critical background image.

We selected the PSF reference star, HD 169305, putting a high priority on being able to use observation sequences as similar as possible on both stars\footnote{This was necessary for the PSF observations, as we used the background observation to correct the dark noise of the target and PSF reference observations, which is not well defined for the short integration ramps we had to use for these bright targets.  Therefore we had to execute an identical observing sequence for the background, the science target, and the PSF reference at 25.5 $\micron$. The similarity requirements were relaxed slightly for the coronagraphic observations.}. The reference star was chosen based on the following criteria: similar brightness as Vega in the mid-infrared, no signs of multiplicity, and proximity to Vega. The first criterion is the most important one due to the brighter-fatter-effect in MIRI detectors \citep{argyriou23_brighter_fatter}. This would result in different point source images for two stars of significantly different signal strength. Proximity between the target and PSF reference improves  observing efficiency and  minimizes variations in the telescope figure and background levels. Although  HD 169305 is a M2III giant, the observed wavelengths are in  the Rayleigh-Jeans regime of the stellar SED, also where there are no strong absorption features. 

For the F2550W imaging mode, the BRIGHTSKY subarray (field of view of 56\arcsec$\times$56\arcsec\ and 0.865 s sampling time) allowed short integrations that minimized the central area affected by saturation. For the coronagraphic mode, the field of view is $\sim$14\arcsec$\times$14\arcsec\ with inner working angles of 0\farcs49 and 2\farcs16 for F1550C and F2300C respectively. The Vega observations were repeated at two sky orientations separated by $\sim$10\arcdeg\ in both imaging and coronagraphic modes (see Table \ref{tab:JWSTobs} for details) to facilitate angular differential, high-contrast imaging reduction.

The raw data (\texttt{uncal} files) were first reduced with the pipeline through stages 1 and 2, producing flux calibrated individual exposures (\texttt{cal} files). We used version  1.11.4 for the F2550W data (and confirmed that a scaling factor of 1.1478 corrects the results to more recent versions). For the other bands, we used versions 1.12.0, and 1.12.3.  of the pipeline with the reference files in CRDS context of \texttt{jwst\_1193.pmap} that include the detector time-dependent sensitivity correction. The calibration factors (PHOTMJSR, conversion factor used to translate DN/s/pixel to MJy/sr) are 3.887, 1.071, and 0.874 for the F1550C, F2300C and F2550W data, respectively\footnote{The current photometric calibration uncertainties are of order 3\% or better at all imaging data (\url{https://jwst-docs.stsci.edu/jwst-calibration-status/miri-calibration-status/miri-imaging-calibration-status}).}. Our final reductions are carried out in a self-calibrated mode (i.e., matching the stellar signals empirically), so the details of the absolute calibration do not affect our results. As future improvements in the absolute flux calibration become available, one can simply apply a scaling factor in the measured fluxes without influencing our conclusions.

Within the pipeline, we kept the majority of settings at their default values, only changing the jump rejection threshold to 5 (from the default value of 4) and turning off the dark correction (\texttt{dark\_current.skip}). The dark subtraction was done using the background images as custom dark frames, which were observed in the same integration settings as in the science targets. Given the short integration ramps (e.g., 5 groups at 25.5 $\micron$), using custom dark frames provided more accurate reductions than using the CRDS dark frames, which are optimized for longer ramps.

\subsection{The F2550W non-coronagraphic imaging reduction}
 
As demonstrated for Fomalhaut \citep{gaspar23_fom} 
and now these Vega observations, traditional PSF subtraction is a powerful high-contrast imaging mode for MIRI. It allows smaller inner working angles 
than the  coronagraphs (dependent on saturation) and also a larger field of view (dependent on subarray and dither pattern used). It provides these benefits at a similar wavelength to the 23 $\micron$ coronagraph. At 15 $\mu$m, it also avoids the  aberrations induced by the phase mask.

For PSF subtraction, after the stage 1 and 2 reductions we performed custom reduction steps using IRAF and idp3 to minimize the instrumental artifacts and those from PSF subtraction as follows. The first step was to generate the background image to subtract from each science and PSF reference image. We initially generated a median combined background image from all four dither positions using 3$\sigma$ rejection.  For both science and reference star data, we found that the background subtracted image in the individual dither positions contained deeper row effects as the dithering sequence advanced. To include this effect in the background image used for corrections, we generated difference images between the first and other dither position data (P2-P1, P3-P1, and P4-P1), and calculated the median row values, within a 2$\sigma$ clipping factor. These row corrections were then added to the previously generated median combined image to produce a dithering-sequence-dependent but median-averaged background correction, which was then subtracted from all science (Vega) and PSF reference (HD 169305) data for each of the dither positions. 
 
\begin{figure}
    \centering
    \includegraphics[width=0.5\textwidth]{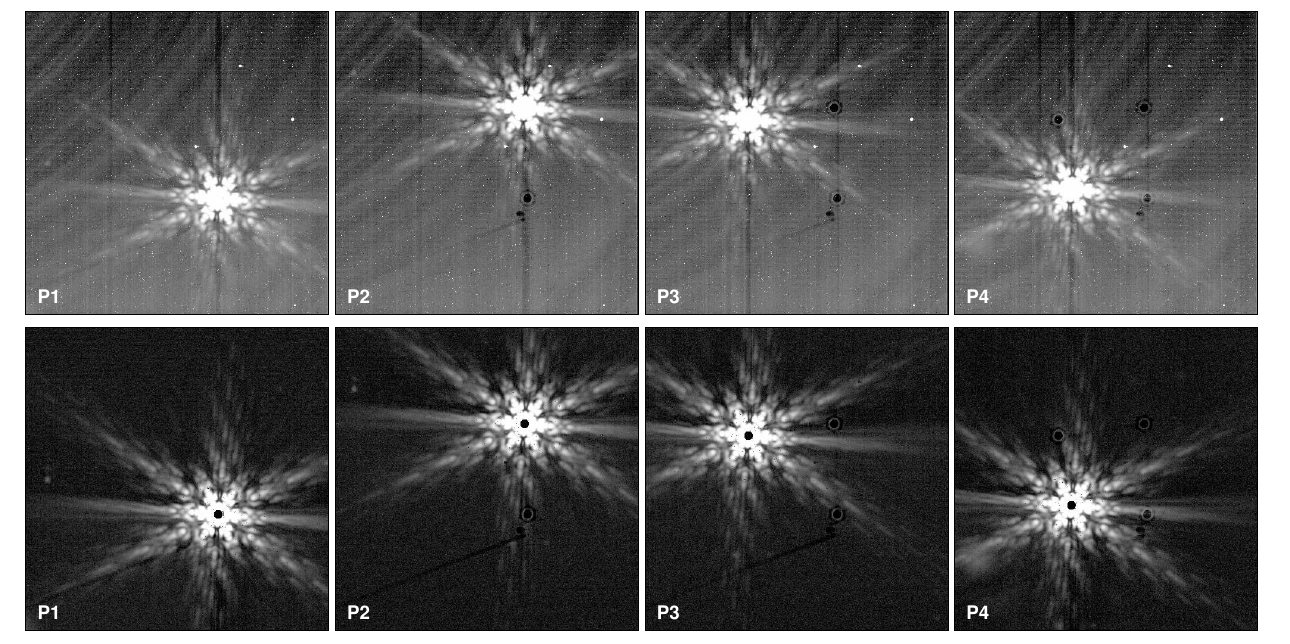}
    \caption{{\it Top row:} The F2550W PSF (HD 169305) images following the stage 1 and 2 pipeline reduction steps, without background or other corrections applied at the four dither positions (P1 to P4). Image scaling is logarithmic between 911 and 2066 MJy sr$^{-1}$. {\it Bottom row:} The same images, but with background correction applied, row and column artifacts removed, and background base level adjusted. The images are scaled logarithmicaly here as well, between -6.9 and 1147 MJy sr$^{-1}$.}
    \label{fig:imageclean}
\end{figure}

While this process cleared the majority of the thermal background signal, there were still row/column artifacts and dither-dependent, pedestal background offsets. To mitigate them, for each of the dither positions we determined the pedestal background level using a box size of 100 pixels$\times$50 pixels in clean areas of the images as far from the target as possible. For positions 1 through 4, the pedestal levels were 0.25, -1.61, -2.82, and -3.02 MJy sr$^{-1}$ for the PSF star, which were subtracted from the individual dither images. The remnant background levels for the Vega observations were somewhat more negative, with values between -8.32 and -11.15 MJy sr$^{-1}$, which were also subtracted and corrected. The saturated core also created flux depression in the detector column direction with $\sim$10 pixel width near the stellar core, which was manually masked and corrected. The background in the Vega data (for both sky orientations) was removed in a similar approach. Figure \ref{fig:imageclean} shows the comparison between the pipeline reduced and our final (background cleaned, zero level adjusted, column artifact corrected) PSF images for the 4 dither positions.

\begin{figure*}
    \centering
    \includegraphics[width=0.76\textwidth]{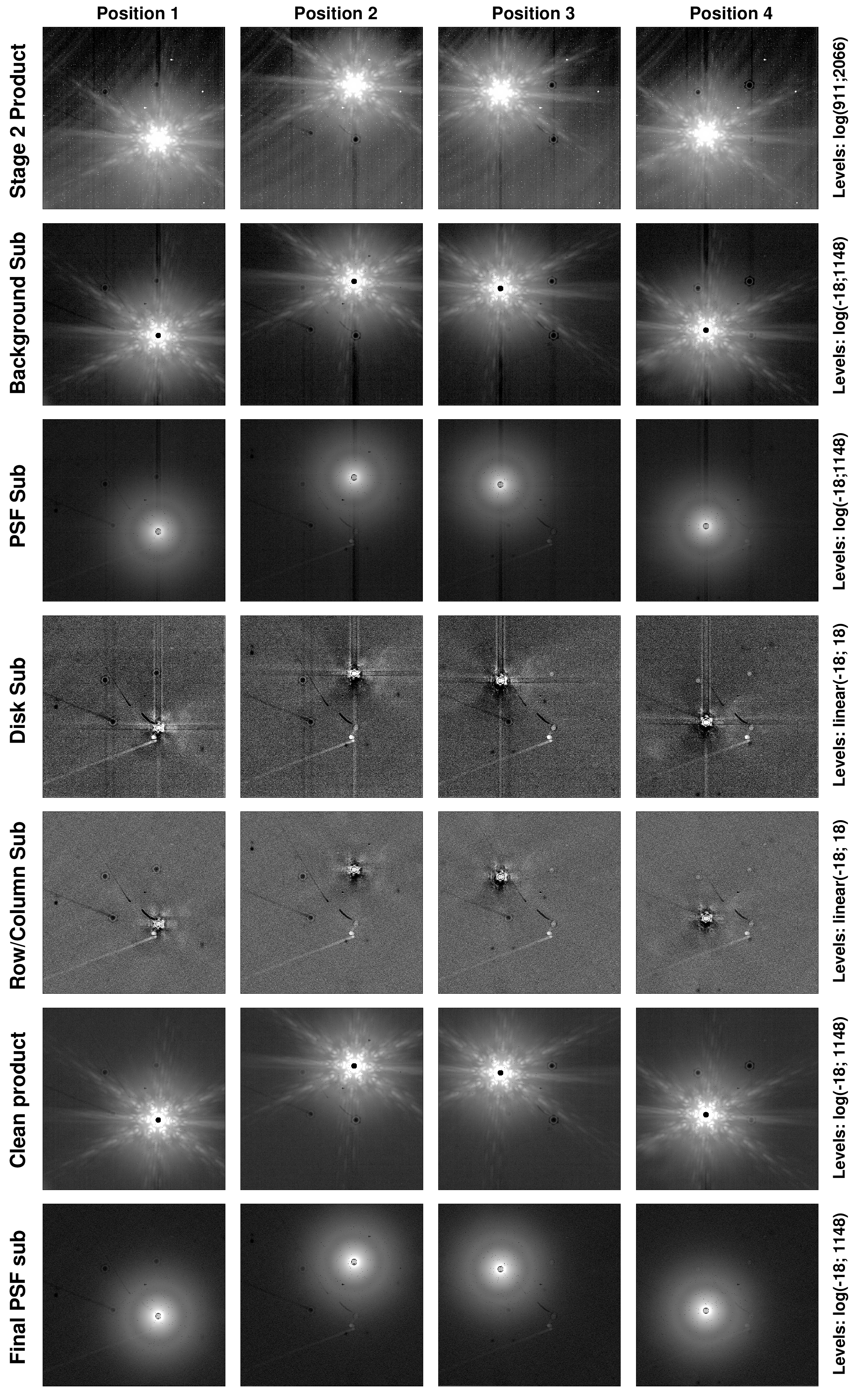}
    \caption{The iterative reduction sequence for the F2550W dataset. The columns show the four dither positions of the Vega data displayed in logarithmic scale with the boundaries (in MJy sr$^{-1}$) given on the right side of each row. Details about the sequence (from top to bottom) are itemized in Section \ref{fy_section}. 
    }
    \label{fig:Vega_Rot1red}
\end{figure*}

The pole-on viewing orientation of the Vega system and radially symmetric structure of the disk simplified the removal of residual background and the row and column artifacts  via an iterative process as showcased in Figure \ref{fig:Vega_Rot1red}. These same steps were also applicable for the reference star. Classical PSF subtraction (target $-$ PSF reference) per dither position was also done during this process. We note that this iterative process assumes a face-on, axis-symmetric geometry, and  therefore might not be applicable to other more complex conditions. The steps are: 

\begin{enumerate}
    \setlength\itemsep{-0.5em}
    \item Determining the position matched PSF pixel offset at each dither position.
    \item Shifting and rotating the science and PSF images to a common grid, scaling and subtracting the PSFs, and median combining the non-distortion corrected disk images using custom masks.
    \item Calculating a median disk radial profile.
    \item Removal of this median disk radial profile and the PSFs from the original background corrected images.
    \item Calculating the column and row corrections on the residual images (see details below).
    \item Removal of the residual column and row corrections from the original background corrected science images.
    \item Repeating the previous four steps until the science images are free of most of the artifacts.
    \item Application of stage 3 distortion corrections to the cleaned PSF and science images.
\end{enumerate}
Any row and/or column artifacts that are not removed previously are ultimately removed  via the PSF subtraction step.

In \cite{gaspar23_fom}, we showed that, with well-reduced images,  the stellar contributions can be subtracted with minimal residuals for F2550W observations using PSF images matched by position in the focal plane. Therefore, following the steps above, we removed the stellar portion of the Vega images using PSF observations taken at the identical positions in the focal plane. In these steps, we used the non-distortion corrected images, as our goal was to estimate the row and column artifacts in the original detector coordinate space. In the third row of Figure \ref{fig:Vega_Rot1red}, we show the four PSF-subtracted Vega disk images obtained at the first rotation following Steps 1 and 2. These images allowed us to estimate a preliminary median disk radial profile, which we were then also able to remove, revealing the residual row and column artifacts (shown in the fourth row of Figure \ref{fig:Vega_Rot1red}). The row and column artifacts were fit as slopes this time (rather than median values) and  separated into four sections in $\pm 20$ pixel-wide cross-shaped bands centered on the saturated core. Once they were removed (see verification image in the fifth row of Figure \ref{fig:Vega_Rot1red}), stage 3 distortion correction steps were applied to the clean images (sixth row in the Figure) 
In the last row of Figure \ref{fig:Vega_Rot1red}, we show how the reduced (background, sky constant, and row/column detector artifact cleaned) images look at the four dither positions, following the position dependent PSF subtractions (note: these are not the distortion corrected images combined for the final image). To actually obtain the final image, steps 1 and 2 were repeated on the distortion corrected images, resulting in the median combined PSF subtracted image as shown in Figure \ref{fig:miri_pr}.

\subsection{The F1550C coronagraphic imaging reduction}

\begin{figure}[!t]
    \centering
    \includegraphics[width=0.5\textwidth]{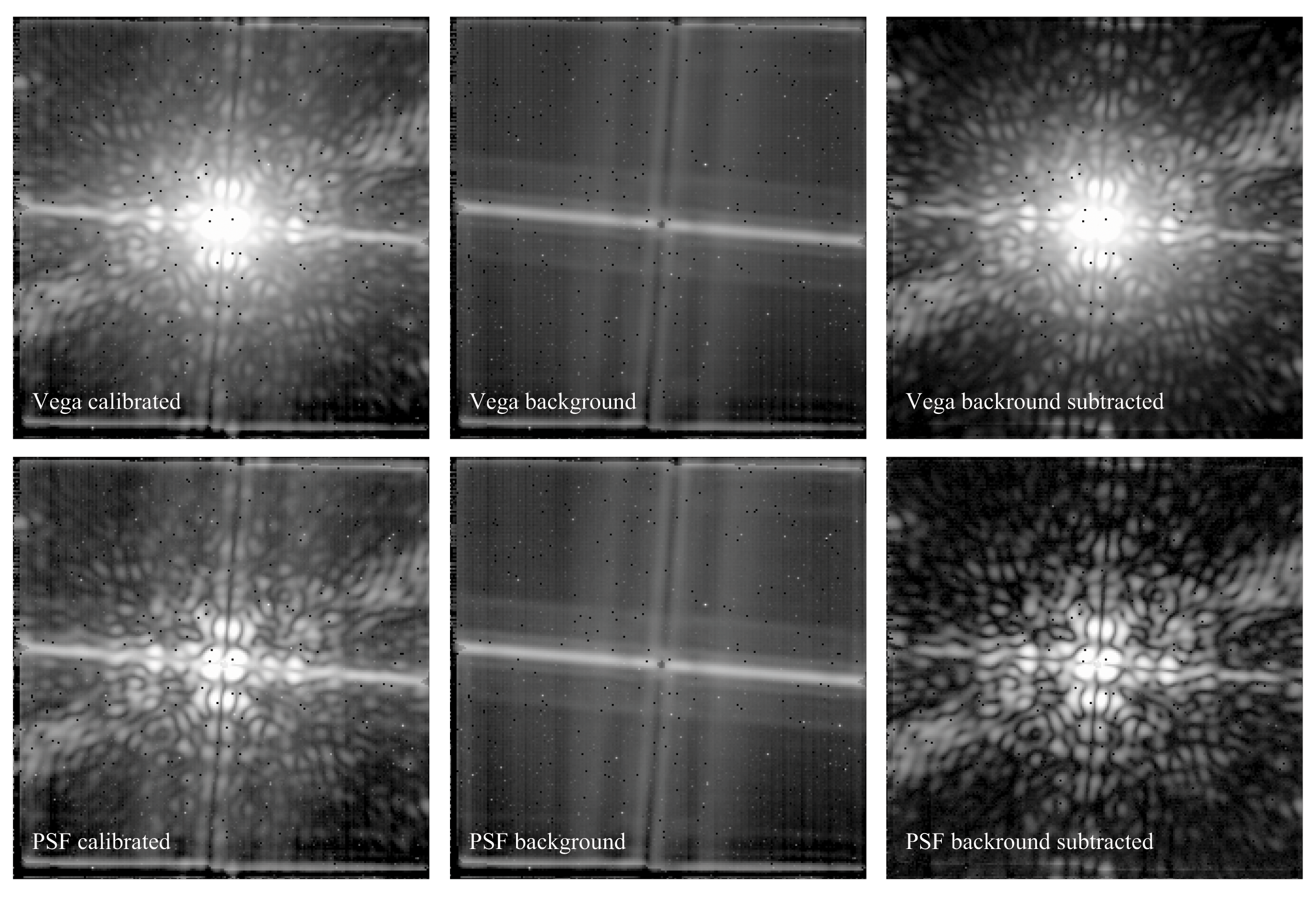}
    \caption{The background subtraction process for the F1550C coronagraphic observations, shown for one of the Vega ({\it top row}) and one of the PSF observations ({\it bottom row}).}
    \label{fig:15bckgsub}
\end{figure}

Following the stage 1 and 2 pipeline reduction steps detailed above, we processed the F1550C coronagraphic images with custom scripts using IRAF and idp3. The observing sequence included background images for both Vega and its PSF source, as they were observed with different group and integration values, requiring custom backgrounds for both. These background observations were dithered in a 2 point pattern to allow the exclusion of possible bright sources in the combined images. Given that only 2 dither pointings were observed, the background images were combined by keeping only the lower value for each pixel,  to reject brighter background sources. The PSF star was observed in the 9-point small grid dithered mode using shifts of around 0.09 pixels, in both F1550C and F2300C. However, only the four quadrant phase mask (4QPM) data (i.e., F1550C) show strong dependence on positioning behind the masks, while the 23.0 $\micron$ Lyot coronagraph PSF is not as sensitive \citep{gaspar23_fom}. Background subtraction is critical for the F1550C data to eliminate the ``glow bar'' artifact. The combined background images were subtracted from the two Vega rotation images and from the 9 point small grid dithered (SGD) PSF images to obtain clean images. In Figure \ref{fig:15bckgsub}, we show the process of the background subtraction for the F1550C images, highlighting it for one of the Vega and one of the PSF images.

\begin{figure}[!t]
    \centering
    \includegraphics[width=0.5\textwidth]{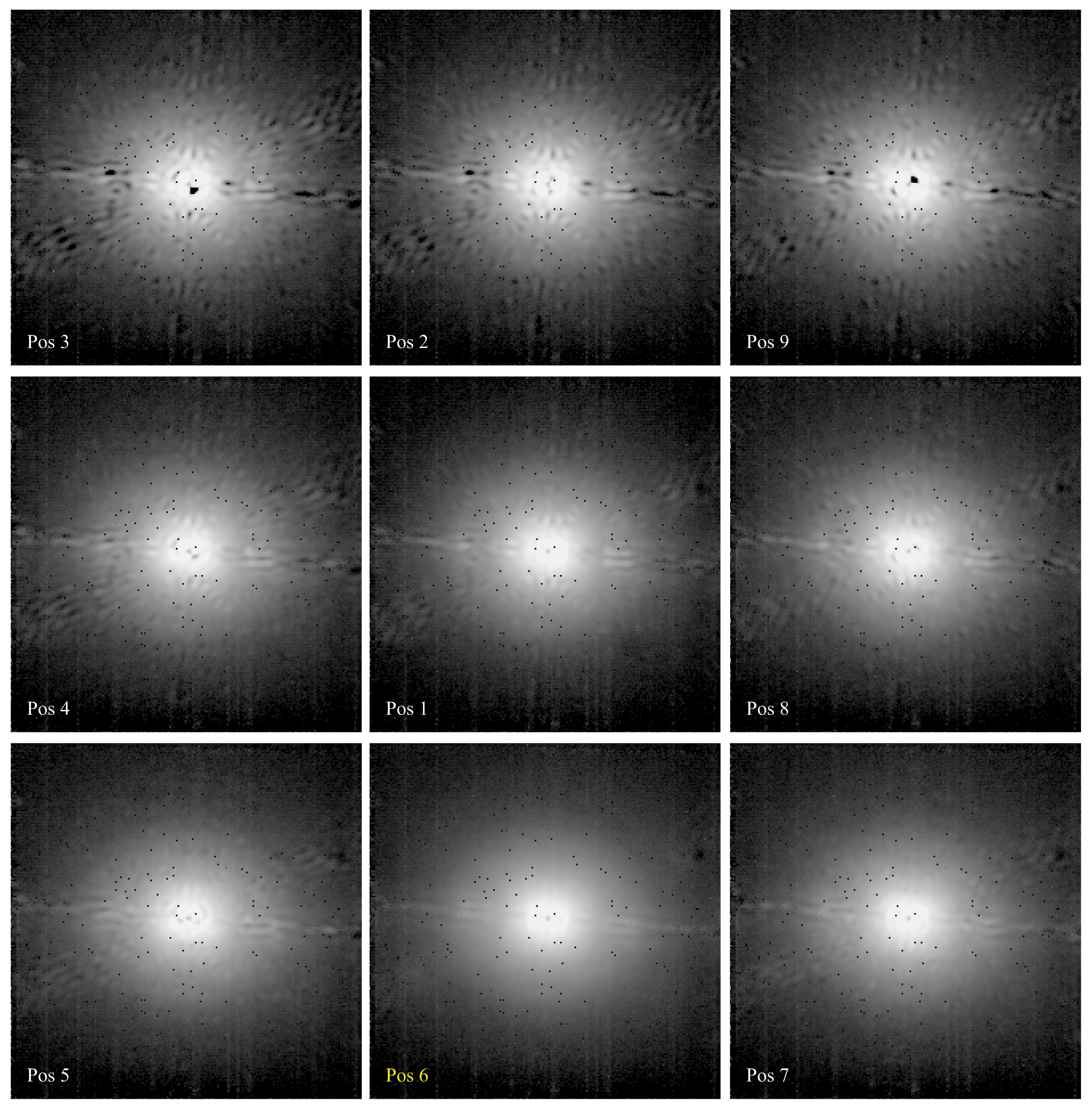}
    \caption{Locating the best PSF, using a scaling factor of 1.43, for the first rotation of Vega observations. The subtractions are shown for all nine SGD PSF positions, laid out in spatial order. For the first rotation, 
    position 6 (marked with yellow font) was clearly the best PSF to use.}
    \label{fig:15psfsub}
\end{figure}

Following background subtraction, the processing of the F1550C coronagraphic images was relatively straightforward. The state-of-the-art KLIP \citep[Karhunen-Lo\`eve Image Projection;][]{soummer12} processing used in the JWST pipeline and in spaceKLIP \citep{spaceklip1,carter23_ers} is the ideal tool to obtain the highest contrast ratio reductions for point-sources, but results in the over-subtraction of extended features, such as circumstellar disks \citep[e.g.,][]{lawson23}. Therefore, we employ classical reference PSF differential imaging (RDI) reduction techniques for the processing, as we did for the Fomalhaut F1550C observations \citep{gaspar23_fom}. The complex coronagraphic PSF of the four quadrant phase mask (4QPM) depends strongly on the positioning of the bright central source on the phase mask. Therefore, we treated the 9 dithered F1550C PSF star images as independent PSFs and subtract all nine PSFs from both Vega observations, after scaling them for an optimum match. We  determine the best one for each rotational dither by visual inspection.  In Figure \ref{fig:15psfsub}, we show the PSF subtractions using each of the nine SGD PSFs for the first rotation of the Vega observations. Visual inspection clearly shows that position 6 yields the least subtraction residuals; for the second rotation position 1 (the central pointing) was the best. We also visually examined subtraction residuals as a function of PSF scaling and determined 1.45 to be the best one, with visible residuals appearing at 1.41 and 1.49 ($\pm 3\%$).

We removed the remaining column artifacts visible in Figure \ref{fig:15psfsub} with an iterative process similar that used for the F2550W dataset. The two Vega observations, obtained at separate rotational dithers, were combined into a single image, allowing us to determine a median disk radial profile. Once the median radial profile was subtracted from the two PSF subtracted images, median column values were calculated and subtracted. These column artifact corrected images were then converted to a geometric distortion corrected plane and combined as shown in Figure \ref{fig:miri_pr}. Due do uneven transmission of the 4QPM the Vega disk seems to be azimuthally asymmetric with the F1550C filter/mask, unlike the F2550W non-coronagraphic image.

\subsection{The F2300C coronagraphic imaging reduction}

\begin{figure}[!t]
    \centering
    \includegraphics[width=0.5\textwidth]{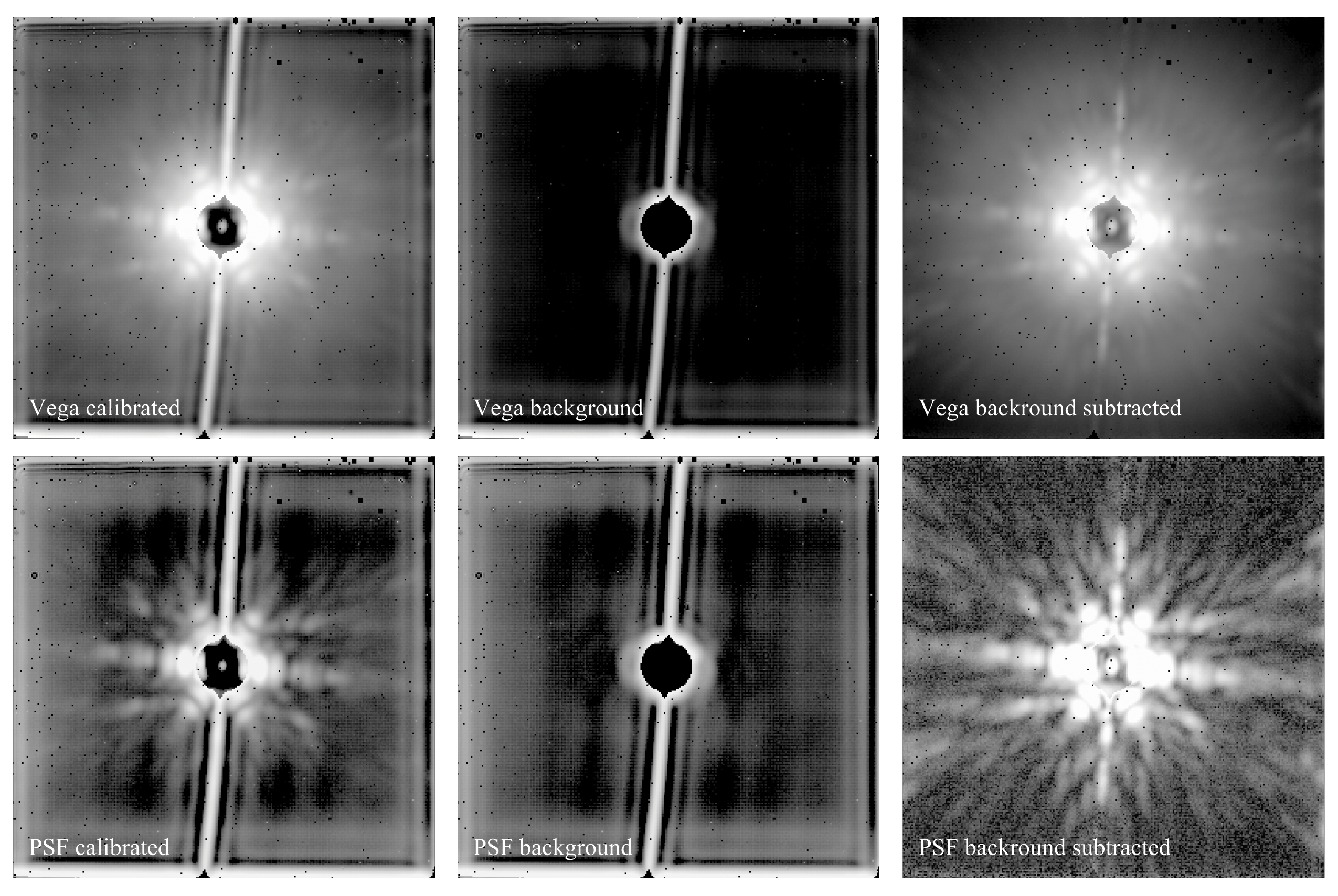}
    \caption{The background subtraction process for the F2300C coronagraphic observations, shown for one of the Vega ({\it top row}) and one of the PSF observations ({\it bottom row}). All images are in logarithmic
    scaling but at different levels to highlight their features.}
    \label{fig:23bckgsub}
\end{figure}

We reduced the F2300C coronagraphic data with the  1.12.3 version of the pipeline. Following basic stage 1 and 2 reductions, as described above, the processing of the calibrated images followed along mostly the same steps as they did for the F1550C 
observations, therefore we detail only the differences.

Background observations were taken for both the science (Vega) and PSF (HD 169305) targets in a two point dither pattern, allowing the exclusion of background sources. As for the F1550C backgrounds, we combined the dithered backgrounds by keeping the lower pixel values. These background images were then subtracted from the two - rotationally dithered - Vega observations and the nine point SGD pattered PSF images. In Figure \ref{fig:23bckgsub}, we show the background subtraction process for the F2300C images, highlighting it for one of the Vega and one of the PSF images.

For the F2300C observations, we used all available PSFs, not just the best aligned one (as we did for the F1550C), as the 23.0 $\micron$ Lyot coronagraph PSF is not as sensitive to positioning as the ones produced by the 4QPMs. To determine alignment, we used the ``false'' peaks in the center of the coronagraphs, seen in Figure \ref{fig:23bckgsub}. The largest offset between a PSF and the target was $\sim 0.2$ px. The best scaling factor for the PSF was determined to be 1.43, just like at 25.5 $\micron$, with acceptable results between 1.39 and 1.47. The images were masked, where necessary, rotated/shifted, PSF subtracted, and finally combined, as shown in Figure \ref{fig:miri_pr}. The image is similar to the non-coronagraphic observations with the F2550W filter, with slightly better noise properties, but a smaller FOV and
a larger useful inner working angle.

\facilities{JWST(MIRI)}


\acknowledgments

We are grateful to Jane Morrison for her assistance on JWST pipeline processing, and also to Dr.\ Tokunaga for digging into his 40-year-old log books to find the correct aperture size used for the measurements presented in Appendix \ref{sec:appendix_otherphotometry}.
Work on this paper was supported in part by grant 80NSSC18K0555, from NASA Goddard Space Flight Center to the University of Arizona. RM acknowledges NASA grant 80NSSC18K0397 and the ``Alien Earths" project (supported by NASA under Agreement No. 80NSSC21K0593).
 Based on observations with
the NASA/ESA/CSA JWST, obtained at the Space Telescope
Science Institute, which is operated by AURA, Inc., under
NASA contract NAS 5-03127. 

The data presented in this article were obtained from the Mikulski Archive for Space Telescopes (MAST) at the Space Telescope Science Institute. These observations are associated with JWST program 1193. The standard pipeline processed Level 2 and Level 3 observations can be accessed via \dataset[10.17909/5m03-8163]{http://dx.doi.org/10.17909/5m03-8163}. We also provide our final reductions of the dataset at https://github.com/merope82/Vega.



\begin{thebibliography}{}
\expandafter\ifx\csname natexlab\endcsname\relax\def\natexlab#1{#1}\fi
\providecommand{\url}[1]{\href{#1}{#1}}
\providecommand{\dodoi}[1]{doi:~\href{http://doi.org/#1}{\nolinkurl{#1}}}
\providecommand{\doeprint}[1]{\href{http://ascl.net/#1}{\nolinkurl{http://ascl.net/#1}}}
\providecommand{\doarXiv}[1]{\href{https://arxiv.org/abs/#1}{\nolinkurl{https://arxiv.org/abs/#1}}}

\bibitem[{{Absil} {et~al.}(2006){Absil}, {di Folco}, {M{\'e}rand}, {Augereau},
  {Coud{\'e} du Foresto}, {Aufdenberg}, {Kervella}, {Ridgway}, {Berger}, {ten
  Brummelaar}, {Sturmann}, {Sturmann}, {Turner}, \& {McAlister}}]{absil06}
{Absil}, O., {di Folco}, E., {M{\'e}rand}, A., {et~al.} 2006, \aap, 452, 237,
  \dodoi{10.1051/0004-6361:20054522}

\bibitem[{{Absil} {et~al.}(2008){Absil}, {di Folco}, {M{\'e}rand}, {Augereau},
  {Coud{\'e} du Foresto}, {Defr{\`e}re}, {Kervella}, {Aufdenberg}, {Desort},
  {Ehrenreich}, {Lagrange}, {Montagnier}, {Olofsson}, {ten Brummelaar},
  {McAlister}, {Sturmann}, {Sturmann}, \& {Turner}}]{absil08}
---. 2008, \aap, 487, 1041, \dodoi{10.1051/0004-6361:200810008}

\bibitem[{{Absil} {et~al.}(2013){Absil}, {Defr{\`e}re}, {Coud{\'e} du Foresto},
  {Di Folco}, {M{\'e}rand}, {Augereau}, {Ertel}, {Hanot}, {Kervella},
  {Mollier}, {Scott}, {Che}, {Monnier}, {Thureau}, {Tuthill}, {ten Brummelaar},
  {McAlister}, {Sturmann}, {Sturmann}, \& {Turner}}]{absil13}
{Absil}, O., {Defr{\`e}re}, D., {Coud{\'e} du Foresto}, V., {et~al.} 2013,
  \aap, 555, A104, \dodoi{10.1051/0004-6361/201321673}

\bibitem[{{Absil} {et~al.}(2021){Absil}, {Marion}, {Ertel}, {Defr{\`e}re},
  {Kennedy}, {Romagnolo}, {Le Bouquin}, {Christiaens}, {Milli}, {Bonsor},
  {Olofsson}, {Su}, \& {Augereau}}]{absil2021}
{Absil}, O., {Marion}, L., {Ertel}, S., {et~al.} 2021, \aap, 651, A45,
  \dodoi{10.1051/0004-6361/202140561}

\bibitem[{{Argyriou} {et~al.}(2023){Argyriou}, {Lage}, {Rieke}, {Gasman},
  {Bouwman}, {Morrison}, {Libralato}, {Dicken}, {Brandl},
  {{\'A}lvarez-M{\'a}rquez}, {Labiano}, {Regan}, \&
  {Ressler}}]{argyriou23_brighter_fatter}
{Argyriou}, I., {Lage}, C., {Rieke}, G.~H., {et~al.} 2023, arXiv e-prints,
  arXiv:2303.13517, \dodoi{10.48550/arXiv.2303.13517}

\bibitem[{{Aumann} {et~al.}(1984){Aumann}, {Gillett}, {Beichman}, {de Jong},
  {Houck}, {Low}, {Neugebauer}, {Walker}, \& {Wesselius}}]{aumann1984}
{Aumann}, H.~H., {Gillett}, F.~C., {Beichman}, C.~A., {et~al.} 1984, \apjl,
  278, L23, \dodoi{10.1086/184214}

\bibitem[{{Backman} \& {Paresce}(1993)}]{backman93_PPIII}
{Backman}, D.~E., \& {Paresce}, F. 1993, in Protostars and Planets III, ed.
  E.~H. {Levy} \& J.~I. {Lunine}, 1253

\bibitem[{{Beichman} {et~al.}(2024){Beichman}, {Bryden}, {Llop Sayson}, \&
  {Ygouf}}]{beichman24_vega}
{Beichman}, C., {Bryden}, G., {Llop Sayson}, J., \& {Ygouf}, M. 2024, \apj, in
  prep.

\bibitem[{{Beichman} {et~al.}(1988){Beichman}, {Neugebauer}, {Habing}, {Clegg},
  \& {Chester}}]{beichman88}
{Beichman}, C.~A., {Neugebauer}, G., {Habing}, H.~J., {Clegg}, P.~E., \&
  {Chester}, T.~J. 1988, in Infrared astronomical satellite (IRAS) catalogs and
  atlases. Volume 1: Explanatory supplement, Vol.~1

\bibitem[{{Bonsor} {et~al.}(2012){Bonsor}, {Augereau}, \&
  {Th{\'e}bault}}]{bonsor12}
{Bonsor}, A., {Augereau}, J.~C., \& {Th{\'e}bault}, P. 2012, \aap, 548, A104,
  \dodoi{10.1051/0004-6361/201220005}

\bibitem[{{Bonsor} {et~al.}(2014){Bonsor}, {Raymond}, {Augereau}, \&
  {Ormel}}]{bonsor14}
{Bonsor}, A., {Raymond}, S.~N., {Augereau}, J.-C., \& {Ormel}, C.~W. 2014,
  \mnras, 441, 2380, \dodoi{10.1093/mnras/stu721}

\bibitem[{{Booth} {et~al.}(2023){Booth}, {Pearce}, {Krivov}, {Wyatt}, {Dent},
  {Hales}, {Lestrade}, {Cruz-S{\'a}enz de Miera}, {Faramaz}, {L{\"o}hne}, \&
  {Chavez-Dagostino}}]{booth23_epseri_clumps}
{Booth}, M., {Pearce}, T.~D., {Krivov}, A.~V., {et~al.} 2023, \mnras, 521,
  6180, \dodoi{10.1093/mnras/stad938}

\bibitem[{{Burns} {et~al.}(1979){Burns}, {Lamy}, \& {Soter}}]{burns79}
{Burns}, J.~A., {Lamy}, P.~L., \& {Soter}, S. 1979, \icarus, 40, 1,
  \dodoi{10.1016/0019-1035(79)90050-2}

\bibitem[{{Bushouse} {et~al.}(2023){Bushouse}, {Eisenhamer}, {Dencheva},
  {Davies}, {Greenfield}, {Morrison}, {Hodge}, {Simon}, {Grumm}, {Droettboom},
  {Slavich}, {Sosey}, {Pauly}, {Miller}, {Jedrzejewski}, {Hack}, {Davis},
  {Crawford}, {Law}, {Gordon}, {Regan}, {Cara}, {MacDonald}, {Bradley},
  {Shanahan}, {Jamieson}, {Teodoro}, {Williams}, \&
  {Pena-Guerrero}}]{bushouse23_jwstpipeline}
{Bushouse}, H., {Eisenhamer}, J., {Dencheva}, N., {et~al.} 2023, {JWST
  Calibration Pipeline}, 1.12.5,  Zenodo, \dodoi{10.5281/zenodo.10022973}

\bibitem[{{Carter} {et~al.}(2023){Carter}, {Hinkley}, {Kammerer}, {Skemer},
  {Biller}, {Leisenring}, {Millar-Blanchaer}, {Petrus}, {Stone}, {Ward-Duong},
  {Wang}, {Girard}, {Hines}, {Perrin}, {Pueyo}, {Balme r}, {Bonavita},
  {Bonnefoy}, {Chauvin}, {Choquet}, {Christiaens}, {Danielski}, {Kennedy},
  {Matthews}, {Miles}, {Patapis}, {Ray}, {Rickman}, {Sallum}, {Stapelfeldt},
  {Whiteford}, {Zhou}, {Absil}, {Boccaletti}, {Booth}, {Bowler}, {Chen},
  {Currie}, {Fortney}, {Grady}, {Greebaum}, {Henning}, {Hoch}, {Janson},
  {Kalas}, {Kenworthy}, {Kervella}, {Krau s}, {Lagage}, {Liu}, {Macintosh},
  {Marino}, {Marley}, {Marois}, {Matthews}, {Mawet}, {McElwain}, {Metchev},
  {Meyer}, {Molliere}, {Moran}, {Morley}, {Mukherjee}, {P antin},
  {Quirrenbach}, {Rebollido}, {Ren}, {Schneider}, {Vasist}, {Worthen}, {Wyatt},
  {Briesemeister}, {Bryan}, {Calissendorff}, {Cantalloube}, {Cugno}, {De
  Furio}, {Dupuy}, {Factor}, {Fahe rty}, {Fitzgerald}, {Franson}, {Gonzales},
  {Hood}, {Howe }, {Kuzuhara}, {Lagrange}, {Lawson}, {Lazzoni}, {Lew}, {Liu},
  {Llop-Sayson}, {Lloyd}, {Martinez}, {Mazoyer}, {Palma-B ifani}, {Quanz},
  {Redai}, {Samland}, {Schlieder}, {Tamura}, {Tan}, {Uyama}, {Vigan}, {Vos},
  {Wagner}, {Wolff}, {Ygouf}, {Zhang}, {Zhang}, \& {Zhang}}]{carter23_ers}
{Carter}, A.~L., {Hinkley}, S., {Kammerer}, J., {et~al.} 2023, \apjl, 951, L20,
  \dodoi{10.3847/2041-8213/acd93e}

\bibitem[{{Ch{\'a}vez Dagostino} {et~al.}(2023){Ch{\'a}vez Dagostino},
  {Marshall}, {Bertone}, {Vega}, \& {S{\'a}nchez
  Arg{\"u}elles}}]{chavez23_LMT_Vega}
{Ch{\'a}vez Dagostino}, M., {Marshall}, J., {Bertone}, E., {Vega}, O., \&
  {S{\'a}nchez Arg{\"u}elles}, D. 2023, in Revista Mexicana de Astronomia y
  Astrofisica Conference Series, Vol.~55, Revista Mexicana de Astronomia y
  Astrofisica Conference Series, 38--40,
  \dodoi{https://doi.org/10.22201/ia.14052059p.2023.55.08}

\bibitem[{{Chen} {et~al.}(2020){Chen}, {Su}, \& {Xu}}]{chen_su_xu20}
{Chen}, C.~H., {Su}, K. Y.~L., \& {Xu}, S. 2020, Nature Astronomy, 4, 328,
  \dodoi{10.1038/s41550-020-1067-6}

\bibitem[{{Chiang} {et~al.}(2009){Chiang}, {Kite}, {Kalas}, {Graham}, \&
  {Clampin}}]{chiang09}
{Chiang}, E., {Kite}, E., {Kalas}, P., {Graham}, J.~R., \& {Clampin}, M. 2009,
  \apj, 693, 734, \dodoi{10.1088/0004-637X/693/1/734}

\bibitem[{{Chittidi} {et~al.}(2024){Chittidi}, {MacGregor}, \&
  {Lovell}}]{chittidi24}
{Chittidi}, J., {MacGregor}, M., \& {Lovell}, J. 2024, in American Astronomical
  Society Meeting Abstracts, Vol. 243, American Astronomical Society Meeting
  Abstracts, 413.07

\bibitem[{{Cohen} {et~al.}(2001){Cohen}, {Walker}, {Jayaraman}, {Barker}, \&
  {Price}}]{cohen01}
{Cohen}, M., {Walker}, R.~G., {Jayaraman}, S., {Barker}, E., \& {Price}, S.~D.
  2001, \aj, 121, 1180, \dodoi{10.1086/318751}

\bibitem[{{Defr{\`e}re} {et~al.}(2011){Defr{\`e}re}, {Absil}, {Augereau}, {di
  Folco}, {Berger}, {Coud{\'e} du Foresto}, {Kervella}, {Le Bouquin},
  {Lebreton}, {Millan-Gabet}, {Monnier}, {Olofsson}, \& {Traub}}]{defrere2011}
{Defr{\`e}re}, D., {Absil}, O., {Augereau}, J.~C., {et~al.} 2011, \aap, 534,
  A5, \dodoi{10.1051/0004-6361/201117017}

\bibitem[{{Deller} \& {Maddison}(2005)}]{deller05}
{Deller}, A.~T., \& {Maddison}, S.~T. 2005, \apj, 625, 398,
  \dodoi{10.1086/429365}

\bibitem[{{Dohnanyi}(1969)}]{dohnanyi69}
{Dohnanyi}, J.~S. 1969, J. Geophys. Res., 74, 2431,
  \dodoi{10.1029/JB074i010p02531}

\bibitem[{{Dong} {et~al.}(2020){Dong}, {Dawson}, {Shannon}, \&
  {Morrison}}]{dong20}
{Dong}, J., {Dawson}, R.~I., {Shannon}, A., \& {Morrison}, S. 2020, \apj, 889,
  47, \dodoi{10.3847/1538-4357/ab64f7}

\bibitem[{{Ertel} {et~al.}(2020){Ertel}, {Defr{\`e}re}, {Hinz}, {Mennesson},
  {Kennedy}, {Danchi}, {Gelino}, {Hill}, {Hoffmann}, {Mazoyer}, {Rieke},
  {Shannon}, {Stapelfeldt}, {Spalding}, {Stone}, {Vaz}, {Weinberger},
  {Willems}, {Absil}, {Arbo}, {Bailey}, {Beichman}, {Bryden}, {Downey},
  {Durney}, {Esposito}, {Gaspar}, {Grenz}, {Haniff}, {Leisenring}, {Marion},
  {McMahon}, {Millan-Gabet}, {Montoya}, {Morzinski}, {Perera}, {Pinna}, {Pott},
  {Power}, {Puglisi}, {Roberge}, {Serabyn}, {Skemer}, {Su}, {Vaitheeswaran}, \&
  {Wyatt}}]{ertel20_hosts}
{Ertel}, S., {Defr{\`e}re}, D., {Hinz}, P., {et~al.} 2020, \aj, 159, 177,
  \dodoi{10.3847/1538-3881/ab7817}

\bibitem[{{Esposito} {et~al.}(2020){Esposito}, {Kalas}, {Fitzgerald},
  {Millar-Blanchaer}, {Duch{\^e}ne}, {Patience}, {Hom}, {Perrin}, {De Rosa},
  {Chiang}, {Czekala}, {Macintosh}, {Graham}, {Ansdell}, {Arriaga}, {Bruzzone},
  {Bulger}, {Chen}, {Cotten}, {Dong}, {Draper}, {Follette}, {Hung}, {Lopez},
  {Matthews}, {Mazoyer}, {Metchev}, {Rameau}, {Ren}, {Rice}, {Song}, {Stahl},
  {Wang}, {Wolff}, {Zuckerman}, {Ammons}, {Bailey}, {Barman}, {Chilcote},
  {Doyon}, {Gerard}, {Goodsell}, {Greenbaum}, {Hibon}, {Hinkley}, {Ingraham},
  {Konopacky}, {Maire}, {Marchis}, {Marley}, {Marois}, {Nielsen},
  {Oppenheimer}, {Palmer}, {Poyneer}, {Pueyo}, {Rajan}, {Rantakyr{\"o}},
  {Ruffio}, {Savransky}, {Schneider}, {Sivaramakrishnan}, {Soummer}, {Thomas},
  \& {Ward-Duong}}]{esposito20}
{Esposito}, T.~M., {Kalas}, P., {Fitzgerald}, M.~P., {et~al.} 2020, \aj, 160,
  24, \dodoi{10.3847/1538-3881/ab9199}

\bibitem[{{Faramaz} {et~al.}(2017){Faramaz}, {Ertel}, {Booth}, {Cuadra}, \&
  {Simmonds}}]{faramaz17_vega}
{Faramaz}, V., {Ertel}, S., {Booth}, M., {Cuadra}, J., \& {Simmonds}, C. 2017,
  \mnras, 465, 2352, \dodoi{10.1093/mnras/stw2846}

\bibitem[{{Friebe} {et~al.}(2022){Friebe}, {Pearce}, \& {L{\"o}hne}}]{friebe22}
{Friebe}, M.~F., {Pearce}, T.~D., \& {L{\"o}hne}, T. 2022, \mnras, 512, 4441,
  \dodoi{10.1093/mnras/stac664}

\bibitem[{{Gardner} {et~al.}(2023){Gardner}, {Mather}, {Abbott}, {Abell},
  {Abernathy}, {Abney}, {Abraham}, {Abraham}, {Abul-Huda}, {Acton}, {Adams},
  {Adams}, {Adler}, {Adriaensen}, {Aguilar}, {Ahmed}, {Ahmed}, {Ahmed},
  {Albat}, {Albert}, {Alberts}, {Aldridge}, {Allen}, {Allen}, {Altenburg},
  {Altunc}, {Alvarez}, {{\'A}lvarez-M{\'a}rquez}, {Alves de Oliveira},
  {Ambrose}, {Anandakrishnan}, {Andersen}, {Anderson}, {Anderson}, {Anderson},
  {Anderson}, {Aprea}, {Archer}, {Arenberg}, {Argyriou}, {Arribas}, {Artigau},
  {Arvai}, {Atcheson}, {Atkinson}, {Averbukh}, {Aymergen}, {Bacinski},
  {Baggett}, {Bagnasco}, {Baker}, {Balzano}, {Banks}, {Baran}, {Barker},
  {Barrett}, {Barringer}, {Barto}, {Bast}, {Baudoz}, {Baum}, {Beatty},
  {Beaulieu}, {Bechtold}, {Beck}, {Beddard}, {Beichman}, {Bellagama}, {Bely},
  {Berger}, {Bergeron}, {Bernier}, {Bertch}, {Beskow}, {Betz}, {Biagetti},
  {Birkmann}, {Bjorklund}, {Blackwood}, {Blazek}, {Blossfeld}, {Bluth},
  {Boccaletti}, {Boegner}, {Bohlin}, {Boia}, {B{\"o}ker}, {Bonaventura},
  {Bond}, {Bosley}, {Boucarut}, {Bouchet}, {Bouwman}, {Bower}, {Bowers},
  {Bowers}, {Boyce}, {Boyer}, {Boyer}, {Boyer}, {Boyer}, {Bradley}, {Brady},
  {Brandl}, {Brannen}, {Breda}, {Bremmer}, {Brennan}, {Bresnahan}, {Bright},
  {Broiles}, {Bromenschenkel}, {Brooks}, {Brooks}, {Brown}, {Brown}, {Brown},
  {Bruce}, {Bryson}, {Bujanda}, {Bullock}, {Bunker}, {Bureo}, {Burt}, {Bush},
  {Bushouse}, {Bussman}, {Cabaud}, {Cale}, {Calhoon}, {Calvani}, {Canipe},
  {Caputo}, {Cara}, {Carey}, {Case}, {Cesari}, {Cetorelli}, {Chance},
  {Chandler}, {Chaney}, {Chapman}, {Charlot}, {Chayer}, {Cheezum}, {Chen},
  {Chen}, {Cherinka}, {Chichester}, {Chilton}, {Chittiraibalan}, {Clampin},
  {Clark}, {Clark}, {Clark}, {Claybrooks}, {Cleveland}, {Cohen}, {Cohen},
  {Col{\'o}n}, {Coleman}, {Colina}, {Comber}, {Comeau}, {Comer}, {Conde Reis},
  {Connolly}, {Conroy}, {Contos}, {Contreras}, {Cook}, {Cooper}, {Cooper},
  {Correia}, {Correnti}, {Cossou}, {Costanza}, {Coulais}, {Cox}, {Coyle},
  {Cracraft}, {Crew}, {Curtis}, {Cusveller}, {Da Costa Maciel}, {Dailey},
  {Daugeron}, {Davidson}, {Davies}, {Davis}, {Davis}, {Day}, {de Chambure}, {de
  Jong}, {De Marchi}, {Dean}, {Decker}, {Delisa}, {Dell}, {Dellagatta},
  {Dembinska}, {Demosthenes}, {Dencheva}, {Deneu}, {DePriest}, {Deschenes},
  {Dethienne}, {Detre}, {Diaz}, {Dicken}, {DiFelice}, {Dillman}, {Disharoon},
  {Dixon}, {Doggett}, {Dominguez}, {Donaldson}, {Doria-Warner}, {Santos},
  {Doty}, {Douglas}, {Doyon}, {Dressler}, {Driggers}, {Driggers}, {Dunn},
  {DuPrie}, {Dupuis}, {Durning}, {Dutta}, {Earl}, {Eccleston}, {Ecobichon},
  {Egami}, {Ehrenwinkler}, {Eisenhamer}, {Eisenhower}, {Eisenstein}, {El
  Hamel}, {Elie}, {Elliott}, {Elliott}, {Engesser}, {Espinoza}, {Etienne},
  {Etxaluze}, {Evans}, {Fabreguettes}, {Falcolini}, {Falini}, {Fatig},
  {Feeney}, {Feinberg}, {Fels}, {Ferdous}, {Ferguson}, {Ferrarese}, {Ferreira},
  {Ferruit}, {Ferry}, {Filippazzo}, {Firre}, {Fix}, {Flagey}, {Flanagan},
  {Fleming}, {Florian}, {Flynn}, {Foiadelli}, {Fontaine}, {Fontanella},
  {Forshay}, {Fortner}, {Fox}, {Framarini}, {Francisco}, {Franck}, {Franx},
  {Franz}, {Friedman}, {Friend}, {Frost}, {Fu}, {Fullerton}, {Gaillard},
  {Galkin}, {Gallagher}, {Galyer}, {Garc{\'\i}a Mar{\'\i}n}, {Gardner},
  {Garland}, {Garrett}, {Gasman}, {G{\'a}sp{\'a}r}, {Gastaud}, {Gaudreau},
  {Gauthier}, {Geers}, {Geithner}, {Gennaro}, {Gerber}, {Gereau}, {Giampaoli},
  {Giardino}, {Gibbons}, {Gilbert}, {Gilman}, {Girard}, {Giuliano}, {Gkountis},
  {Glasse}, {Glassmire}, {Glauser}, {Glazer}, {Goldberg}, {Golimowski},
  {Gonzaga}, {Gordon}, {Gordon}, {Goudfrooij}, {Gough}, {Graham}, {Grau},
  {Green}, {Greene}, {Greene}, {Greenfield}, {Greenhouse}, {Greve}, {Greville},
  {Grimaldi}, {Groe}, {Groebner}, {Grumm}, {Grundy}, {G{\"u}del}, {Guillard},
  {Guldalian}, {Gunn}, {Gurule}, {Gutman}, {Guy}, {Guyot}, {Hack}, {Haderlein},
  {Hagan}, {Hagedorn}, {Hainline}, {Haley}, {Hami}, {Hamilton}, {Hammann},
  {Hammel}, {Hanley}, {Hansen}, {Hardy}, {Harnisch}, {Harr}, {Harris}, {Hart},
  {Hartig}, {Hasan}, {Hashim}, {Hashimoto}, {Haskins}, {Hawkins}, {Hayden},
  {Hayden}, {Healy}, {Hecht}, {Heeg}, {Hejal}, {Helm}, {Hengemihle}, {Henning},
  {Henry}, {Henry}, {Henshaw}, {Hernandez}, {Herrington}, {Heske}, {Hesman},
  {Hickey}, {Hilbert}, {Hines}, {Hinz}, {Hirsch}, {Hitcho}, {Hodapp}, {Hodge},
  {Hoffman}, {Holfeltz}, {Holler}, {Hoppa}, {Horner}, {Howard}, {Howard},
  {Huber}, {Hunkeler}, {Hunter}, {Hunter}, {Hurd}, {Hurst}, {Hutchings},
  {Hylan}, {Ignat}, {Illingworth}, {Irish}, {Isaacs}, {Jackson}, {Jaffe},
  {Jahic}, {Jahromi}, {Jakobsen}, {James}, {James}, {James}, {Jamieson},
  {Jandra}, {Jayawardhana}, {Jedrzejewski}, {Jeffers}, {Jensen}, {Joanne},
  {Johns}, {Johnson}, {Johnson}, {Johnson}, {Johnson}, {Johnson}, {Johnson},
  {Johnstone}, {Jollet}, {Jones}, {Jones}, {Jones}, {Jones}, {Jones}, {Jordan},
  {Jordan}, {Jue}, {Jurkowski}, {Justis}, {Justtanont}, {Kaleida}, {Kalirai},
  {Kalmanson}, {Kaltenegger}, {Kammerer}, {Kan}, {Kanarek}, {Kao}, {Karakla},
  {Karl}, {Kassin}, {Kauffman}, {Kavanagh}, {Kelley}, {Kelly}, {Kendrew},
  {Kennedy}, {Kenny}, {Keski-Kuha}, {Keyes}, {Khan}, {Kidwell}, {Kimble},
  {King}, {King}, {Kinzel}, {Kirk}, {Kirkpatrick}, {Klaassen}, {Klingemann},
  {Klintworth}, {Knapp}, {Knight}, {Knollenberg}, {Knutsen}, {Koehler},
  {Koekemoer}, {Kofler}, {Kontson}, {Kovacs}, {Kozhurina-Platais}, {Krause},
  {Kriss}, {Krist}, {Kristoffersen}, {Krogel}, {Krueger}, {Kulp}, {Kumari},
  {Kwan}, {Kyprianou}, {Labador}, {Labiano}, {Lafreni{\`e}re}, {Lagage},
  {Laidler}, {Laine}, {Laird}, {Lajoie}, {Lallo}, {Lam}, {LaMassa}, {Lambros},
  {Lampenfield}, {Lander}, {Langston}, {Larson}, {Larson}, {LaVerghetta},
  {Law}, {Lawrence}, {Lee}, {Lee}, {Lee}, {Leisenring}, {Leveille}, {Levenson},
  {Levi}, {Levine}, {Lewis}, {Lewis}, {Lewis}, {Libralato}, {Lidon},
  {Liebrecht}, {Lightsey}, {Lilly}, {Lim}, {Lim}, {Ling}, {Link}, {Link},
  {Lipinski}, {Liu}, {Lo}, {Lobmeyer}, {Logue}, {Long}, {Long}, {Long}, {Long},
  {L{\'o}pez-Caniego}, {Lotz}, {Love-Pruitt}, {Lubskiy}, {Luers}, {Luetgens},
  {Luevano}, {Lui}, {Lund}, {Lundquist}, {Lunine}, {L{\"u}tzgendorf}, {Lynch},
  {MacDonald}, {MacDonald}, {Macias}, {Macklis}, {Maghami}, {Maharaja},
  {Maiolino}, {Makrygiannis}, {Malla}, {Malumuth}, {Manjavacas}, {Marini},
  {Marrione}, {Marston}, {Martel}, {Martin}, {Martin}, {Martinez}, {Maschmann},
  {Masci}, {Masetti}, {Maszkiewicz}, {Matthews}, {Matuskey}, {McBrayer},
  {McCarthy}, {McCaughrean}, {McClare}, {McClare}, {McCloskey}, {McClurg},
  {McCoy}, {McElwain}, {McGregor}, {McGuffey}, {McKay}, {McKenzie}, {McLean},
  {McMaster}, {McNeil}, {De Meester}, {Mehalick}, {Meixner}, {Mel{\'e}ndez},
  {Menzel}, {Menzel}, {Merz}, {Mesterharm}, {Meyer}, {Meyett}, {Meza},
  {Midwinter}, {Milam}, {Miller}, {Miller}, {Miskey}, {Misselt}, {Mitchell},
  {Mohan}, {Montoya}, {Moran}, {Morishita}, {Moro-Mart{\'\i}n}, {Morrison},
  {Morrison}, {Morse}, {Moschos}, {Moseley}, {Mosier}, {Mosner}, {Mountain},
  {Muckenthaler}, {Mueller}, {Mueller}, {Muhiem}, {M{\"u}hlmann}, {Mullally},
  {Mullen}, {Munger}, {Murphy}, {Murray}, {Muzerolle}, {Mycroft}, {Myers},
  {Myers}, {Myers}, {Myers}, {Myrick}, {Nagle}, {Nayak}, {Naylor}, {Neff},
  {Nelan}, {Nella}, {Nguyen}, {Nguyen}, {Nickson}, {Nidhiry}, {Niedner},
  {Nieto-Santisteban}, {Nikolov}, {Nishisaka}, {Noriega-Crespo}, {Nota},
  {O'Mara}, {Oboryshko}, {O'Brien}, {Ochs}, {Offenberg}, {Ogle}, {Ohl},
  {Olmsted}, {Osborne}, {O'Shaughnessy}, {{\"O}stlin}, {O'Sullivan}, {Otor},
  {Ottens}, {Ouellette}, {Outlaw}, {Owens}, {Pacifici}, {Page}, {Paranilam},
  {Park}, {Parrish}, {Paschal}, {Patapis}, {Patel}, {Patrick}, {Pattishall},
  {Paul}, {Paul}, {Pauly}, {Pavlovsky}, {Pe{\~n}a-Guerrero}, {Pedder}, {Peek},
  {Pelham}, {Penanen}, {Perriello}, {Perrin}, {Perrine}, {Perrygo}, {Peslier},
  {Petach}, {Peterson}, {Pfarr}, {Pierson}, {Pietraszkiewicz}, {Pilchen},
  {Pipher}, {Pirzkal}, {Pitman}, {Player}, {Plesha}, {Plitzke}, {Pohner},
  {Poletis}, {Pollizzi}, {Polster}, {Pontius}, {Pontoppidan}, {Porges},
  {Potter}, {Prescott}, {Proffitt}, {Pueyo}, {Quispe Neira}, {Radich}, {Rager},
  {Rameau}, {Ramey}, {Ramos Alarcon}, {Rampini}, {Rapp}, {Rashford},
  {Rauscher}, {Ravindranath}, {Rawle}, {Rawlings}, {Ray}, {Regan}, {Rehm},
  {Rehm}, {Reid}, {Reis}, {Renk}, {Reoch}, {Ressler}, {Rest}, {Reynolds},
  {Richon}, {Richon}, {Ridgaway}, {Riedel}, {Rieke}, {Rieke}, {Rifelli},
  {Rigby}, {Riggs}, {Ringel}, {Ritchie}, {Rix}, {Robberto}, {Robinson},
  {Robinson}, {Robinson}, {Rock}, {Rodriguez}, {Rodr{\'\i}guez del Pino},
  {Roellig}, {Rohrbach}, {Roman}, {Romelfanger}, {Romo}, {Rosales}, {Rose},
  {Roteliuk}, {Roth}, {Rothwell}, {Rouzaud}, {Rowe}, {Rowlands}, {Roy},
  {Royer}, {Rui}, {Rumler}, {Rumpl}, {Russ}, {Ryan}, {Ryan}, {Saad}, {Sabata},
  {Sabatino}, {Sabbi}, {Sabelhaus}, {Sabia}, {Sahu}, {Saif}, {Salvignol},
  {Samara-Ratna}, {Samuelson}, {Sanders}, {Sappington}, {Sargent}, {Sauer},
  {Savadkin}, {Sawicki}, {Schappell}, {Scheffer}, {Scheithauer}, {Scherer},
  {Schiff}, {Schlawin}, {Schmeitzky}, {Schmitz}, {Schmude}, {Schneider},
  {Schreiber}, {Schroeven-Deceuninck}, {Schultz}, {Schwab}, {Schwartz},
  {Scoccimarro}, {Scott}, {Scott}, {Seaton}, {Seely}, {Seery}, {Seidleck},
  {Sembach}, {Shanahan}, {Shaughnessy}, {Shaw}, {Shay}, {Sheehan}, {Sheth},
  {Shih}, {Shivaei}, {Siegel}, {Sienkiewicz}, {Simmons}, {Simon}, {Sirianni},
  {Sivaramakrishnan}, {Slade}, {Sloan}, {Slocum}, {Slowinski}, {Smith},
  {Smith}, {Smith}, {Smith}, {Smith}, {Smith}, {Smolik}, {Soderblom}, {Sohn},
  {Sokol}, {Sonneborn}, {Sontag}, {Sooy}, {Soummer}, {Southwood}, {Spain},
  {Sparmo}, {Speer}, {Spencer}, {Sprofera}, {Stallcup}, {Stanley},
  {Stansberry}, {Stark}, {Starr}, {Stassi}, {Steck}, {Steeley}, {Stephens},
  {Stephenson}, {Stewart}, {Stiavelli}, {}, {Strada}, {Straughn}, {Streetman},
  {Strickland}, {Strobele}, {Stuhlinger}, {Stys}, {Such}, {Sukhatme},
  {Sullivan}, {Sullivan}, {Sumner}, {Sun}, {Sunnquist}, {Swade}, {Swam},
  {Swenton}, {Swoish}, {Tam Litten}, {Tamas}, {Tao}, {Taylor}, {Taylor}, {te
  Plate}, {Van Tea}, {Teague}, {Telfer}, {Temim}, {Texter}, {Thatte},
  {Thompson}, {Thompson}, {Thomson}, {Thronson}, {Tierney}, {Tikkanen},
  {Tinnin}, {Tippet}, {Todd}, {Tran}, {Trauger}, {Trejo}, {Vinh Truong},
  {Tsukamoto}, {Tufail}, {Tumlinson}, {Tustain}, {Tyra}, {Ubeda}, {Underwood},
  {Uzzo}, {Vaclavik}, {Valenduc}, {Valenti}, {Van Campen}, {van de Wetering},
  {Van Der Marel}, {van Haarlem}, {Vandenbussche}, {van Dishoeck},
  {Vanterpool}, {Vernoy}, {Vila Costas}, {Volk}, {Voorzaat}, {Voyton}, {Vydra},
  {Waddy}, {Waelkens}, {Wahlgren}, {Walker}, {Wander}, {Warfield}, {Warner},
  {Wasiak}, {Wasiak}, {Wehner}, {Weiler}, {Weilert}, {Weiss}, {Wells}, {Welty},
  {Wheate}, {Wheeler}, {White}, {Whitehouse}, {Whiteleather}, {Whitman},
  {Williams}, {Willmer}, {Willott}, {Willoughby}, {Wilson}, {Wilson}, {Wilson},
  {Windhorst}, {Wislowski}, {Wolfe}, {Wolfe}, {Wolff}, {Wondel}, {Woo},
  {Woods}, {Worden}, {Workman}, {Wright}, {Wu}, {Wu}, {Wun}, {Wymer},
  {Yadetie}, {Yan}, {Yang}, {Yates}, {Yeager}, {Yerger}, {Young}, {Young},
  {Yu}, {Yu}, {Zak}, {Zeidler}, {Zepp}, {Zhou}, {Zincke}, {Zonak}, \&
  {Zondag}}]{gardner23_jwst}
{Gardner}, J.~P., {Mather}, J.~C., {Abbott}, R., {et~al.} 2023, \pasp, 135,
  068001, \dodoi{10.1088/1538-3873/acd1b5}

\bibitem[{{G{\'a}sp{\'a}r} {et~al.}(2012){G{\'a}sp{\'a}r}, {Psaltis}, {Rieke},
  \& {{\"O}zel}}]{gaspar12b}
{G{\'a}sp{\'a}r}, A., {Psaltis}, D., {Rieke}, G.~H., \& {{\"O}zel}, F. 2012,
  \apj, 754, 74, \dodoi{10.1088/0004-637X/754/1/74}

\bibitem[{{G{\'a}sp{\'a}r} {et~al.}(2023){G{\'a}sp{\'a}r}, {Wolff}, {Rieke},
  {Leisenring}, {Morrison}, {Su}, {Ward-Duong}, {Aguilar}, {Ygouf}, {Beichman},
  {Llop-Sayson}, \& {Bryden}}]{gaspar23_fom}
{G{\'a}sp{\'a}r}, A., {Wolff}, S.~G., {Rieke}, G.~H., {et~al.} 2023, Nature
  Astronomy, \dodoi{10.1038/s41550-023-01962-6}

\bibitem[{{Holland} {et~al.}(2017){Holland}, {Matthews}, {Kennedy}, {Greaves},
  {Wyatt}, {Booth}, {Bastien}, {Bryden}, {Butner}, {Chen}, {Chrysostomou},
  {Davies}, {Dent}, {Di Francesco}, {Duch{\^e}ne}, {Gibb}, {Friberg}, {Ivison},
  {Jenness}, {Kavelaars}, {Lawler}, {Lestrade}, {Marshall}, {Moro-Martin},
  {Pani{\'c}}, {Phillips}, {Serjeant}, {Schieven}, {Sibthorpe}, {Vican},
  {Ward-Thompson}, {van der Werf}, {White}, {Wilner}, \&
  {Zuckerman}}]{holland17_SONS}
{Holland}, W.~S., {Matthews}, B.~C., {Kennedy}, G.~M., {et~al.} 2017, \mnras,
  470, 3606, \dodoi{10.1093/mnras/stx1378}

\bibitem[{{Hughes} {et~al.}(2018){Hughes}, {Duch{\^e}ne}, \&
  {Matthews}}]{hughes18}
{Hughes}, A.~M., {Duch{\^e}ne}, G., \& {Matthews}, B.~C. 2018, \araa, 56, 541,
  \dodoi{10.1146/annurev-astro-081817-052035}

\bibitem[{{Hughes} {et~al.}(2012){Hughes}, {Wilner}, {Mason}, {Carpenter},
  {Plambeck}, {Chiang}, {Andrews}, {Williams}, {Hales}, {Su}, {Chiang},
  {Dicker}, {Korngut}, \& {Devlin}}]{hughes12}
{Hughes}, A.~M., {Wilner}, D.~J., {Mason}, B., {et~al.} 2012, \apj, 750, 82,
  \dodoi{10.1088/0004-637X/750/1/82}

\bibitem[{{Hurt} \& {MacGregor}(2023)}]{hurt_macgregor23}
{Hurt}, S.~A., \& {MacGregor}, M.~A. 2023, arXiv e-prints, arXiv:2304.07446,
  \dodoi{10.48550/arXiv.2304.07446}

\bibitem[{{Hurt} {et~al.}(2021){Hurt}, {Quinn}, {Latham}, {Vanderburg},
  {Esquerdo}, {Calkins}, {Berlind}, {Angus}, {Latham}, \& {Zhou}}]{hurt21}
{Hurt}, S.~A., {Quinn}, S.~N., {Latham}, D.~W., {et~al.} 2021, \aj, 161, 157,
  \dodoi{10.3847/1538-3881/abdec8}

\bibitem[{{Ishihara} {et~al.}(2010){Ishihara}, {Onaka}, {Kataza}, {Salama},
  {Alfageme}, {Cassatella}, {Cox}, {Garc{\'\i}a-Lario}, {Stephenson}, {Cohen},
  {Fujishiro}, {Fujiwara}, {Hasegawa}, {Ita}, {Kim}, {Matsuhara}, {Murakami},
  {M{\"u}ller}, {Nakagawa}, {Ohyama}, {Oyabu}, {Pyo}, {Sakon}, {Shibai},
  {Takita}, {Tanab{\'e}}, {Uemizu}, {Ueno}, {Usui}, {Wada}, {Watarai},
  {Yamamura}, \& {Yamauchi}}]{akari_irc_catalog}
{Ishihara}, D., {Onaka}, T., {Kataza}, H., {et~al.} 2010, \aap, 514, A1,
  \dodoi{10.1051/0004-6361/200913811}

\bibitem[{{Kammerer} {et~al.}(2022){Kammerer}, {Girard}, {Carter}, {Perrin},
  {Cooper}, {Thatte}, {Vandal}, {Leisenring}, {Wang}, {Balmer}, {Sivarama
  krishnan}, {Pueyo}, {Ward-Duong}, {Sunnquist}, \& {Adams Redai}}]{spaceklip1}
{Kammerer}, J., {Girard}, J., {Carter}, A.~L., {et~al.} 2022, in Society of
  Photo-Optical Instrumentation Engineers (SPIE) Conference Series, Vol. 12180,
  Space Telescopes and Instrumentation 2022: Optical, Infrared, and Millimeter
  Wave, ed. L.~E. {Coyle}, S.~{Matsuura}, \& M.~D. {Perrin}, 121803N,
  \dodoi{10.1117/12.2628865}

\bibitem[{{Kirchschlager} {et~al.}(2017){Kirchschlager}, {Wolf}, {Krivov},
  {Mutschke}, \& {Brunngr{\"a}ber}}]{kirchschlager2017}
{Kirchschlager}, F., {Wolf}, S., {Krivov}, A.~V., {Mutschke}, H., \&
  {Brunngr{\"a}ber}, R. 2017, \mnras, 467, 1614, \dodoi{10.1093/mnras/stx202}

\bibitem[{{Kobayashi} {et~al.}(2011){Kobayashi}, {Kimura}, {Watanabe},
  {Yamamoto}, \& {M{\"u}ller}}]{kobayashi11_sublimationtemperature}
{Kobayashi}, H., {Kimura}, H., {Watanabe}, S.~i., {Yamamoto}, T., \&
  {M{\"u}ller}, S. 2011, Earth, Planets and Space, 63, 1067,
  \dodoi{10.5047/eps.2011.03.012}

\bibitem[{{Kobayashi} {et~al.}(2009){Kobayashi}, {Watanabe}, {Kimura}, \&
  {Yamamoto}}]{kobayashi09}
{Kobayashi}, H., {Watanabe}, S.-i., {Kimura}, H., \& {Yamamoto}, T. 2009,
  \icarus, 201, 395, \dodoi{10.1016/j.icarus.2009.01.002}

\bibitem[{{Kuchner} \& {Stark}(2010)}]{kuchner_start_2010}
{Kuchner}, M.~J., \& {Stark}, C.~C. 2010, \aj, 140, 1007,
  \dodoi{10.1088/0004-6256/140/4/1007}

\bibitem[{{Laureijs} {et~al.}(2002){Laureijs}, {Jourdain de Muizon}, {Leech},
  {Siebenmorgen}, {Dominik}, {Habing}, {Trams}, \& {Kessler}}]{laureijs02}
{Laureijs}, R.~J., {Jourdain de Muizon}, M., {Leech}, K., {et~al.} 2002, \aap,
  387, 285, \dodoi{10.1051/0004-6361:20020366}

\bibitem[{{Lawson} {et~al.}(2023){Lawson}, {Schlieder}, {Leisenring}, {Bogat},
  {Beichman}, {Bryden}, {G{\'a}sp{\'a}r}, {Groff}, {McElwain}, {Meyer},
  {Barclay}, {Calissendorff}, {De Furio}, {Ygouf}, {Boccaletti}, {Greene},
  {Krist}, {Plavchan}, {Rieke}, {Roellig}, {Stansberry}, {Wisniewski}, \&
  {Young}}]{lawson23}
{Lawson}, K., {Schlieder}, J.~E., {Leisenring}, J.~M., {et~al.} 2023, \aj, 166,
  150, \dodoi{10.3847/1538-3881/aced08}

\bibitem[{{Li} \& {Draine}(2001)}]{li_draine01}
{Li}, A., \& {Draine}, B.~T. 2001, \apj, 554, 778, \dodoi{10.1086/323147}

\bibitem[{{Liou} \& {Zook}(1999)}]{liou99}
{Liou}, J., \& {Zook}, H.~A. 1999, AJ, 118, 580, \dodoi{10.1086/300938}

\bibitem[{{Lipatov} \& {Brandt}(2020)}]{lipatov20}
{Lipatov}, M., \& {Brandt}, T.~D. 2020, \apj, 901, 100,
  \dodoi{10.3847/1538-4357/aba8f5}

\bibitem[{{Lissauer} \& {Stewart}(1993)}]{lissauer93}
{Lissauer}, J.~J., \& {Stewart}, G.~R. 1993, in Protostars and Planets III, ed.
  E.~H. {Levy} \& J.~I. {Lunine}, 1061

\bibitem[{{L{\"o}hne} {et~al.}(2017){L{\"o}hne}, {Krivov}, {Kirchschlager},
  {Sende}, \& {Wolf}}]{lohne17}
{L{\"o}hne}, T., {Krivov}, A.~V., {Kirchschlager}, F., {Sende}, J.~A., \&
  {Wolf}, S. 2017, \aap, 605, A7, \dodoi{10.1051/0004-6361/201630297}

\bibitem[{{MacGregor} {et~al.}(2017){MacGregor}, {Matr{\`a}}, {Kalas},
  {Wilner}, {Pan}, {Kennedy}, {Wyatt}, {Duchene}, {Hughes}, {Rieke}, {Clampin},
  {Fitzgerald}, {Graham}, {Holland}, {Pani{\'c}}, {Shannon}, \&
  {Su}}]{macgregor17_fom}
{MacGregor}, M.~A., {Matr{\`a}}, L., {Kalas}, P., {et~al.} 2017, \apj, 842, 8,
  \dodoi{10.3847/1538-4357/aa71ae}

\bibitem[{{Marshall} {et~al.}(2022){Marshall}, {Chavez-Dagostino},
  {Sanchez-Arguelles}, {Matr{\`a}}, {del Burgo}, {Kemper}, {Bertone}, {Dent},
  {Vega}, {Wilson}, {G{\'o}mez-Ruiz}, \& {Monta{\~n}a}}]{marshall22_vega_lmt}
{Marshall}, J.~P., {Chavez-Dagostino}, M., {Sanchez-Arguelles}, D., {et~al.}
  2022, \mnras, 514, 3815, \dodoi{10.1093/mnras/stac1510}

\bibitem[{{Matr{\`a}} {et~al.}(2019){Matr{\`a}}, {Wyatt}, {Wilner}, {Dent},
  {Marino}, {Kennedy}, \& {Milli}}]{matra19b}
{Matr{\`a}}, L., {Wyatt}, M.~C., {Wilner}, D.~J., {et~al.} 2019, \aj, 157, 135,
  \dodoi{10.3847/1538-3881/ab06c0}

\bibitem[{{Matr{\`a}} {et~al.}(2020){Matr{\`a}}, {Dent}, {Wilner}, {Marino},
  {Wyatt}, {Marshall}, {Su}, {Chavez}, {Hales}, {Hughes}, {Greaves}, \&
  {Corder}}]{matra20}
{Matr{\`a}}, L., {Dent}, W. R.~F., {Wilner}, D.~J., {et~al.} 2020, \apj, 898,
  146, \dodoi{10.3847/1538-4357/aba0a4}

\bibitem[{{McElwain} {et~al.}(2023){McElwain}, {Feinberg}, {Perrin}, {Clampin},
  {Mountain}, {Lallo}, {Lajoie}, {Kimble}, {Bowers}, {Stark}, {Acton},
  {Atkinson}, {Barinek}, {Barto}, {Basinger}, {Beck}, {Bergkoetter}, {Bluth},
  {Boucarut}, {Brady}, {Brooks}, {Brown}, {Byard}, {Carey}, {Carrasquilla},
  {Chae}, {Chaney}, {Chayer}, {Chonis}, {Cohen}, {Cole}, {Comeau}, {Coon},
  {Coppock}, {Coyle}, {Dean}, {Dziak}, {Eisenhower}, {Flagey}, {Franck},
  {Gallagher}, {Gilman}, {Glassman}, {Green}, {Grieco}, {Haase},
  {Hadjimichael}, {Hagopian}, {Hahn}, {Hartig}, {Havey}, {Hayden}, {Hellekson},
  {Hicks}, {Holfeltz}, {Howard}, {Huguet}, {Jahne}, {Johnson}, {Johnston},
  {Jurling}, {Kegley}, {Kennard}, {Keski-Kuha}, {Knight}, {Kulp}, {Levi},
  {Levine}, {Lightsey}, {Luetgens}, {Mather}, {Matthews}, {McKay}, {Mehalick},
  {Mel{\'e}ndez}, {Mosier}, {Murphy}, {Nelan}, {Niedner}, {Nol}, {Ohara},
  {Ohl}, {Olczak}, {Osborne}, {Park}, {Perrygo}, {Pueyo}, {Redding}, {Regan},
  {Reynolds}, {Rifelli}, {Rigby}, {Sabatke}, {Saif}, {Scorse}, {Seo}, {Shi},
  {Sigrist}, {Smith}, {Smith}, {Smith}, {Sohn}, {Stahl}, {Telfer}, {Terlecki},
  {Texter}, {Van Buren}, {Van Campen}, {Vila}, {Voyton}, {Waldman}, {Walker},
  {Weiser}, {Wells}, {West}, {Whitman}, {Wolf}, \& {Zielinski}}]{mcelwain2023}
{McElwain}, M.~W., {Feinberg}, L.~D., {Perrin}, M.~D., {et~al.} 2023, \pasp,
  135, 058001, \dodoi{10.1088/1538-3873/acada0}

\bibitem[{{Mennesson} {et~al.}(2014){Mennesson}, {Millan-Gabet}, {Serabyn},
  {Colavita}, {Absil}, {Bryden}, {Wyatt}, {Danchi}, {Defr{\`e}re}, {Dor{\'e}},
  {Hinz}, {Kuchner}, {Ragland}, {Scott}, {Stapelfeldt}, {Traub}, \&
  {Woillez}}]{mennesson2014}
{Mennesson}, B., {Millan-Gabet}, R., {Serabyn}, E., {et~al.} 2014, \apj, 797,
  119, \dodoi{10.1088/0004-637X/797/2/119}

\bibitem[{{Meshkat} {et~al.}(2018){Meshkat}, {Nilsson}, {Aguilar}, {Vasisht},
  {Oppenheimer}, {Su}, {Cady}, {Lockhart}, {Matthews}, {Dekany}, {Leisenring},
  {Ygouf}, {Mawet}, {Pueyo}, \& {Beichman}}]{meshkat18_vega}
{Meshkat}, T., {Nilsson}, R., {Aguilar}, J., {et~al.} 2018, \aj, 156, 214,
  \dodoi{10.3847/1538-3881/aae14f}

\bibitem[{{Monnier} {et~al.}(2012){Monnier}, {Che}, {Zhao}, {Ekstr{\"o}m},
  {Maestro}, {Aufdenberg}, {Baron}, {Georgy}, {Kraus}, {McAlister}, {Pedretti},
  {Ridgway}, {Sturmann}, {Sturmann}, {ten Brummelaar}, {Thureau}, {Turner}, \&
  {Tuthill}}]{monnier12_chara}
{Monnier}, J.~D., {Che}, X., {Zhao}, M., {et~al.} 2012, \apjl, 761, L3,
  \dodoi{10.1088/2041-8205/761/1/L3}

\bibitem[{{Moro-Mart{\'\i}n} \& {Malhotra}(2002)}]{moromartin02}
{Moro-Mart{\'\i}n}, A., \& {Malhotra}, R. 2002, \aj, 124, 2305,
  \dodoi{10.1086/342849}

\bibitem[{{Morrison} \& {Malhotra}(2015)}]{morrison_malhotra15}
{Morrison}, S., \& {Malhotra}, R. 2015, \apj, 799, 41,
  \dodoi{10.1088/0004-637X/799/1/41}

\bibitem[{{Myrvang} {et~al.}(2018){Myrvang}, {Baumann}, {Mann}, \&
  {Stamm}}]{myrvang18}
{Myrvang}, M., {Baumann}, C., {Mann}, I., \& {Stamm}, J. 2018, in EGU General
  Assembly Conference Abstracts, EGU General Assembly Conference Abstracts,
  9533

\bibitem[{{Najita} {et~al.}(2022){Najita}, {Kenyon}, \& {Bromley}}]{najita22}
{Najita}, J.~R., {Kenyon}, S.~J., \& {Bromley}, B.~C. 2022, \apj, 925, 45,
  \dodoi{10.3847/1538-4357/ac37b6}

\bibitem[{{Ozernoy} {et~al.}(2000){Ozernoy}, {Gorkavyi}, {Mather}, \&
  {Taidakova}}]{ozernoy00}
{Ozernoy}, L.~M., {Gorkavyi}, N.~N., {Mather}, J.~C., \& {Taidakova}, T.~A.
  2000, \apjl, 537, L147, \dodoi{10.1086/312779}

\bibitem[{{Pan} \& {Sari}(2005)}]{pan_sari05}
{Pan}, M., \& {Sari}, R. 2005, \icarus, 173, 342,
  \dodoi{10.1016/j.icarus.2004.09.004}

\bibitem[{{Pan} \& {Schlichting}(2012)}]{pan_schilichting12}
{Pan}, M., \& {Schlichting}, H.~E. 2012, \apj, 747, 113,
  \dodoi{10.1088/0004-637X/747/2/113}

\bibitem[{{Pawellek} {et~al.}(2014){Pawellek}, {Krivov}, {Marshall},
  {Montesinos}, {{\'A}brah{\'a}m}, {Mo{\'o}r}, {Bryden}, \&
  {Eiroa}}]{pawellek14}
{Pawellek}, N., {Krivov}, A.~V., {Marshall}, J.~P., {et~al.} 2014, \apj, 792,
  65, \dodoi{10.1088/0004-637X/792/1/65}

\bibitem[{{Pearce} {et~al.}(2022){Pearce}, {Launhardt}, {Ostermann}, {Kennedy},
  {Gennaro}, {Booth}, {Krivov}, {Cugno}, {Henning}, {Quirrenbach}, {Barcucci},
  {Matthews}, {Ruh}, \& {Stone}}]{pearce22}
{Pearce}, T.~D., {Launhardt}, R., {Ostermann}, R., {et~al.} 2022, \aap, 659,
  A135, \dodoi{10.1051/0004-6361/202142720}

\bibitem[{{Pearce} {et~al.}(2024){Pearce}, {Krivov}, {Sefilian}, {Jankovic}, {
  L{\"o}hne}, {Morgner}, {Wyatt}, {Booth}, \& {Marino}}]{pearce24_sharpness}
{Pearce}, T.~D., {Krivov}, A.~V., {Sefilian}, A.~A., {et~al.} 2024, \mnras,
  527, 3876, \dodoi{10.1093/mnras/stad3462}

\bibitem[{{Perrin} {et~al.}(2014){Perrin}, {Sivaramakrishnan}, {Lajoie},
  {Elliott}, {Pueyo}, {Ravindranath}, \& {Albert}}]{webbpsf_tool}
{Perrin}, M.~D., {Sivaramakrishnan}, A., {Lajoie}, C.-P., {et~al.} 2014, in
  Society of Photo-Optical Instrumentation Engineers (SPIE) Conference Series,
  Vol. 9143, Space Telescopes and Instrumentation 2014: Optical, Infrared, and
  Millimeter Wave, ed. J.~{Oschmann}, Jacobus~M., M.~{Clampin}, G.~G. {Fazio},
  \& H.~A. {MacEwen}, 91433X, \dodoi{10.1117/12.2056689}

\bibitem[{{Price} {et~al.}(2004){Price}, {Paxson}, {Engelke}, \&
  {Murdock}}]{price04}
{Price}, S.~D., {Paxson}, C., {Engelke}, C., \& {Murdock}, T.~L. 2004, \aj,
  128, 889, \dodoi{10.1086/422024}

\bibitem[{{Raymond} \& {Bonsor}(2014)}]{raymond_bonsor14}
{Raymond}, S.~N., \& {Bonsor}, A. 2014, \mnras, 442, L18,
  \dodoi{10.1093/mnrasl/slu048}

\bibitem[{{Reche} {et~al.}(2008){Reche}, {Beust}, {Augereau}, \&
  {Absil}}]{reche08}
{Reche}, R., {Beust}, H., {Augereau}, J.~C., \& {Absil}, O. 2008, \aap, 480,
  551, \dodoi{10.1051/0004-6361:20077934}

\bibitem[{{Reg{\'a}ly} {et~al.}(2018){Reg{\'a}ly}, {Dencs}, {Mo{\'o}r}, \&
  {Kov{\'a}cs}}]{regaly18_cavity}
{Reg{\'a}ly}, Z., {Dencs}, Z., {Mo{\'o}r}, A., \& {Kov{\'a}cs}, T. 2018,
  \mnras, 473, 3547, \dodoi{10.1093/mnras/stx2604}

\bibitem[{{Rieke} {et~al.}(2023){Rieke}, {Engelke}, {Su}, \&
  {Casagrande}}]{rieke23_absflxIII}
{Rieke}, G.~H., {Engelke}, C., {Su}, K., \& {Casagrande}, L. 2023, \aj, 165,
  99, \dodoi{10.3847/1538-3881/ac9f1b}

\bibitem[{{Rieke} {et~al.}(2016){Rieke}, {G{\'a}sp{\'a}r}, \&
  {Ballering}}]{rieke16_magnetictrapping}
{Rieke}, G.~H., {G{\'a}sp{\'a}r}, A., \& {Ballering}, N.~P. 2016, \apj, 816,
  50, \dodoi{10.3847/0004-637X/816/2/50}

\bibitem[{{Rieke} {et~al.}(2022){Rieke}, {Su}, {Sloan}, \&
  {Schlawin}}]{rieke22_sirius}
{Rieke}, G.~H., {Su}, K., {Sloan}, G.~C., \& {Schlawin}, E. 2022, \aj, 163, 45,
  \dodoi{10.3847/1538-3881/ac3b5d}

\bibitem[{{Rieke} {et~al.}(2015){Rieke}, {Wright}, {B{\"o}ker}, {Bouwman},
  {Colina}, {Glasse}, {Gordon}, {Greene}, {G{\"u}del}, {Henning}, {Justtanont},
  {Lagage}, {Meixner}, {N{\o}rgaard-Nielsen}, {Ray}, {Ressler}, {van Dishoeck},
  \& {Waelkens}}]{rieke15_miri}
{Rieke}, G.~H., {Wright}, G.~S., {B{\"o}ker}, T., {et~al.} 2015, \pasp, 127,
  584, \dodoi{10.1086/682252}

\bibitem[{{Rigley} \& {Wyatt}(2020)}]{rigley_wyatt20_prdrag}
{Rigley}, J.~K., \& {Wyatt}, M.~C. 2020, \mnras, 497, 1143,
  \dodoi{10.1093/mnras/staa2029}

\bibitem[{{Rigley} \& {Wyatt}(2022)}]{rigley_wyatt22_cometfragmentation}
---. 2022, \mnras, 510, 834, \dodoi{10.1093/mnras/stab3482}

\bibitem[{{Schneider} {et~al.}(2014){Schneider}, {Grady}, {Hines}, {Stark},
  {Debes}, {Carson}, {Kuchner}, {Perrin}, {Weinberger}, {Wisniewski},
  {Silverstone}, {Jang-Condell}, {Henning}, {Woodgate}, {Serabyn},
  {Moro-Martin}, {Tamura}, {Hinz}, \& {Rodigas}}]{schneider14}
{Schneider}, G., {Grady}, C.~A., {Hines}, D.~C., {et~al.} 2014, \aj, 148, 59,
  \dodoi{10.1088/0004-6256/148/4/59}

\bibitem[{{Sefilian} {et~al.}(2023){Sefilian}, {Rafikov}, \&
  {Wyatt}}]{sefilian23}
{Sefilian}, A.~A., {Rafikov}, R.~R., \& {Wyatt}, M.~C. 2023, \apj, 954, 100,
  \dodoi{10.3847/1538-4357/ace68e}

\bibitem[{{Sibthorpe} {et~al.}(2010){Sibthorpe}, {Vandenbussche}, {Greaves},
  {Pantin}, {Olofsson}, {Acke}, {Barlow}, {Blommaert}, {Bouwman}, {Brandeker},
  {Cohen}, {De Meester}, {Dent}, {di Francesco}, {Dominik}, {Fridlund}, {Gear},
  {Glauser}, {Gomez}, {Hargrave}, {Harvey}, {Henning}, {Heras}, {Hogerheijde},
  {Holland}, {Ivison}, {Leeks}, {Lim}, {Liseau}, {Matthews}, {Naylor},
  {Pilbratt}, {Polehampton}, {Regibo}, {Royer}, {Sicilia-Aguilar}, {Swinyard},
  {Waelkens}, {Walker}, \& {Wesson}}]{sibthorpe10}
{Sibthorpe}, B., {Vandenbussche}, B., {Greaves}, J.~S., {et~al.} 2010, \aap,
  518, L130, \dodoi{10.1051/0004-6361/201014574}

\bibitem[{{Smith} \& {Terrile}(1984)}]{smith84}
{Smith}, B.~A., \& {Terrile}, R.~J. 1984, Science, 226, 1421,
  \dodoi{10.1126/science.226.4681.1421}

\bibitem[{{Soummer} {et~al.}(2012){Soummer}, {Pueyo}, \& {Larkin}}]{soummer12}
{Soummer}, R., {Pueyo}, L., \& {Larkin}, J. 2012, \apjl, 755, L28,
  \dodoi{10.1088/2041-8205/755/2/L28}

\bibitem[{{Stark} \& {Kuchner}(2008)}]{stark08_resonantsignature}
{Stark}, C.~C., \& {Kuchner}, M.~J. 2008, \apj, 686, 637,
  \dodoi{10.1086/591442}

\bibitem[{{Stuber} {et~al.}(2023){Stuber}, {L{\"o}hne}, \& {Wolf}}]{stuber23}
{Stuber}, T.~A., {L{\"o}hne}, T., \& {Wolf}, S. 2023, \aap, 669, A3,
  \dodoi{10.1051/0004-6361/202243240}

\bibitem[{{Su} {et~al.}(2016){Su}, {Rieke}, {Defr{\'e}re}, {Wang}, {Lai},
  {Wilner}, {van Lieshout}, \& {Lee}}]{su16_fomalhaut}
{Su}, K. Y.~L., {Rieke}, G.~H., {Defr{\'e}re}, D., {et~al.} 2016, \apj, 818,
  45, \dodoi{10.3847/0004-637X/818/1/45}

\bibitem[{{Su} {et~al.}(2022){Su}, {Rieke}, {Marengo}, \&
  {Schlawin}}]{su22_SpitzerPSF}
{Su}, K. Y.~L., {Rieke}, G.~H., {Marengo}, M., \& {Schlawin}, E. 2022, \aj,
  163, 46, \dodoi{10.3847/1538-3881/ac3b5e}

\bibitem[{{Su} {et~al.}(2005){Su}, {Rieke}, {Misselt}, {Stansberry},
  {Moro-Martin}, {Stapelfeldt}, {Werner}, {Trilling}, {Bendo}, {Gordon},
  {Hines}, {Wyatt}, {Holland}, {Marengo}, {Megeath}, \& {Fazio}}]{su05}
{Su}, K.~Y.~L., {Rieke}, G.~H., {Misselt}, K.~A., {et~al.} 2005, \apj, 628,
  487, \dodoi{10.1086/430819}

\bibitem[{{Su} {et~al.}(2006){Su}, {Rieke}, {Stansberry}, {Bryden},
  {Stapelfeldt}, {Trilling}, {Muzerolle}, {Beichman}, {Moro-Martin}, {Hines},
  \& {Werner}}]{su06}
{Su}, K.~Y.~L., {Rieke}, G.~H., {Stansberry}, J.~A., {et~al.} 2006, \apj, 653,
  675, \dodoi{10.1086/508649}

\bibitem[{{Su} {et~al.}(2013){Su}, {Rieke}, {Malhotra}, {Stapelfeldt},
  {Hughes}, {Bonsor}, {Wilner}, {Balog}, {Watson}, {Werner}, \&
  {Misselt}}]{su13}
{Su}, K.~Y.~L., {Rieke}, G.~H., {Malhotra}, R., {et~al.} 2013, \apj, 763, 118,
  \dodoi{10.1088/0004-637X/763/2/118}

\bibitem[{{Tabeshian} \& {Wiegert}(2016)}]{tabeshian16}
{Tabeshian}, M., \& {Wiegert}, P.~A. 2016, \apj, 818, 159,
  \dodoi{10.3847/0004-637X/818/2/159}

\bibitem[{{Thebault} {et~al.}(2012){Thebault}, {Kral}, \&
  {Ertel}}]{thebault12_planetsignature_collisionallyactivedisks}
{Thebault}, P., {Kral}, Q., \& {Ertel}, S. 2012, \aap, 547, A92,
  \dodoi{10.1051/0004-6361/201219962}

\bibitem[{{Thureau} {et~al.}(2014){Thureau}, {Greaves}, {Matthews}, {Kennedy},
  {Phillips}, {Booth}, {Duch{\^e}ne}, {Horner}, {Rodriguez}, {Sibthorpe}, \&
  {Wyatt}}]{thureau14}
{Thureau}, N.~D., {Greaves}, J.~S., {Matthews}, B.~C., {et~al.} 2014, \mnras,
  445, 2558, \dodoi{10.1093/mnras/stu1864}

\bibitem[{{Tokunaga}(1984)}]{tokunaga84}
{Tokunaga}, A.~T. 1984, \aj, 89, 172, \dodoi{10.1086/113497}

\bibitem[{{van Lieshout} {et~al.}(2014){van Lieshout}, {Dominik}, {Kama}, \&
  {Min}}]{vanlieshout14_dustsublimation}
{van Lieshout}, R., {Dominik}, C., {Kama}, M., \& {Min}, M. 2014, \aap, 571,
  A51, \dodoi{10.1051/0004-6361/201322090}

\bibitem[{{Walker} \& {Heinrichsen}(2000)}]{walker2000}
{Walker}, H.~J., \& {Heinrichsen}, I. 2000, \icarus, 143, 147,
  \dodoi{10.1006/icar.1999.6235}

\bibitem[{{Wisdom}(1980)}]{wisdom80}
{Wisdom}, J. 1980, \aj, 85, 1122, \dodoi{10.1086/112778}

\bibitem[{{Wolff} {et~al.}(2024){Wolff}, {Gaspar}, {Rieke}, {Leisenring}, \&
  {Others}}]{wolff24}
{Wolff}, S.~G., {Gaspar}, A., {Rieke}, G.~H., {Leisenring}, J.~M., \& {Others}.
  2024, \apj, submitted

\bibitem[{{Wright} {et~al.}(2023){Wright}, {Rieke}, {Glasse}, {Ressler},
  {Garc{\'\i}a Mar{\'\i}n}, {Aguilar}, {Alberts}, {{\'A}lvarez-M{\'a}rquez},
  {Argyriou}, {Banks}, {Baudoz}, {Boccaletti}, {Bouchet}, {Bouwman}, {Brandl},
  {Breda}, {Bright}, {Cale}, {Colina}, {Cossou}, {Coulais}, {Cracraft}, {De
  Meester}, {Dicken}, {Engesser}, {Etxaluze}, {Fox}, {Friedman}, {Fu},
  {Gasman}, {G{\'a}sp{\'a}r}, {Gastaud}, {Geers}, {Glauser}, {Gordon},
  {Greene}, {Greve}, {Grundy}, {G{\"u}del}, {Guillard}, {Haderlein},
  {Hashimoto}, {Henning}, {Hines}, {Holler}, {Detre}, {Jahromi}, {James},
  {Jones}, {Justtanont}, {Kavanagh}, {Kendrew}, {Klaassen}, {Krause},
  {Labiano}, {Lagage}, {Lambros}, {Larson}, {Law}, {Lee}, {Libralato}, {Lorenzo
  Alverez}, {Meixner}, {Morrison}, {Mueller}, {Murray}, {Mycroft}, {Myers},
  {Nayak}, {Naylor}, {Nickson}, {Noriega-Crespo}, {{\"O}stlin}, {O'Sullivan},
  {Ottens}, {Patapis}, {Penanen}, {Pietraszkiewicz}, {Ray}, {Regan},
  {Roteliuk}, {Royer}, {Samara-Ratna}, {Samuelson}, {Sargent}, {Scheithauer},
  {Schneider}, {Schreiber}, {Shaughnessy}, {Sheehan}, {Shivaei}, {Sloan},
  {Tamas}, {Teague}, {Temim}, {Tikkanen}, {Tustain}, {van Dishoeck},
  {Vandenbussche}, {Weilert}, {Whitehouse}, \& {Wolff}}]{wright23_miri}
{Wright}, G.~S., {Rieke}, G.~H., {Glasse}, A., {et~al.} 2023, \pasp, 135,
  048003, \dodoi{10.1088/1538-3873/acbe66}

\bibitem[{{Wyatt} \& {Jackson}(2016)}]{wyatt_jackson16}
{Wyatt}, M.~C., \& {Jackson}, A.~P. 2016, Space Sci. Rev.,
  \dodoi{10.1007/s11214-016-0248-1}

\end{thebibliography}
\end{document}